\documentclass[11pt,english]{article}
\usepackage[latin9]{inputenc}
\usepackage{float}
\usepackage{amsmath}
\usepackage{amssymb}
\usepackage{graphicx}
\usepackage{setspace}
\usepackage{subscript}
\makeatletter

\providecommand{\tabularnewline}{\\}

\@ifundefined{date}{}{\date{}}
\usepackage{esint}
\usepackage{hyperref}
\usepackage{latexsym}
\usepackage{graphicx}\usepackage{bm}\usepackage{longtable}

\usepackage{xcolor}

\@addtoreset{equation}{section}

\makeatother

\usepackage{babel}
\begin{document}
\title{Holographic Schwinger effect and electric instability with anisotropy}
\maketitle
\begin{center}
Si-wen Li\footnote{Email: siwenli@dlmu.edu.cn}, Sen-kai Luo\footnote{Email: luosenkai@dlmu.edu.cn},
Hao-qian Li\footnote{Email: lihaoqian@dlmu.edu.cn},
\par\end{center}

\begin{center}
\emph{Department of Physics, School of Science,}\\
\emph{Dalian Maritime University, }\\
\emph{Dalian 116026, China}\\
\par\end{center}

\vspace{8mm}

\begin{abstract}
According to the gauge-gravity duality, we systematically study the
Schwinger effect and electric instability with anisotropy in a top-down
holographic approach. The anisotropic black brane and bubble (soliton)
background in IIB supergravity are employed and the dual theories
in these backgrounds are expected to be anisotropic theory at finite
temperature and anisotropic theory with confinement respectively.
Then performing the potential analysis, we find due to the anisotropy,
the potential barrier behaves oppositely with parallel and perpendicular
electric fields, and this behavior agrees with the previous study
about the quark potential with anisotropy in this system. Afterwards,
we evaluate the pair production rate by solving the equation of motion
for a fundamental string numerically which reveals the consistent
behavior with the potential analysis. Furthermore, the probe D7-brane
as flavor is introduced into the bulk in order to investigate the
electric instability. The vacuum decay rate can be obtained by evaluating
the imaginary part of the D7-brane action which again agrees with
our potential analysis. Solving the associated constraint of gauge
field strength on the flavor brane, we finally obtain the V-A curve
displaying the distinct behavior of the conductivity in parallel and
perpendicular direction which is in agreement with some bottom-up
and phenomenologically holographic approaches in anisotropic fluid.
Accordingly, we believe this work may be remarkable to study the electric
features in strongly coupled anisotropic system.
\end{abstract}
\newpage{}

\tableofcontents{}

\section{Introduction}

There are many interesting phenomena in the heavy-ion collision experiment,
especially under the strong electromagnetic field created by the collision
of the charged particles. The Schwinger effect is one of the most
significant phenomenon in the heavy-ion collision since it relates
to the particle production rate. Specifically, due to the high speed
of the charged particles, an extremely strong electromagnetic field
would be generated in the collision so that virtual pairs of particles
in the vacuum are able to be excited by the electromagnetic field
to become real particles \cite{key-1,key-2}. In this sense, the Schwinger
effect may be remarkable to understand the process of particle creation
and thermalization in the heavy-ion collision. On the other hand,
many researches reveal that the product in the heavy-ion collision,
i.e. the quark-gluon plasma (QGP), is strongly coupled \cite{key-3,key-4}
with anisotropic pressure \cite{key-5,key-6,key-7,key-8}. So when
the collision occurs, it forms highly dense and anisotropic situation
with extremely strong electromagnetic field, therefore the Schwinger
effect would appear and be affected by the anisotropy. In this sense,
investigating the Schwinger effect in an anisotropic dense situation
would be very constructive to the experiment of heavy-ion collision.

However the calculations in \cite{key-1,key-2} imply Schwinger effect
is non-perturbative, thus it is very challenging to evaluate Schwinger
effect in the framework of quantum field theory (QFT) strictly. Fortunately,
the gauge-gravity duality and AdS/CFT may provide an alternative method
to study the strongly coupled QFT \cite{key-9,key-10}, in particular,
to evaluate Schwinger effect in holography \cite{key-11,key-12,key-13,key-14,key-15,key-16,key-17,key-18,key-19}.
Since the equivalence between the type IIB super string theory on
$\mathrm{AdS_{5}}\times S^{5}$ and the four-dimensional $\mathcal{N}=4$
$SU\left(N_{c}\right)$ super Yang-Mills theory on $N_{c}$ D3-branes
is the most famous example in gauge-gravity duality, the authors of
\cite{key-20,key-21} construct the solution in type IIB supergravity
then obtain anisotropic black brane solution and bubble solution as
gravity soliton in order to study anisotropic gauge theory at finite
temperature and with confinement respectively. Accordingly, one of
the motivation of this work is to investigate holographically the
Schwinger effect with anisotropy in the backgrounds proposed in \cite{key-20,key-21}.
And it would also be interesting to compare our work with the holographic
anisotropic Schwinger effect in some bottom-up approaches e.g. \cite{key-21+1}.
So in the first part of this work, we compute the critical electric
field for Schwinger effect, perform the potential analysis then evaluate
the pair production rate numerically by solving the equation of motion
for a fundamental string in bulk with respect to the black brane background
and bubble background. Importantly, due to the anisotropy in the background
geometry, we need to deal with the cases with parallel and perpendicular
electric field respectively. Our numerical results display that, in
the presence of the anisotropy, the potential barrier is suppressed
with the parallel electric field while it is enhanced with the perpendicular
electric field. Thus the pair production rate is increased/decreased
by the anisotropy with the parallel/perpendicular electric field correspondingly.
And this conclusion qualitatively agrees with the analysis of the
quark potential in this system \cite{key-21} and partly, with the
bottom-up approach in \cite{key-21+1}.

In addition, we consider the probe D7-branes as flavor embedding into
the bulk geometry to explore the vacuum electric instability in holography
as \cite{key-22,key-23}. Since the action of the flavor brane corresponds
to the effective flavored Lagrangian in the dual theory, it is allowed
to evaluate the vacuum decay rate by $\left\langle 0\left|U\left(t\right)\right|0\right\rangle \sim e^{i\int\mathcal{L}d^{4}x}$
(where $U\left(t\right)$ refers to the time-evolution operator) for
Schwinger effect once an imaginary part of the Lagrangian can be obtained
\cite{key-23} \footnote{There are also some works with such a holographic setup by including
the magnetic field, or in the D4/D8 approach, as \cite{key-24,key-25,key-26}}. Keeping these in hand, we turn on the parallel and perpendicular
electric field on the flavor brane with respect to the anisotropy,
then evaluate the vacuum decay rate in the black brane and bubble
background respectively. Afterwards, we solve the constraint of the
electric charge, stable current and electric field, thus obtain the
relation of the stable current and electric field numerically as the
V-A curve in this holographic approach. The V-A curve illustrates
the conductivity in perpendicular direction is always smaller than
the conductivity in the parallel direction in an anisotropic situation.
And this conclusion is in good agreement with the phenomenological
approaches in the anisotropic background \cite{key-27,key-28}. The
interpretation of this conductivity behavior can be obtained by analyzing
the transport properties of the anisotropic fluid as \cite{key-20,key-27}.
Therefore, this work provides a parallel top-down holographic approach
to study the electric feature in strongly coupled anisotropic system.

The outline of this manuscript is as follows. In Section 2, we review
the anisotropic solutions in IIB supergravity proposed in \cite{key-20,key-21}.
In Section 3, we evaluate the critical electric field then perform
the potential analysis for Schwinger effect. In Section 4, we numerically
solve the equation of motion for a fundamental string in bulk with
circular trajectory at boundary as the Wilson loop, then evaluate
the pair production by computing the onshell action of the string.
In Section 5, we introduce probe D7-brane as flavor, evaluate the
imaginary part of its Lagrangian, obtain the V-A curve by solving
the associated constraints. Summary and discussion are given in the
final section. Besides, we list the analytical formulas of the functions
$\mathcal{F},\mathcal{B},\phi$ (presented in the bulk metric) in
the appendix which are very useful for our calculation.

\section{The anisotropic solution in IIB supergravity}

In this section, we will briefly review the anisotropic solutions
in IIB supergravity which are the black brane solution and the bubble
solution respectively, then outline the dual theory in both gravity
backgrounds. 

\subsection{The black brane solution}

Let us start with the action of type IIB supergravity. In string frame,
it reads,

\begin{equation}
S_{\mathrm{IIB}}=\frac{1}{2\kappa_{10}^{2}}\int d^{10}x\sqrt{-g}\left[e^{-2\phi}\left(\mathcal{R}+4\partial_{M}\phi\partial^{M}\phi\right)-\frac{1}{2}F_{1}^{2}-\frac{1}{4\cdot5!}F_{5}^{2}\right].\label{eq:1}
\end{equation}
This action contains the dynamics of graviton in $g_{\mu\nu}$, dilaton
$\phi$ and Ramond-Ramond forms $C_{0,4}$ with their field strength
$F_{1,5}=dC_{0,4}$ which are the massless bosonic excitations in
the IIB superstring theory. We note that the index $M$ runs over
0 to 9 and $\kappa_{10}$ is the ten-dimensional gravitational coupling
constant $2\kappa_{10}^{2}=\left(2\pi\right)^{7}l_{s}^{8}$. Solving
the equations of motion associated to (\ref{eq:1}), an anisotropic
solution with non-vanished axion field is obtained in \cite{key-20}
as,

\begin{align}
ds^{2} & =\frac{L^{2}}{u^{2}}\left(-\mathcal{F}\mathcal{B}dt^{2}+dx^{2}+dy^{2}+\mathcal{H}dz^{2}+\frac{du^{2}}{\mathcal{F}}\right)+L^{2}\mathcal{Z}d\Omega_{5}^{2},\nonumber \\
F_{1} & =d\chi,\ \chi=az,\ F_{5}=dC_{4}=\frac{4}{L}\left(\Omega_{S^{5}}+\star\Omega_{S^{5}}\right),\nonumber \\
\mathcal{H} & =e^{-\phi},\ \mathcal{Z}=e^{\frac{1}{2}\phi},\label{eq:2}
\end{align}
where $\chi=C_{0}$ is named as the axion field and $\Omega_{5}$
refers to the unit volume form of a five-sphere $S^{5}$. This solution
describes the dynamics of $N_{c}$ coincident D3-branes with $N_{\mathrm{D7}}$
D7-branes dissolved in the bulk in the large $N_{c}$ limit and the
configuration of the D-branes in this system is given in Table \ref{tab:1}.
\begin{table}[h]
\begin{centering}
\begin{tabular}{|c|c|c|c|c|c|c|}
\hline 
Black brane solution & $t$ & $x$ & $y$ & $z$ & $u$ & $\Omega_{5}$\tabularnewline
\hline 
\hline 
$N_{c}$ D3-branes & - & - & - & - &  & \tabularnewline
\hline 
$N_{\mathrm{D7}}$ D7-branes & - & - & - &  &  & -\tabularnewline
\hline 
\end{tabular}
\par\end{centering}
\caption{\label{tab:1} The configuration of the D-branes in the black brane
solution. ``-'' represents the D-brane extends along the direction.}
\end{table}
 The regular functions $\mathcal{F},\mathcal{B},\phi$ depend on the
radial coordinate $u$ and anisotropy $a$ which in general are non-analytical
functions. Noticeably, there is a horizon located at $u=u_{H}$ i.e.
$\mathcal{F}\left(u_{H}\right)=0$ and the holographic boundary is
located at $u=0$. Remarkably, this solution exhibits the anisotropy
in $z$ direction because $g_{xx}=g_{yy}\neq g_{zz}$. The parameters
in this solution are given as,

\begin{equation}
L^{4}=4\pi g_{s}N_{c}l_{s}^{4}=\lambda l_{s}^{4},a=\frac{\lambda n_{\mathrm{D7}}}{4\pi N_{c}},\label{eq:3}
\end{equation}
where $g_{s},L,\lambda$ refers to the string coupling, the radius
of the bulk and the 't Hooft coupling constant respectively. The constant
$n_{\mathrm{D7}}=dN_{\mathrm{D7}}/dz$ is the distribution density
of the $N_{\mathrm{D7}}$ D7-branes which are distributed along $z$
direction. Since the dynamic of the axion field $\chi$, which is
magnetically couples to $N_{\mathrm{D7}}$ D7-branes, is involved
in the bulk, in the large $N_{c}$ limit we have $N_{c}\rightarrow\infty$
while $N_{\mathrm{D7}}/N_{c}\propto a$ is fixed in this system. The
Hawking temperature $T$ can be obtained by avoiding the conical singularities
in the bulk. Resultantly its formula is given as,

\begin{equation}
\delta t_{E}=\frac{4\pi}{\mathcal{F}_{1}\sqrt{\mathcal{B}\left(u_{H}\right)}}=\frac{1}{T},\ \mathcal{F}_{1}=-\frac{d\mathcal{F}}{du}\bigg|_{u=u_{H}}.\label{eq:4}
\end{equation}

It would be difficult to do calculations with (\ref{eq:2}) since
in general the functions $\mathcal{F},\mathcal{B},\phi$ are not analytical.
However, fortunately if the temperature $T$ is sufficiently high
i.e. $T\gg a$, the functions $\mathcal{F},\mathcal{B},\phi$ can
be written analytically as the series of $a$ which are given in the
appendix. On the other hand, since Schwinger effect in the heavy-ion
collision experiment usually occurs in the situation with very high
temperature, it is suitable to employ the analytical formulas of $\mathcal{F},\mathcal{B},\phi$.
And analytical formulas of $\mathcal{F},\mathcal{B},\phi$ would also
simplify our calculation greatly.

\subsection{The bubble solution}

In the gauge-gravity duality, the bubble solution, as the gravity
soliton, can be obtained by performing the double Wick rotation and
compactification to the black brane solution as it is discussed in
the famous \cite{key-29,key-30}. Accordingly, the bubble solution
associated to (\ref{eq:2}) reads \cite{key-21},

\begin{align}
ds^{2} & =\frac{L^{2}}{u^{2}}\left(-dt^{2}+dx^{2}+\mathcal{H}dy^{2}+\mathcal{F}\mathcal{B}dz^{2}+\frac{du^{2}}{\mathcal{F}}\right)+L^{2}\mathcal{Z}d\Omega_{5}^{2},\nonumber \\
F_{1} & =d\chi,\ \chi=ay,\ F_{5}=dC_{4}=\frac{4}{L}\left(\Omega_{S^{5}}+\star\Omega_{S^{5}}\right),\nonumber \\
\mathcal{H} & =e^{-\phi},\ \mathcal{Z}=e^{\frac{1}{2}\phi},\label{eq:5}
\end{align}
where $z$ direction is compactified on a circle $S^{1}$ with a period
$z\sim z+\delta z$,

\begin{equation}
\delta z=\frac{4\pi}{\mathcal{F}_{1}\sqrt{\mathcal{B}\left(u_{KK}\right)}}=\frac{2\pi}{M_{KK}}.\label{eq:6}
\end{equation}
To distinguish the bubble solution (\ref{eq:5}) from the black brane
solution (\ref{eq:2}), we have renamed $u_{H}$ as $u_{KK}$ and
the configuration of the D-branes in the bubble solution is given
in Table \ref{tab:2}. 
\begin{table}
\begin{centering}
\begin{tabular}{|c|c|c|c|c|c|c|}
\hline 
Bubble solution & $t$ & $x$ & $y$ & $\left(z\right)$ & $u$ & $\Omega_{5}$\tabularnewline
\hline 
\hline 
$N_{c}$ D3-branes & - & - & - & - &  & \tabularnewline
\hline 
$N_{\mathrm{D7}}$ D7-branes & - & - &  & - &  & -\tabularnewline
\hline 
\end{tabular}
\par\end{centering}
\caption{\label{tab:2} The configuration of the D-branes in the bubble solution
(\ref{eq:5}).}
\end{table}
 The bubble solution (\ref{eq:5}) describes the Cigar-like bulk geometry
defining only in $u\in\left(0,u_{KK}\right)$, so there is not a horizon
in this solution. Since $g_{tt}$ never goes to zero, the dual theory
would exhibit confinement according to the behavior of the Wilson
loop in this geometry. As the type IIB supergravity theory corresponds
to the $\mathcal{N}=4$ super Yang-Mills theory on D3-brane in holography,
the super Yang-Mills theory on D3-brane would become effectively three-dimensional
below the energy scale $M_{KK}$. Moreover, the conformal symmetry
and supersymmetry in the super Yang-Mills theory would break down
once the periodic and anti-periodic boundary condition is imposed
respectively to the bosonic and fermionic fields along $S^{1}$. In
a word, the dual theory becomes three-dimensional confining Yang-Mills
theory after performing the compactification in \cite{key-29}. Note
that in the bubble solution, the dual theory is defined at zero temperature
limit since the period of $t$ is infinity. And the bubble solution
would be very useful to study the Schwinger effect in a confining
system as \cite{key-14,key-16,key-17,key-19}.

A worthy notable feature here is that if we remain to employ the analytical
formulas of $\mathcal{F},\mathcal{B},\phi$ in the bubble solution,
it refers to the case that the size of the compactified direction
$z$ trends to be vanished so that the dual theory becomes exactly
three-dimensional. The reason is that $z$ direction in the bubble
solution (\ref{eq:5}) corresponds exactly to the $t$ direction in
the black brane solution (\ref{eq:2}) by the double Wick rotation.
That is why they satisfy same condition (\ref{eq:4}), (\ref{eq:6})
in order to avoid conical singularities. Therefore, one may find the
metric with high temperature $T\gg a$ in the black brane solution
corresponds exactly to the case of $M_{KK}\gg a$ in the bubble solution.
That means the dual theory will become exactly three-dimensional if
we take $M_{KK}\rightarrow\infty$ (i.e. $\delta z\rightarrow0$).
Accordingly, we can remain to employ the analytical formulas of $\mathcal{F},\mathcal{B},\phi$
in the bubble solution because the dual theory is expected definitely
to be three-dimensional via holography.

\subsection{The effective action in the dual theory and its anisotropy}

Based on the gauge-gravity duality, the action for the dual theory
can be examined by introducing a probe D3-brane located at the holographic
boundary $u=u_{0}$ with $u_{0}\rightarrow0$. Consider the action
for such a probe D3-brane which is \footnote{The bosonic action is enough since the supersymmetry has been broken
down in the background (\ref{eq:2}) (\ref{eq:5}).},

\begin{equation}
S_{\mathrm{D3}}=-T_{3}\mathrm{Tr}\int d^{3}xdze^{-\phi}\sqrt{-\det\left(g_{\mu\nu}+\mathcal{F}_{\mu\nu}\right)}+\frac{1}{2}T_{3}\mathrm{Tr}\int\chi\mathcal{F}\wedge\mathcal{F},\label{eq:7}
\end{equation}
where $T_{3}=\left(2\pi\right)^{-3}l_{s}^{-4}g_{s}^{-1}$ is the tension
of the D3-brane and $g_{\mu\nu},\mathcal{F}_{\mu\nu}=2\pi\alpha^{\prime}F_{\mu\nu}$
is the induced metric, the gauge field strength respectively on the
probe D3-brane. Impose the black brane solution (\ref{eq:2}) to (\ref{eq:7})
and expand the action up to quadratic order, we obtain,

\begin{equation}
S_{\mathrm{D3}}\simeq-\frac{N_{c}}{4\lambda}\mathrm{Tr}\int_{\mathbb{R}^{3+1}}d^{4}xF_{\mu\nu}^{2}+\frac{1}{32\pi^{2}}\mathrm{Tr}\int_{\mathbb{R}^{3+1}}\theta F\wedge F,\label{eq:8}
\end{equation}
where $\theta=\theta\left(z\right)=az$ is the $z$-depended theta
angle. Notice that the metric is anisotropic, so the dual theory in
the black brane solution is expected to be four-dimensional anisotropic
Yang-Mills theory with a theta term at finite temperature. 

Furthermore, when the bubble solution is imposed to (\ref{eq:7}),
the $z$-dependence can be integrated out by part leading to,

\begin{align}
S_{\mathrm{D3}} & \simeq-\frac{N_{c}}{4\lambda_{3}}\mathrm{Tr}\int_{\mathbb{R}^{2+1}}d^{3}xF_{ab}^{2}-\frac{N_{\mathrm{D7}}}{4\pi}\mathrm{Tr}\int_{\mathbb{R}^{2+1}}\omega_{3},\ a,b=0,1,2,
\end{align}
where 

\begin{equation}
\omega_{3}=A\wedge dA+\frac{2}{3}A\wedge A\wedge A,
\end{equation}
is the Chern-Simons three-form and $\lambda_{3}=\lambda M_{KK}/\left(2\pi\right)$
is the three-dimensional 't Hooft coupling constant because the dual
theory in the bubble solution is expected to be three-dimensional.
Altogether, the dual theory in the bubble solution is three-dimensional
anisotropic and confining Yang-Mills theory at zero temperature limit
with a Chern-Simons term.

Besides, based on the fluid-gravity correspondence, the supergravity
solution of the metric (\ref{eq:2}) (\ref{eq:5}) also describe the
hydrodynamics of the dual theory with anisotropic pressure \cite{key-20}\footnote{As most discussion about gauge-gravity duality and fluid-gravity correspondence,
the black brane background describes the strongly coupled plasma in
holography e.g. QGP. However, the fluid description of the dual theory
corresponding to the bubble background would be less clear since it
is a confining theory. Nonetheless, we can obtain the stress tensor
of the dual theory corresponding to the bubble background which takes
the same formula as the black brane, since the bubble background is
obtained by a double Wick rotation on the black brane background.
So the bubble background also describe anisotropic situation in the
dual theory.}. This could be confirmed by reviewing the formula of the holographically
renormalized stress tensor of the dual theory which is given as \cite{key-20,key-29+1},

\begin{equation}
\left\langle T_{\mu\nu}\right\rangle =\frac{2}{\kappa_{10}^{2}}\left[g_{\mu\nu}^{\left(4\right)}-\mathrm{Tr}g^{\left(4\right)}\eta_{\mu\nu}+\frac{c_{\mathrm{sch}}}{2}\tilde{g}_{\mu\nu}^{\left(4\right)}+...\right],
\end{equation}
where the dots represent the terms independent on the metric and $c_{\mathrm{sch}}$
is a schedule-dependent number. Here $g_{\mu\nu}^{\left(4\right)},\tilde{g}_{\mu\nu}^{\left(4\right)}$
refer to the expansion coefficients of the asymptotic metric near
the boundary on the Fefferman-Graham coordinate. And their exact formulas
are given in \cite{key-20,key-29+1}. As the diagonal components of
$T_{\mu\nu}$ represent the energy $E$ and the spatial pressure $P_{i}$
in the dual theory, we have 

\begin{equation}
\mathrm{diag}\left\langle T_{\mu\nu}\right\rangle =\left(E,P_{i}\right).
\end{equation}
So due to the anisotropy presented in the bulk metric (\ref{eq:2})
(\ref{eq:5}), the pressure in the stress tensor of the dual theory
is also anisotropic. Thus it can be calculated by using (\ref{eq:2})
(\ref{eq:5}) as \cite{key-20} (up to $\mathcal{O}\left(a^{4}\right)$
terms),

\begin{align}
P_{\parallel} & =\frac{N_{c}^{2}}{2\pi^{2}}\left(-\frac{1}{4}\mathcal{F}_{4}-\frac{5}{28}\mathcal{B}_{4}+\frac{611}{16128}a^{4}-\frac{c_{\mathrm{sch}}}{96}a^{4}\right),\nonumber \\
P_{\perp} & =\frac{N_{c}^{2}}{2\pi^{2}}\left(-\frac{1}{4}\mathcal{F}_{4}-\frac{13}{28}\mathcal{B}_{4}+\frac{2227}{16128}a^{4}+\frac{c_{\mathrm{sch}}}{32}a^{4}\right),\label{eq:13}
\end{align}
where in the black brane background, ``$\parallel$'' refers to
the $x,y$ direction and ``$\perp$'' refers to the $z$ direction,
in the bubble background, ``$\parallel$'' refers to the $x$ direction
and ``$\perp$'' refers to the $y$ direction. And $\mathcal{F}_{4},\mathcal{B}_{4}$
are numerical numbers which have to be determined by the equations
of motion for the bulk fields \cite{key-20}. Therefore the pressure
of the dual theory is obviously anisotropic because of $P_{\parallel}\neq P_{\perp}$.

\section{Potential analysis }

In this section, we are going to evaluate the holographic potential
for Schwinger effect according to the AdS/CFT dictionary \cite{key-11,key-12,key-14,key-16}.
Specifically, we need to compute the critical electric field, its
associated separation and the potential behavior with the anisotropy
in the black brane and bubble background respectively. Particularly,
since the background geometry is anisotropic, we need to analyze the
case that the electric field is parallel (denoted by ``$\parallel$'')
and perpendicular (denoted by ``$\perp$'') to the $N_{\mathrm{D7}}$
D7-branes in the bulk. 

\subsection{The critical electric field}

The critical electric field in the dual theory can be evaluated by
considering the classical action for a probe D3-brane located at $u=u_{0}$
with $u_{0}\rightarrow0$ which has been given in (\ref{eq:7}). Accordingly,
it would be easy to understand the following constraint

\begin{equation}
\det\left(g_{\mu\nu}+\mathcal{F}_{\mu\nu}\right)<0,
\end{equation}
should be satisfied in (\ref{eq:7}), otherwise the classical action
will become imaginary. This constraint reduces to a critical value
of the electric field. On the other hand, since the metric is anisotropic,
we turn on the nonzero components of the field strength as $F_{0i}=E_{i},i=1,2,3$
with $E_{\perp}^{2}=E_{3}^{2},E_{\parallel}^{2}=E_{1}^{2}+E_{2}^{2}$
in the black brane background. In the bubble background, we set the
nonzero components of the field strength as $F_{01}=E_{\parallel},F_{02}=E_{\perp}$.
Keeping these in mind then imposing the background geometry (\ref{eq:2})
and (\ref{eq:5}) to the action (\ref{eq:7}) respectively, so in
the black brane background, the action could be written as, 

\begin{align}
S_{\mathrm{D3}} & =-T_{\mathrm{D3}}\int d^{4}xe^{-\phi}\sqrt{-\det\left(g_{\mu\nu}+2\pi\alpha^{\prime}F_{\mu\nu}\right)}\nonumber \\
 & =-T_{\mathrm{D3}}\int d^{4}xe^{-\phi\left(u_{0}\right)}\frac{L^{2}}{u_{0}^{2}}\sqrt{\frac{\mathcal{B}\left(u_{0}\right)\mathcal{F}\left(u_{0}\right)\mathcal{H}\left(u_{0}\right)L^{4}}{u_{0}^{4}}-\left(2\pi\alpha^{\prime}\right)^{2}E_{\perp}^{2}-\left(2\pi\alpha^{\prime}\right)^{2}E_{\parallel}^{2}\mathcal{H}\left(u_{0}\right)},
\end{align}
thus the critical electric field is evaluated as\footnote{We evaluate the critical electric field in parallel and perpendicular
case respectively. That means when we evaluate $E_{\parallel}^{c}$,
$E_{\perp}$ is turned off and vice versa. },

\begin{equation}
E_{\parallel}^{c}=\frac{\sqrt{\mathcal{B}\left(u_{0}\right)\mathcal{F}\left(u_{0}\right)}}{2\pi\alpha^{\prime}}\frac{L^{2}}{u_{0}^{2}},\ E_{\perp}^{c}=\frac{\sqrt{\mathcal{B}\left(u_{0}\right)\mathcal{F}\left(u_{0}\right)\mathcal{H}\left(u_{0}\right)}}{2\pi\alpha^{\prime}}\frac{L^{2}}{u_{0}^{2}}.\label{eq:16}
\end{equation}
Similarly, in the bubble background action (\ref{eq:7}) becomes,

\begin{align}
S_{\mathrm{D3}} & =-T_{\mathrm{D3}}\int d^{4}xe^{-\phi}\sqrt{-\det\left(g_{\mu\nu}+2\pi\alpha^{\prime}F_{\mu\nu}\right)}\nonumber \\
 & =-T_{\mathrm{D3}}\int d^{4}xe^{-\phi\left(u_{0}\right)}\mathcal{B}\left(u_{0}\right)^{1/2}\mathcal{F}\left(u_{0}\right)^{1/2}\frac{L^{2}}{u_{0}^{2}}\sqrt{\frac{L^{4}}{u_{0}^{4}}\mathcal{H}\left(u_{0}\right)-\left(2\pi\alpha^{\prime}E_{\perp}\right)^{2}-\left(2\pi\alpha^{\prime}E_{\parallel}\right)^{2}\mathcal{H}\left(u_{0}\right)},
\end{align}
then the critical electric field is evaluated as,

\begin{equation}
E_{\perp}^{c}=\frac{1}{2\pi\alpha^{\prime}}\frac{L^{2}}{u_{0}^{2}}\sqrt{\mathcal{H}\left(u_{0}\right)},\ E_{\parallel}^{c}=\frac{1}{2\pi\alpha^{\prime}}\frac{L^{2}}{u_{0}^{2}}.\label{eq:18}
\end{equation}

\subsection{The separation and the holographic potential}

\begin{figure}[h]
\begin{centering}
\includegraphics[scale=0.3]{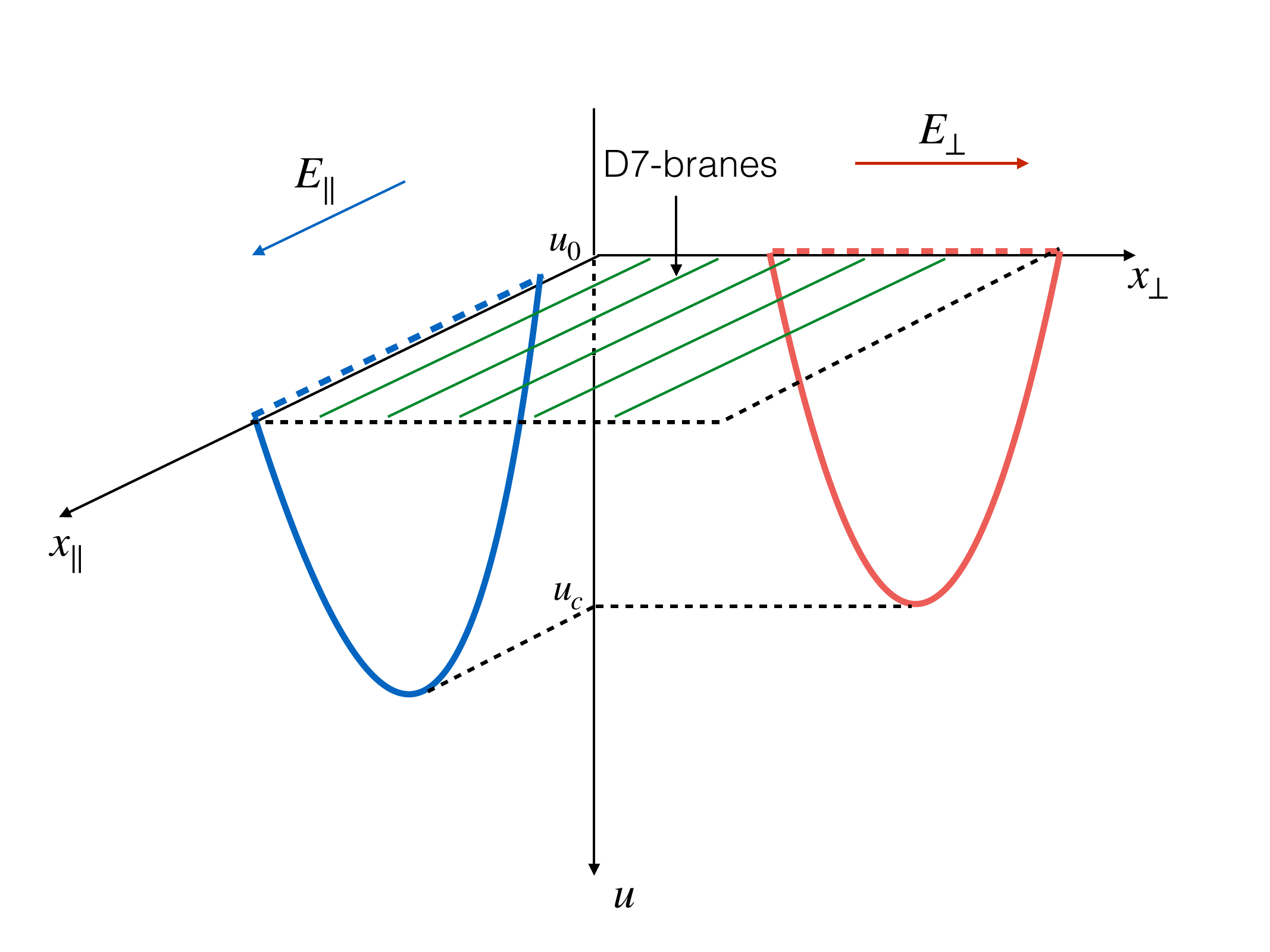}
\par\end{centering}
\caption{\label{fig:1} A fundamental string stretched in the $\left\{ x_{\parallel},u\right\} $
(denoted by the blue line) and $\left\{ x_{\perp},u\right\} $ (denoted
by the red line) plane respectively. The green line refers to the
$N_{\mathrm{D7}}$ dissolved D7-branes in the bulk.}

\end{figure}

\begin{figure}[H]
\begin{centering}
\includegraphics[scale=0.33]{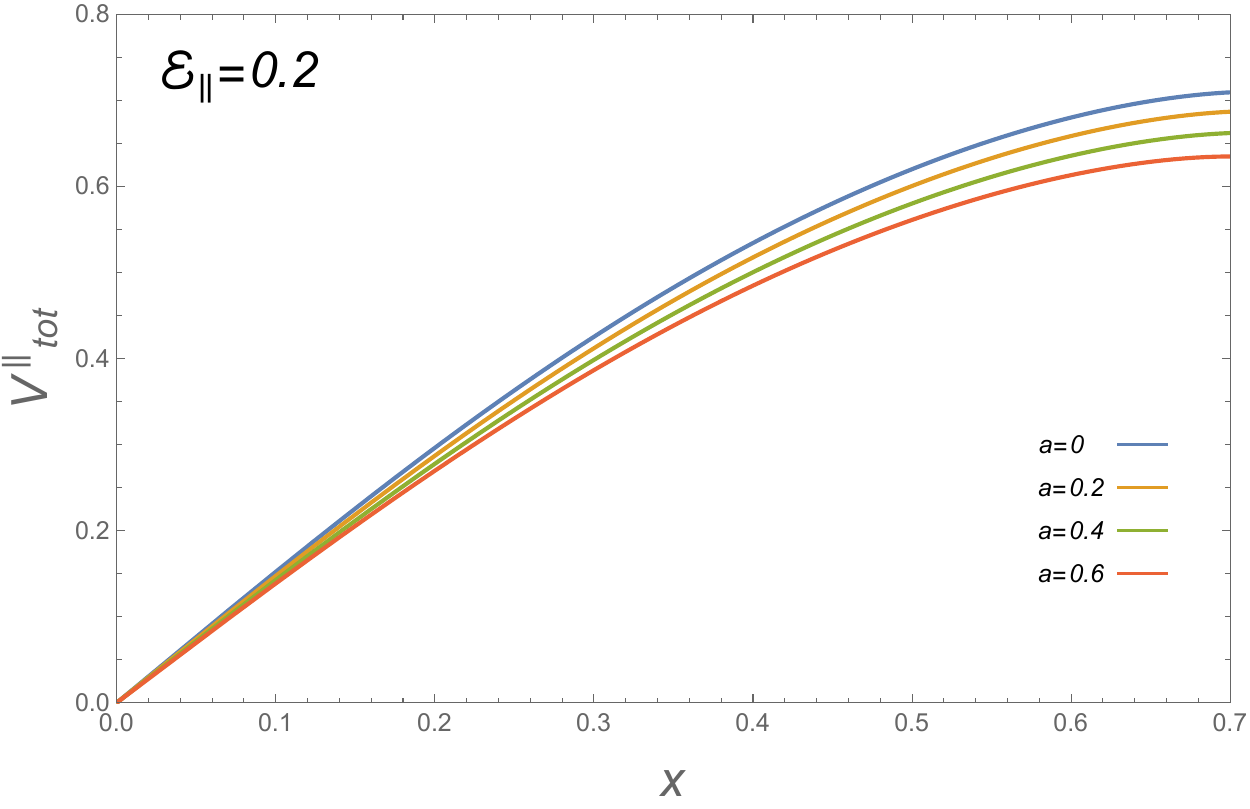}\includegraphics[scale=0.33]{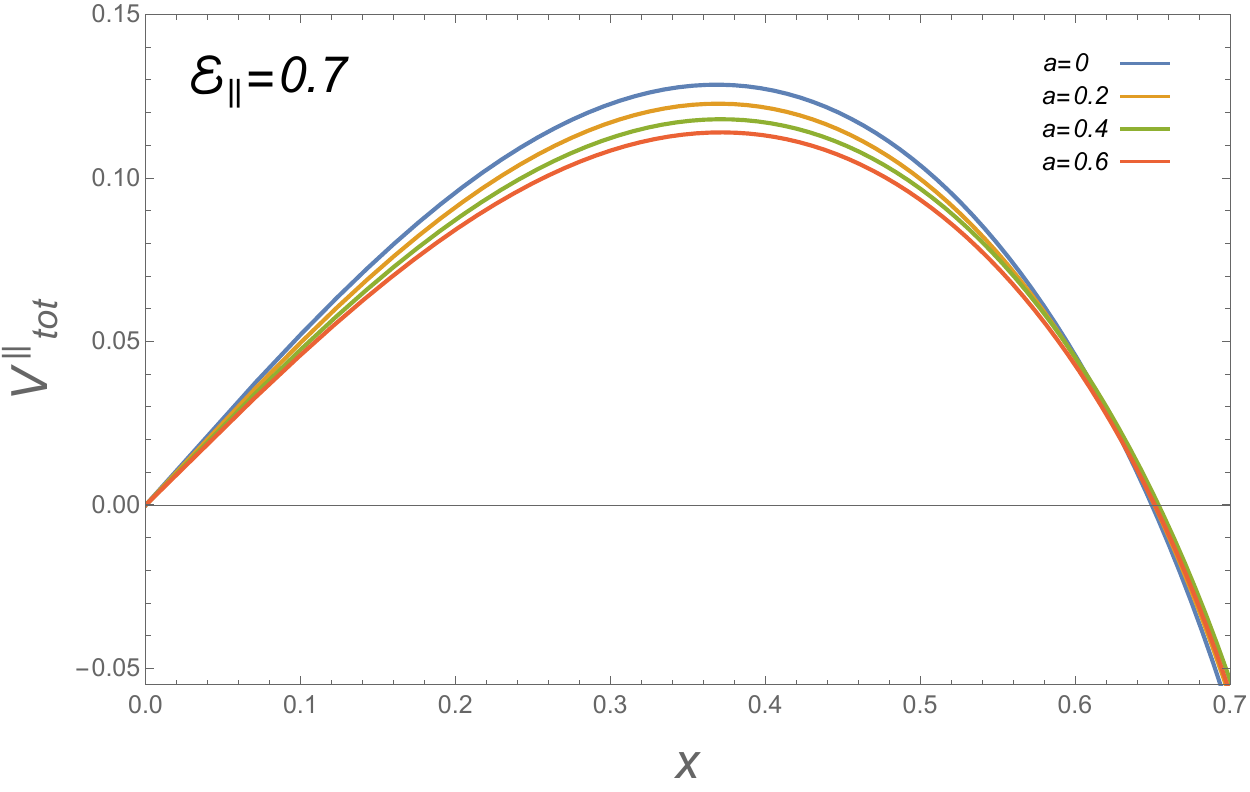}
\par\end{centering}
\begin{centering}
\includegraphics[scale=0.33]{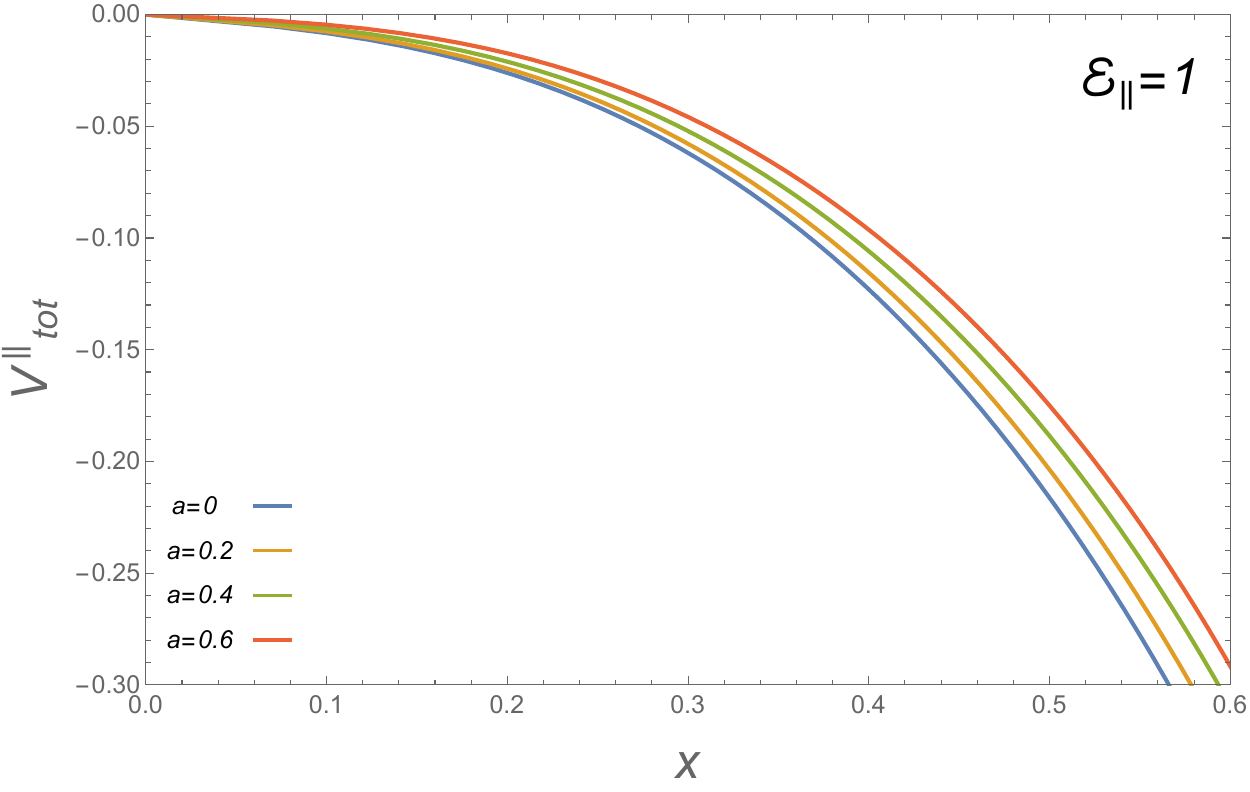}\includegraphics[scale=0.33]{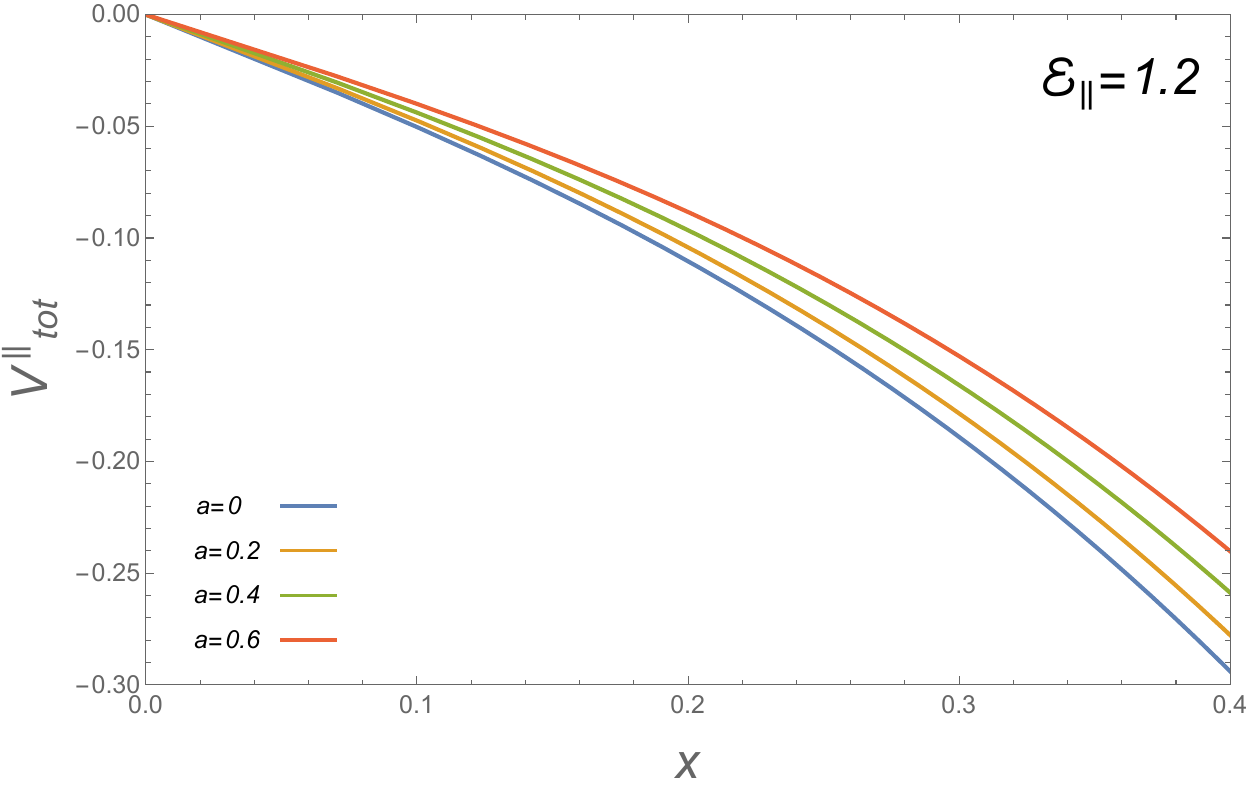}
\par\end{centering}
\caption{\label{fig:4} The holographic potential $V_{\mathrm{tot}}^{\parallel}$
for Schwinger effect as a function of $x$ with various $a$ and fixed
$\mathcal{E}_{\parallel}$ in the black brane background.}
\end{figure}

\begin{figure}[H]
\begin{centering}
\includegraphics[scale=0.33]{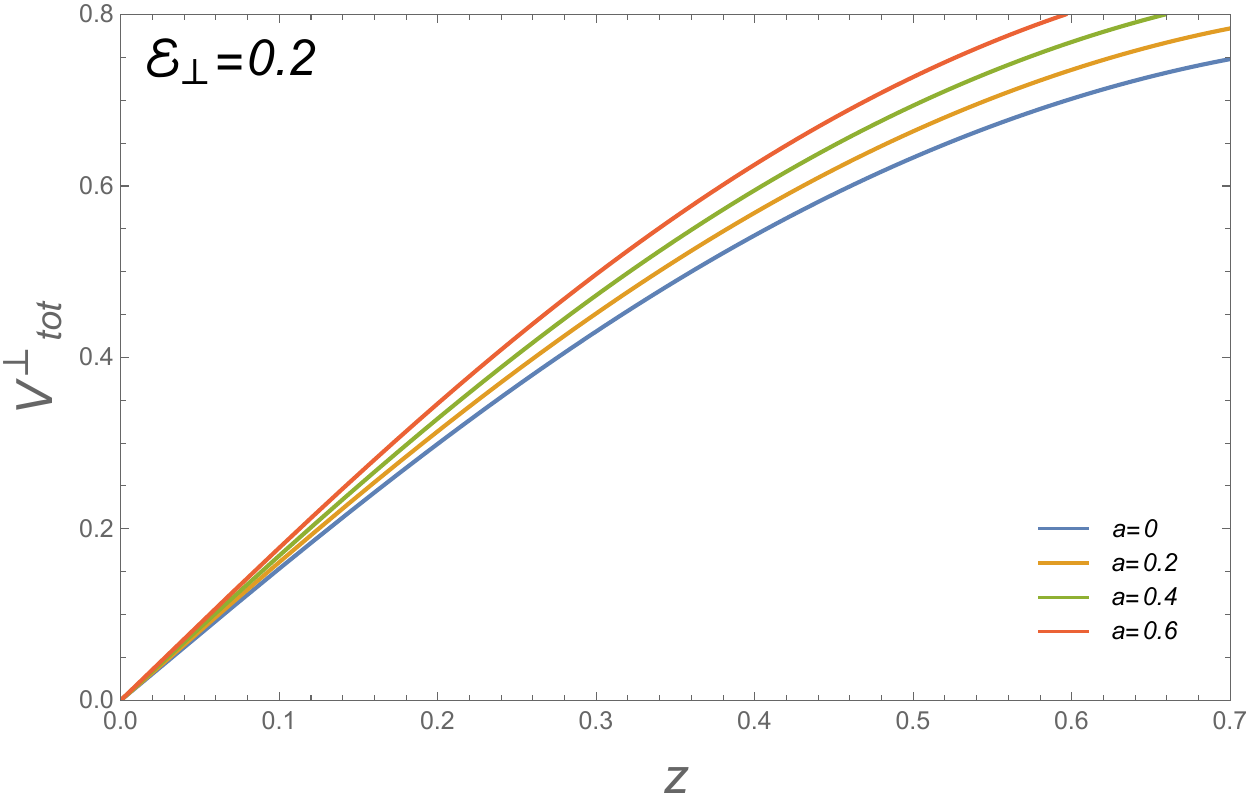}\includegraphics[scale=0.33]{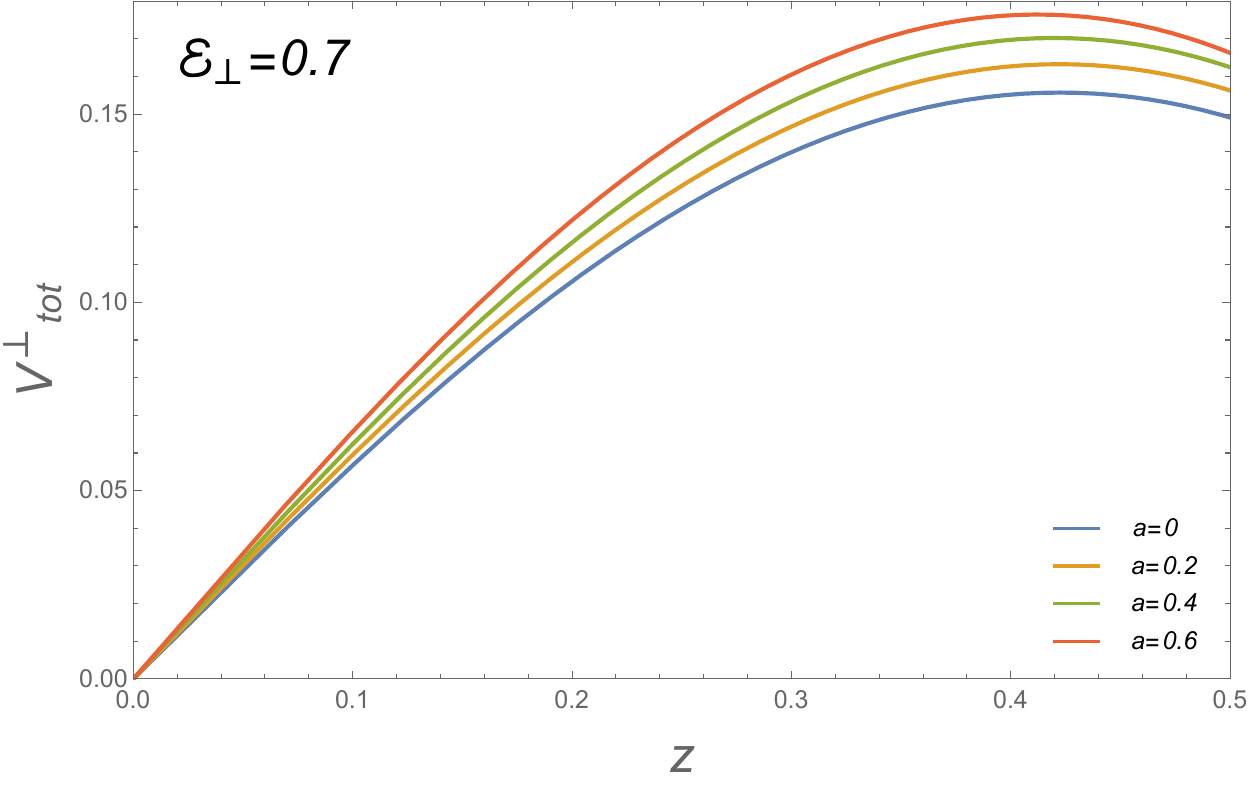}
\par\end{centering}
\begin{centering}
\includegraphics[scale=0.33]{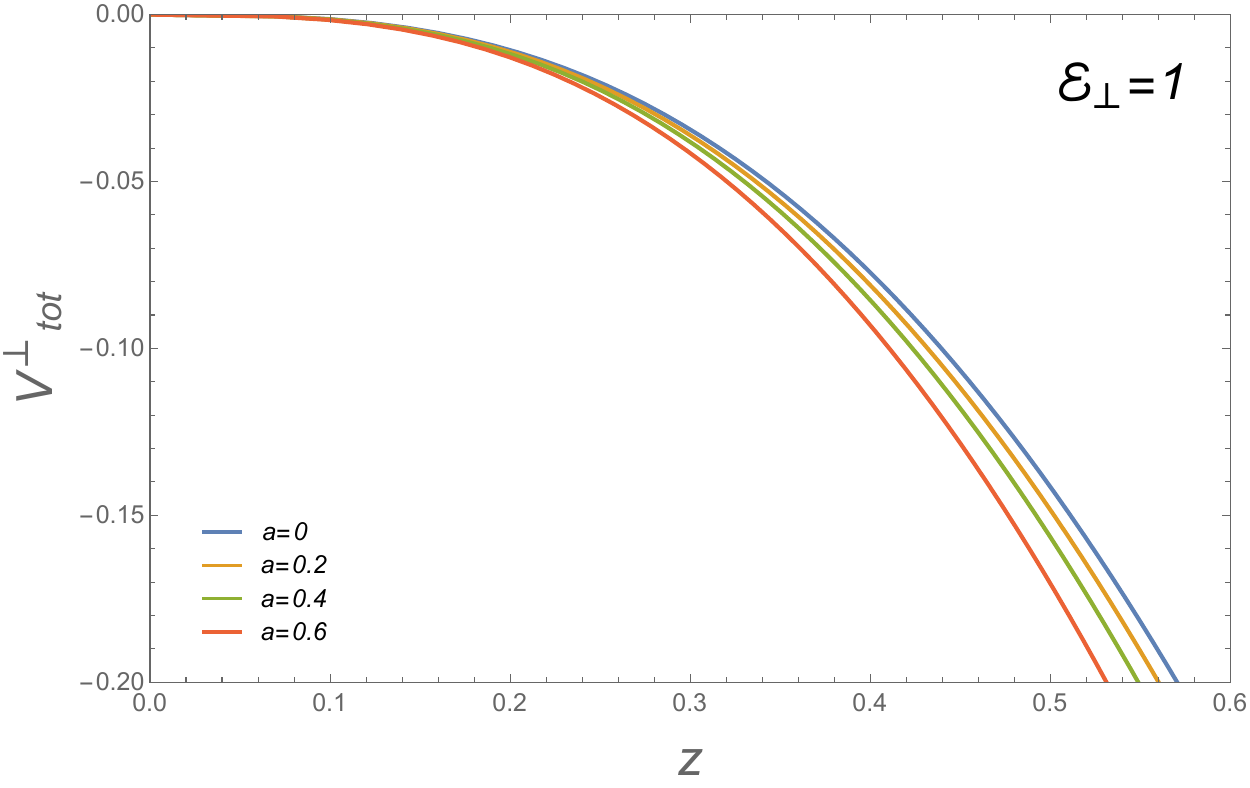}\includegraphics[scale=0.33]{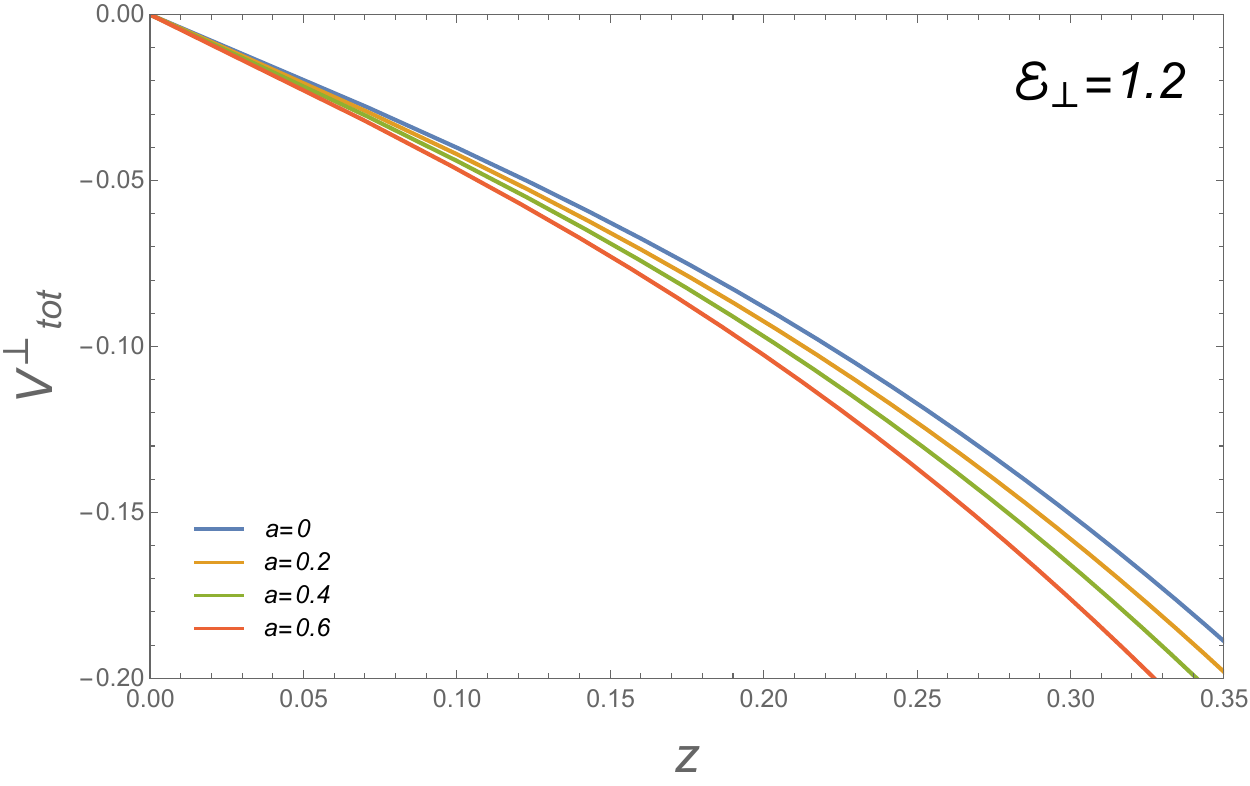}
\par\end{centering}
\caption{\label{fig:5} The holographic potential $V_{\mathrm{tot}}^{\perp}$
for Schwinger effect as a function of $z$ with various $a$ and fixed
$\mathcal{E}_{\perp}$ in the black brane background.}
\end{figure}

\begin{figure}[H]
\begin{centering}
\includegraphics[scale=0.33]{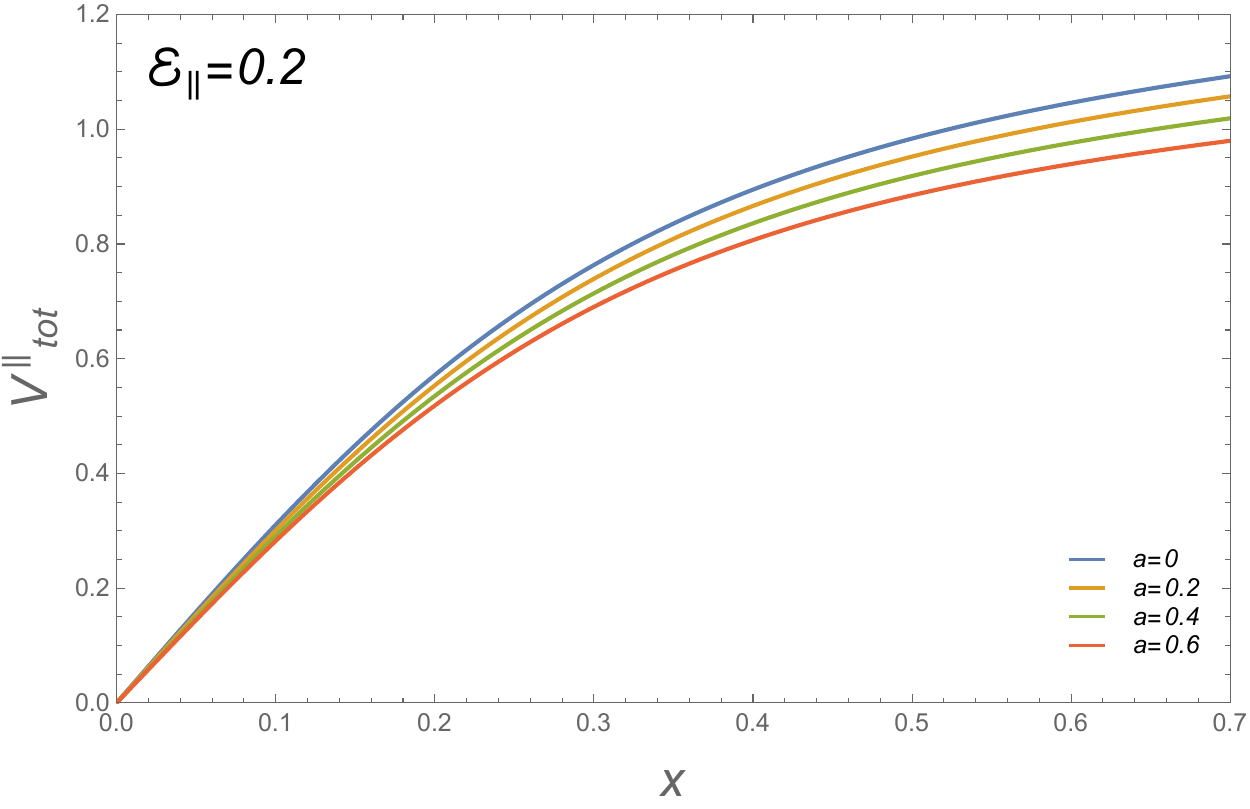}\includegraphics[scale=0.33]{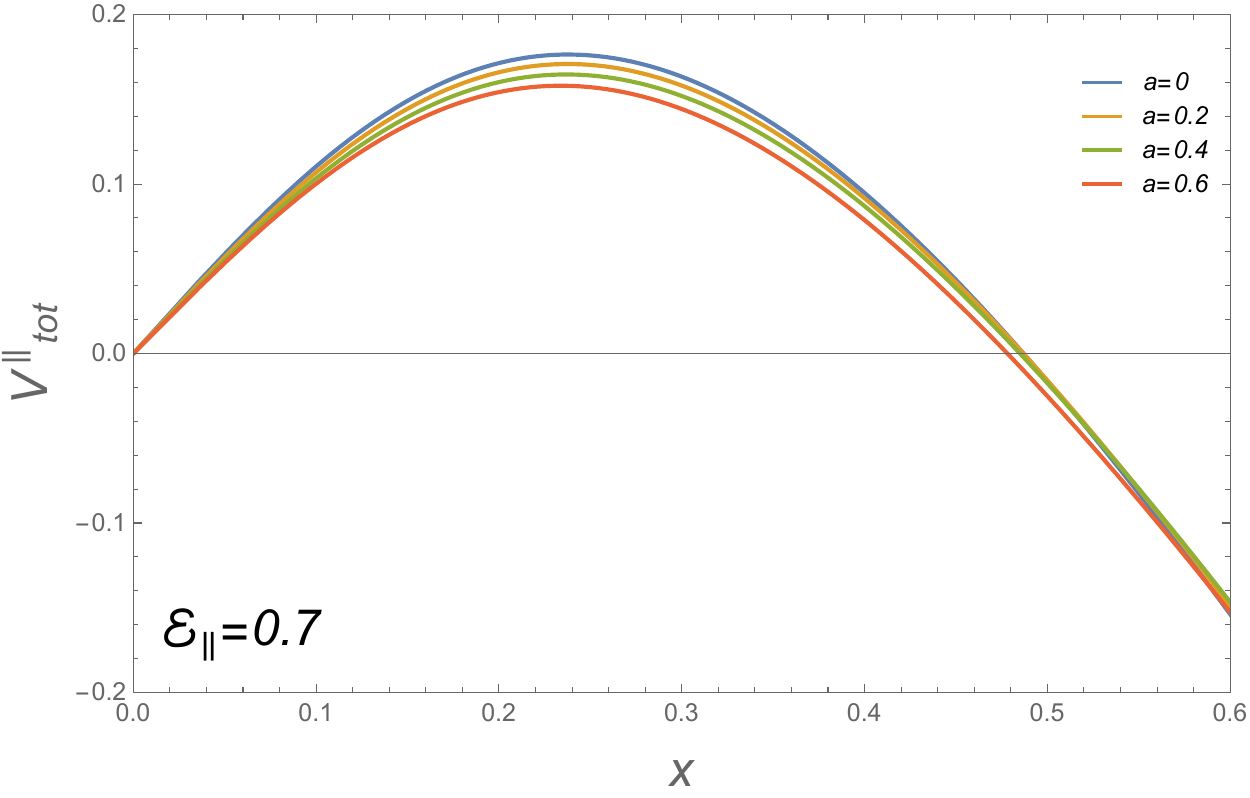}
\par\end{centering}
\begin{centering}
\includegraphics[scale=0.33]{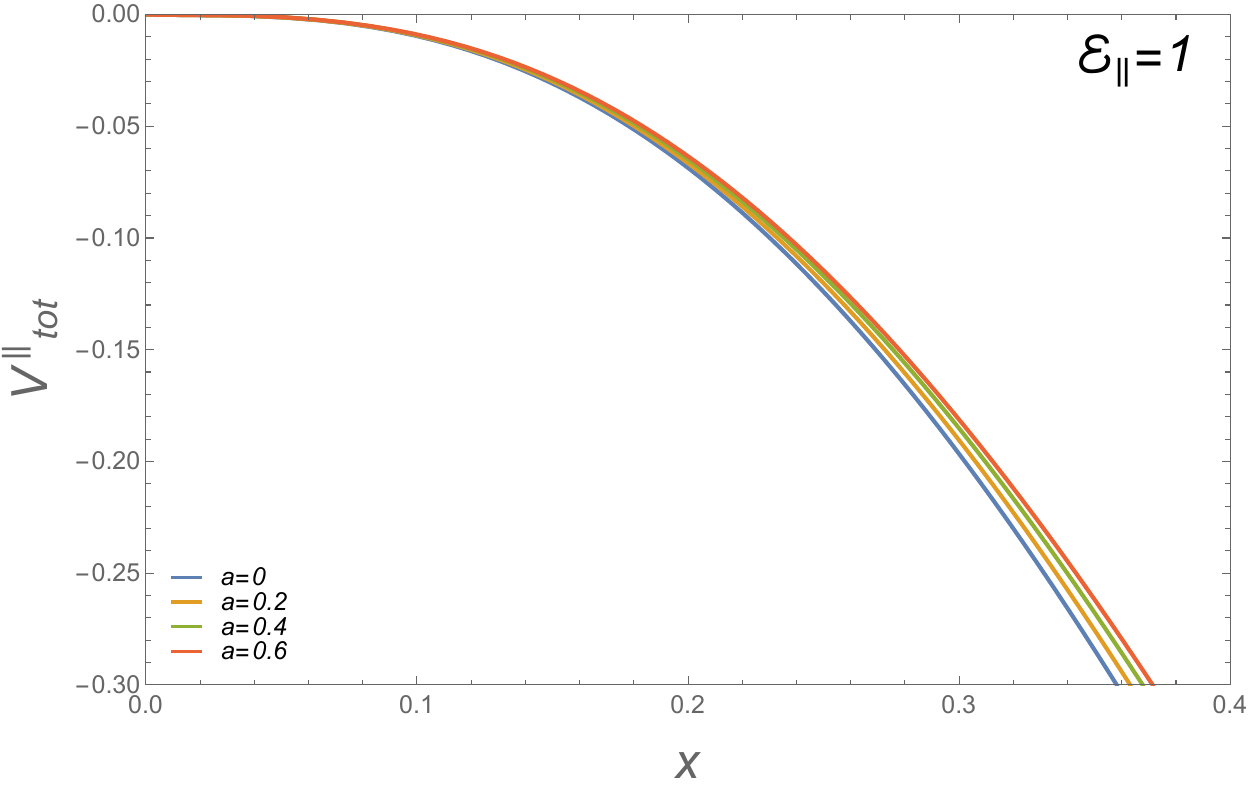}\includegraphics[scale=0.33]{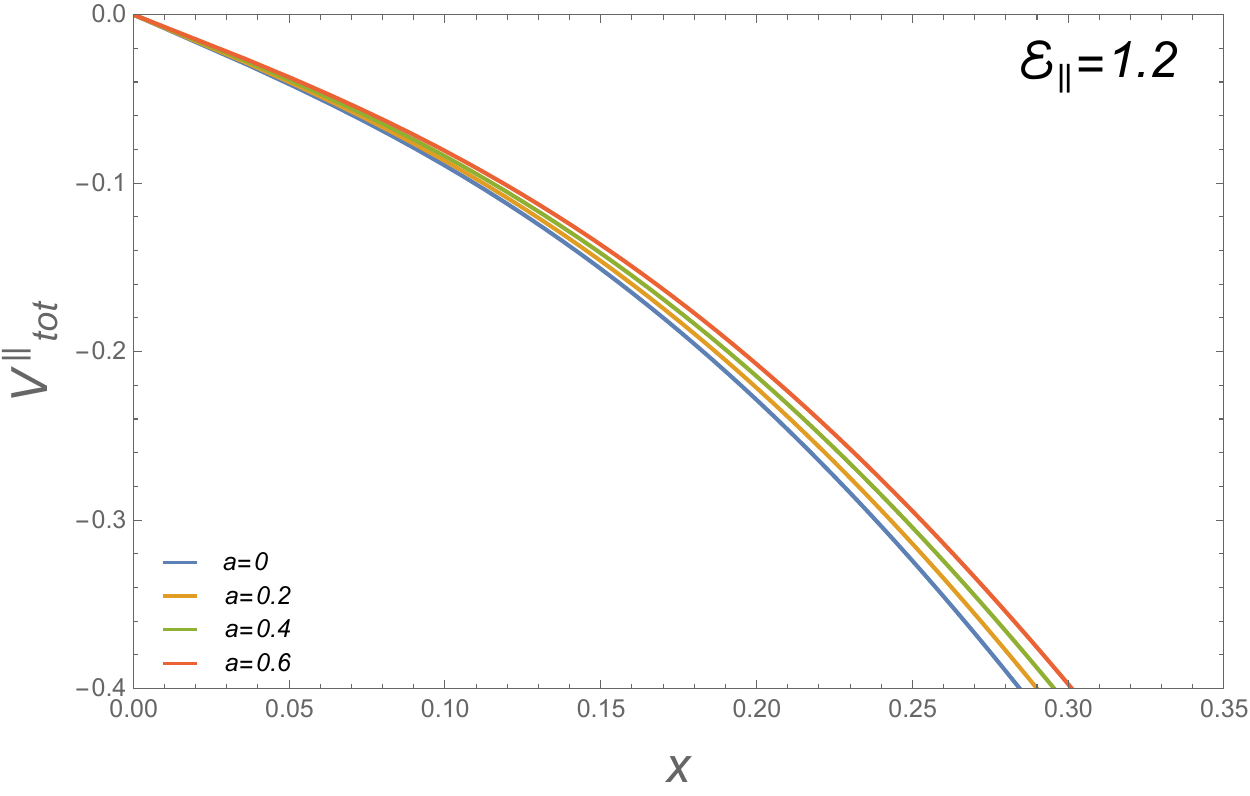}
\par\end{centering}
\caption{\label{fig:6} The holographic potential $V_{\mathrm{tot}}^{\parallel}$
for Schwinger effect as a function of $x$ with various $a$ and fixed
$\mathcal{E}_{\parallel}$ in the bubble background.}
\end{figure}

\begin{figure}[H]
\begin{centering}
\includegraphics[scale=0.33]{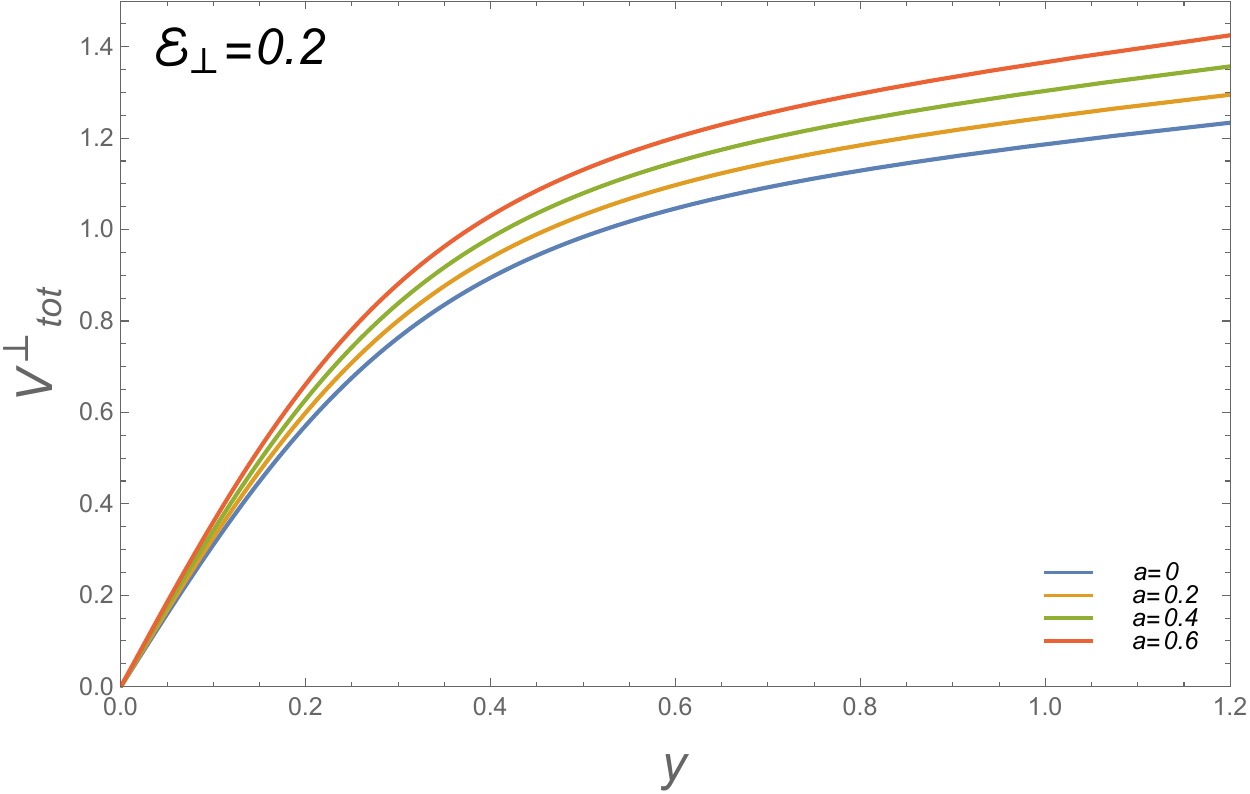}\includegraphics[scale=0.33]{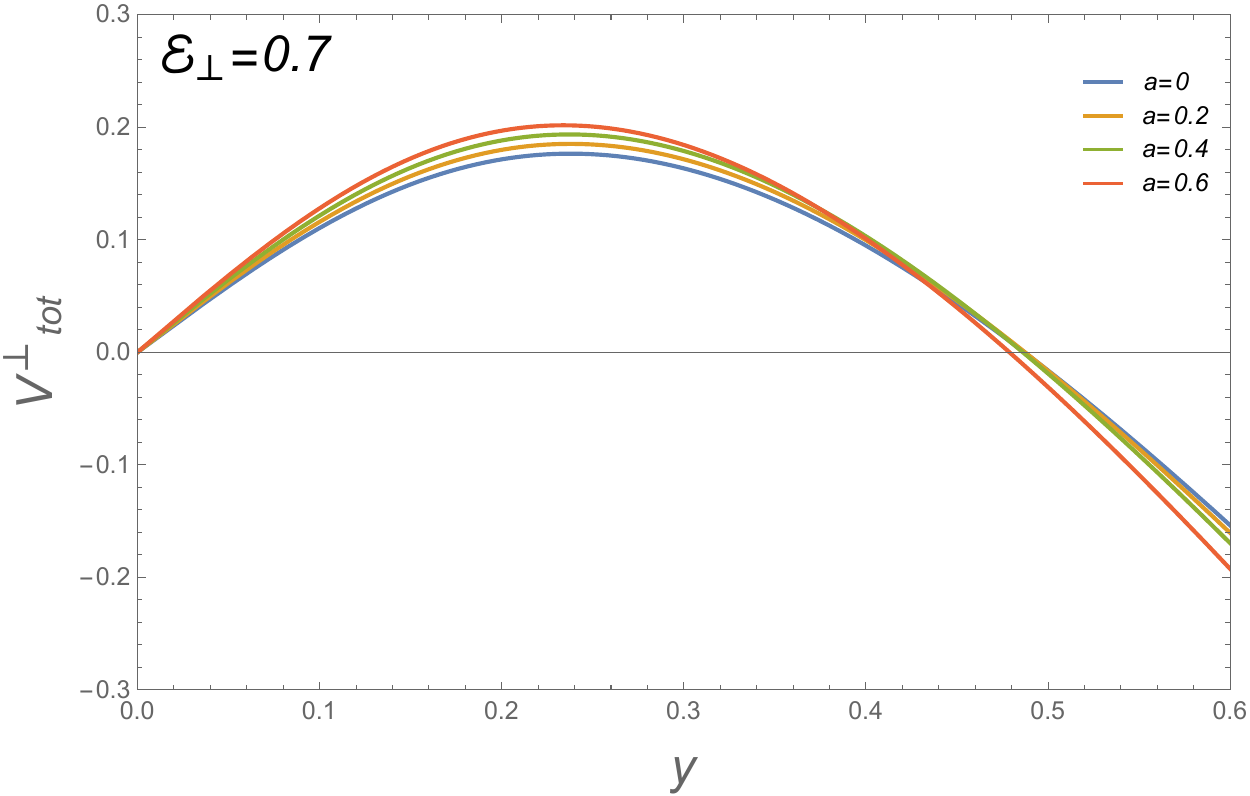}
\par\end{centering}
\begin{centering}
\includegraphics[scale=0.33]{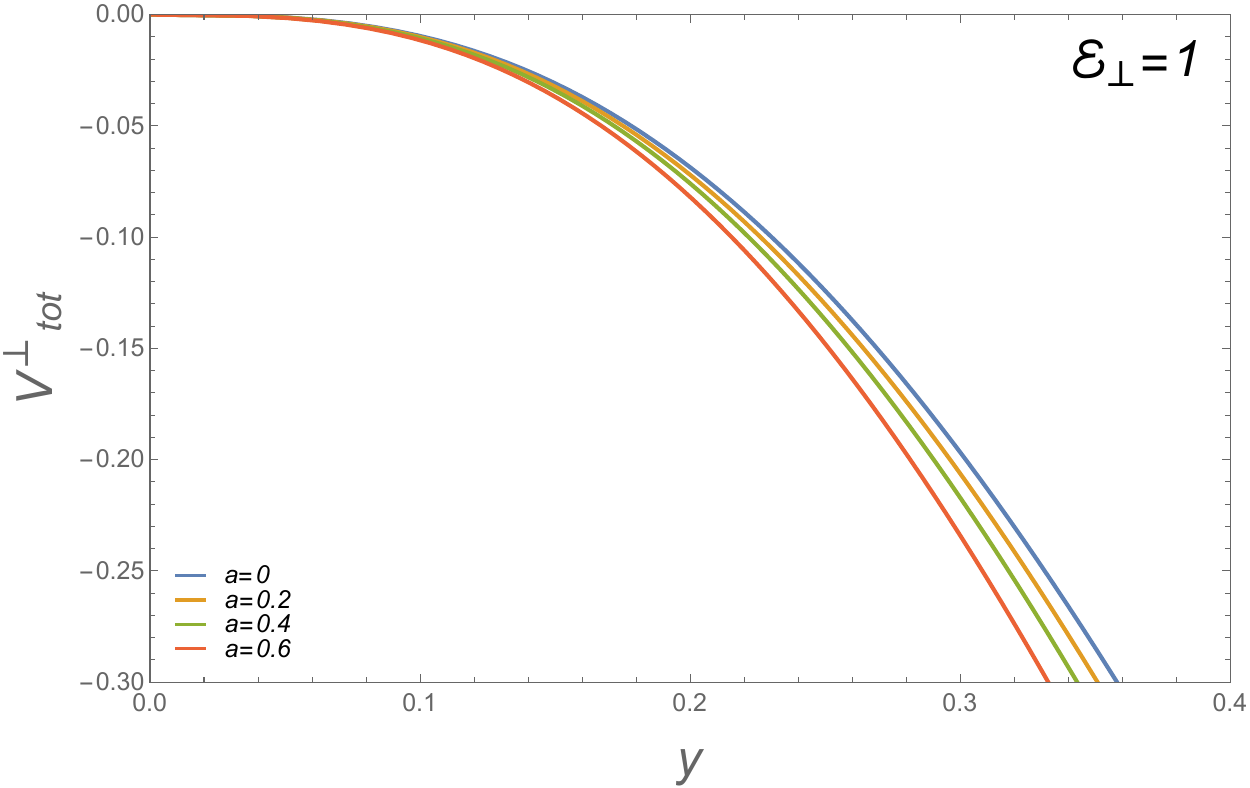}\includegraphics[scale=0.33]{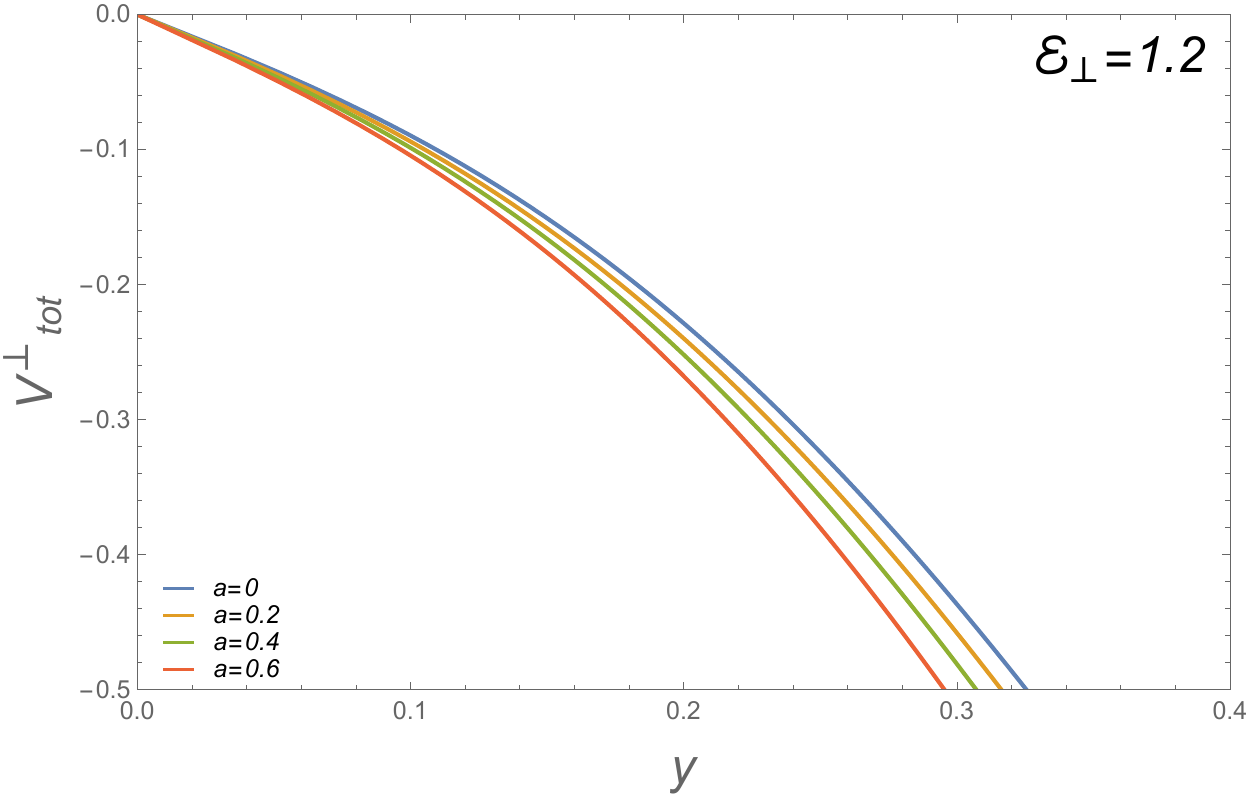}
\par\end{centering}
\caption{\label{fig:7} The holographic potential $V_{\mathrm{tot}}^{\perp}$
for Schwinger effect as a function of $y$ with various $a$ and fixed
$\mathcal{E}_{\perp}$ in the bubble background.}
\end{figure}

In order to evaluate the holographic potential for Schwinger effect,
we need to compute the total energy of a pair of the fundamental particles
which is recognized as the vacuum expectation value (VEV) of the rectangular
Wilson loop. According to the AdS/CFT dictionary \cite{key-31}, the
VEV corresponds to the world-sheet area or namely the onshell Nambu-Goto
(NG) action of a fundamental string. However, the choice of the static
gauge in the parallel and perpendicular case would be a little different
as,

\begin{equation}
\tau=t,x_{\parallel,\perp}=\sigma,u=u\left(\sigma\right),\ \mathrm{others\ are\ constants},
\end{equation}
since the background metric is anisotropic. It implies we could consider
a fundamental string stretched in the $\left\{ x_{\parallel},u\right\} $
and $\left\{ x_{\perp},u\right\} $ plane respectively as it is illustrated
in Figure \ref{fig:1}. 
 So the induced metric on the world-sheet is,

\begin{align}
ds^{2} & =g_{tt}dt^{2}+g_{\parallel,\perp}dx_{\parallel,\perp}^{2}+g_{uu}du^{2}\nonumber \\
 & =g_{tt}dt^{2}+\left(g_{\parallel,\perp}+g_{uu}u^{\prime2}\right)dx_{\parallel,\perp}^{2},\label{eq:20}
\end{align}
where ``$\prime$'' refers to the derivative with respect to $x_{\parallel,\perp}$.
Then the NG Lagrangian $\mathcal{L}_{NG}$ of the string can be obtained
as,

\begin{equation}
\mathcal{L}_{NG}=\sqrt{-\det\left(g_{\alpha\beta}\right)}=\sqrt{g_{tt}\left(g_{\parallel,\perp}+g_{uu}u^{\prime2}\right)},
\end{equation}
and its associated Hamiltonian $\mathcal{H}_{NG}$ is,

\begin{equation}
\mathcal{H}_{NG}=u^{\prime}\frac{\partial\mathcal{L}_{NG}}{\partial u^{\prime}}-\mathcal{L}_{NG},
\end{equation}
which is a constant since the Lagrangian $\mathcal{L}_{NG}$ does
not depend on $\sigma$. Using the condition $\mathcal{H}_{NG}=\mathcal{H}_{NG}|_{u=u_{c}},u^{\prime}|_{u=u_{c}}=0$,
the derivative of $u^{\prime}$ and the formula of the separation
$x_{\parallel,\perp}$ can be solved respectively. Afterwards, the
potential energy (PE) including the static energy (SE) $V_{\mathrm{PE+SE}}$
of the string can be evaluated by

\begin{equation}
V_{\mathrm{PE+SE}}^{\parallel,\perp}=2T_{f}\int_{0}^{x_{\parallel,\perp}/2}\mathcal{L}_{NG}dx_{\parallel,\perp}.\label{eq:23}
\end{equation}
To obtain the total potential energy, we need to compare (\ref{eq:23})
with the associated electric potential energy $E_{\parallel,\perp}x_{\parallel,\perp}$
as,

\begin{equation}
V_{\mathrm{tot}}^{\parallel,\perp}=2T_{f}\int_{0}^{x_{\parallel,\perp}/2}\mathcal{L}_{NG}dx_{\parallel,\perp}-E_{\parallel,\perp}x_{\parallel,\perp}.\label{eq:24}
\end{equation}
Keeping the above formulas in mind, let impose the background geometry
(\ref{eq:2}) and (\ref{eq:5}) to (\ref{eq:20}) - (\ref{eq:24}).
Recall our notations, in the black brane solution, we have $x_{\parallel}=x,x_{\perp}=z$
and in bubble solution $x_{\parallel}=x,x_{\perp}=y$, then the relevant
formulas are listed as follows, in the black brane solution, 
\begin{figure}[h]
\begin{centering}
\includegraphics[scale=0.35]{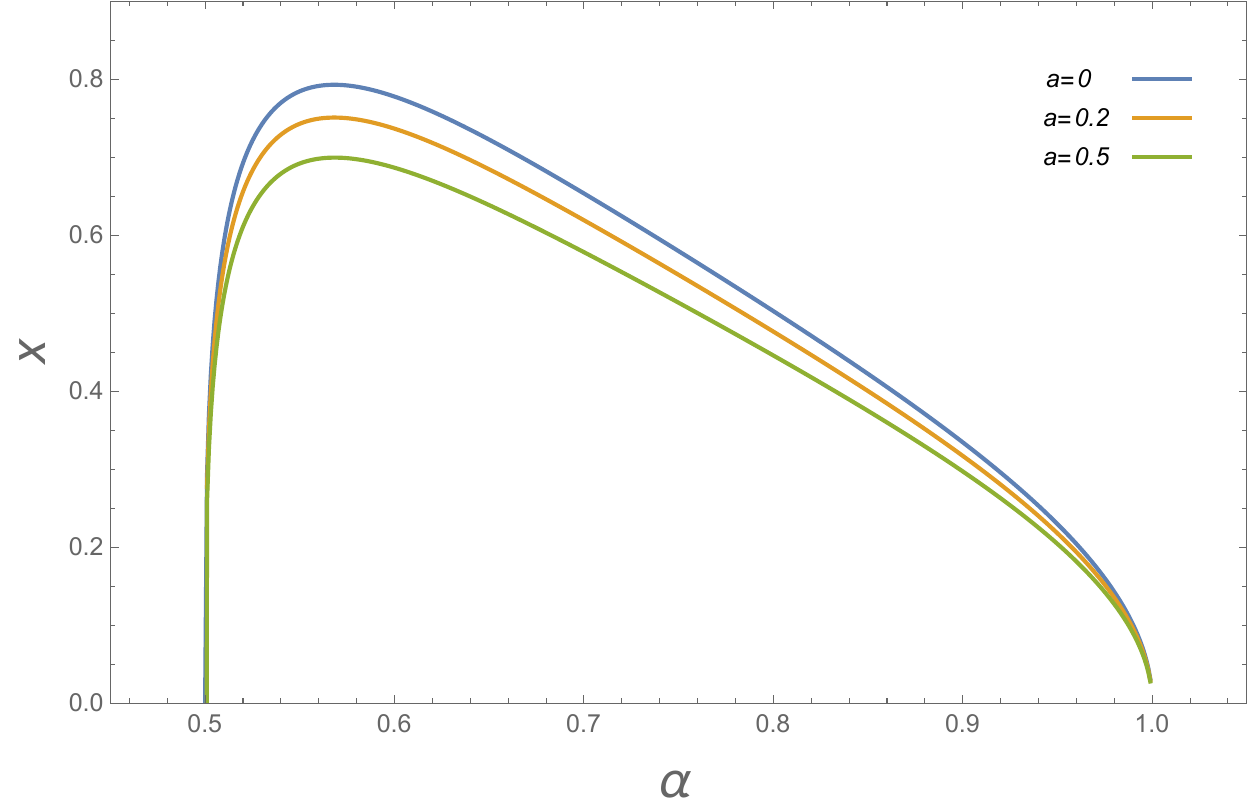}\includegraphics[scale=0.35]{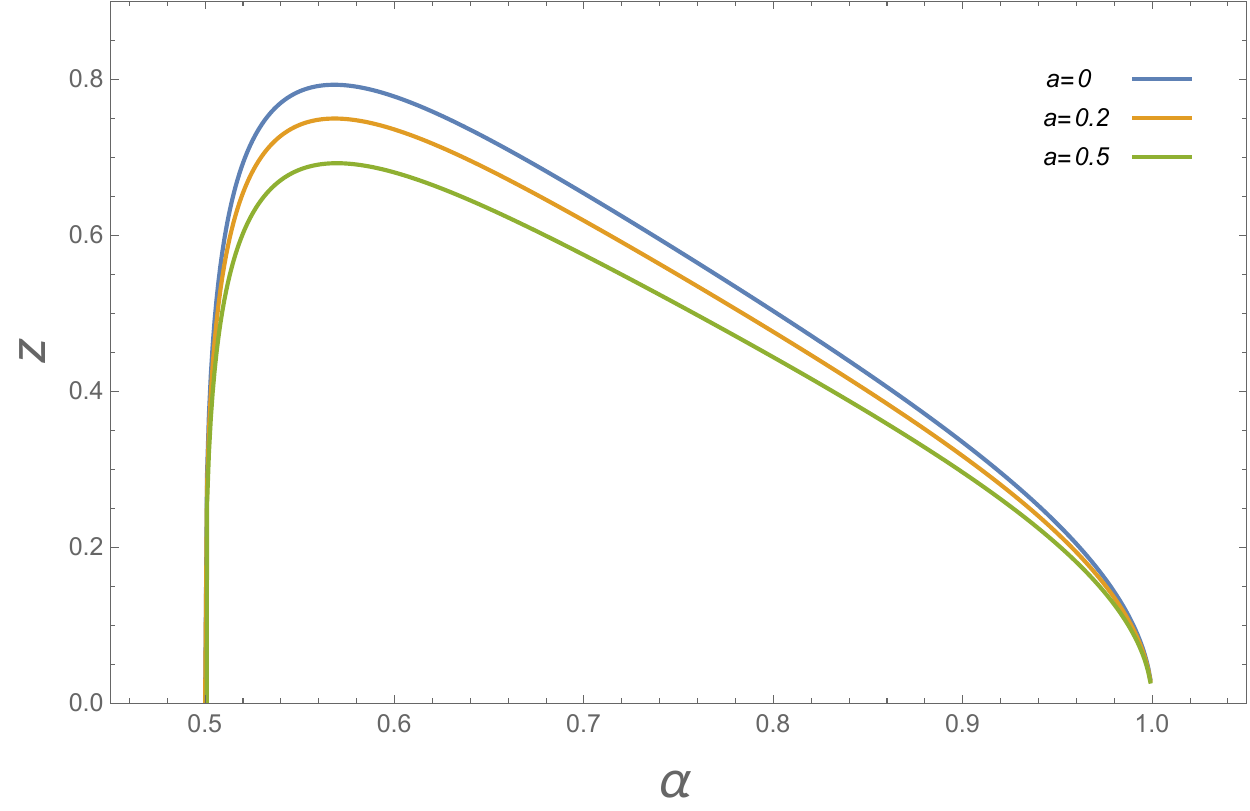}
\par\end{centering}
\begin{centering}
\includegraphics[scale=0.35]{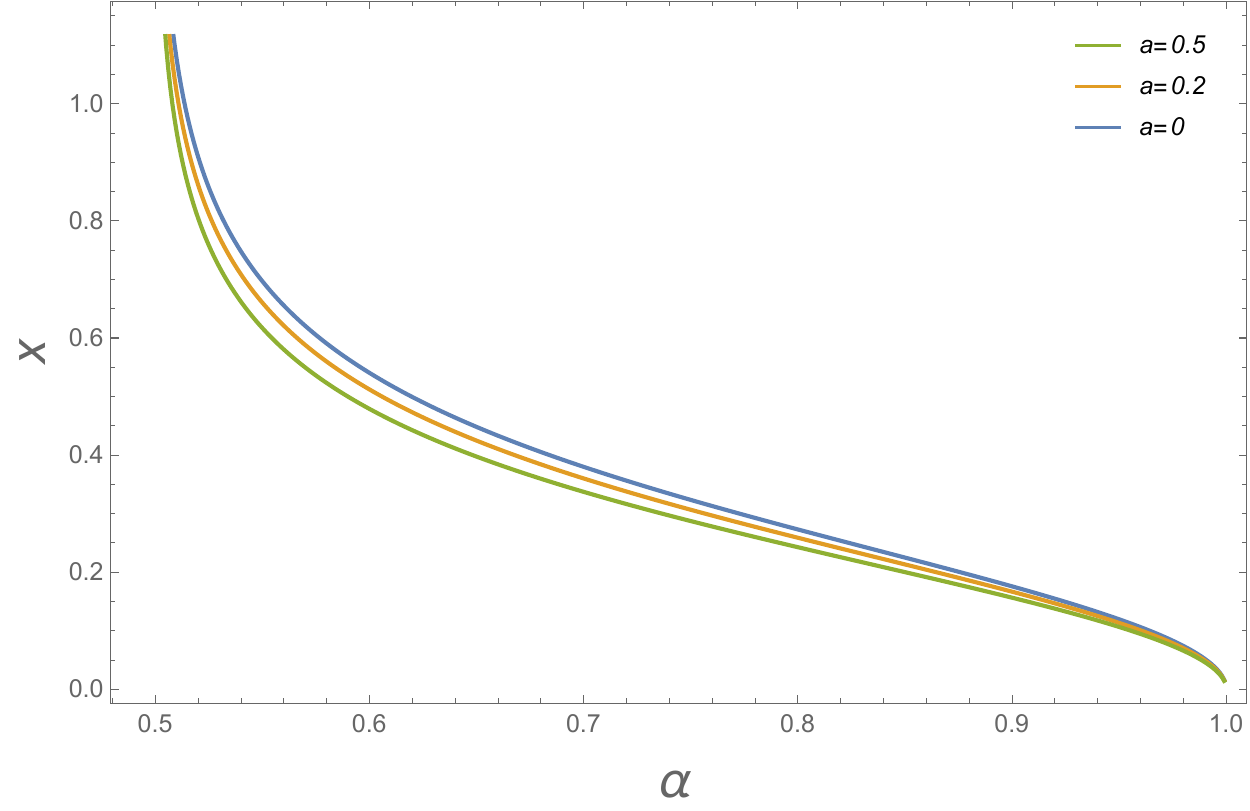}\includegraphics[scale=0.35]{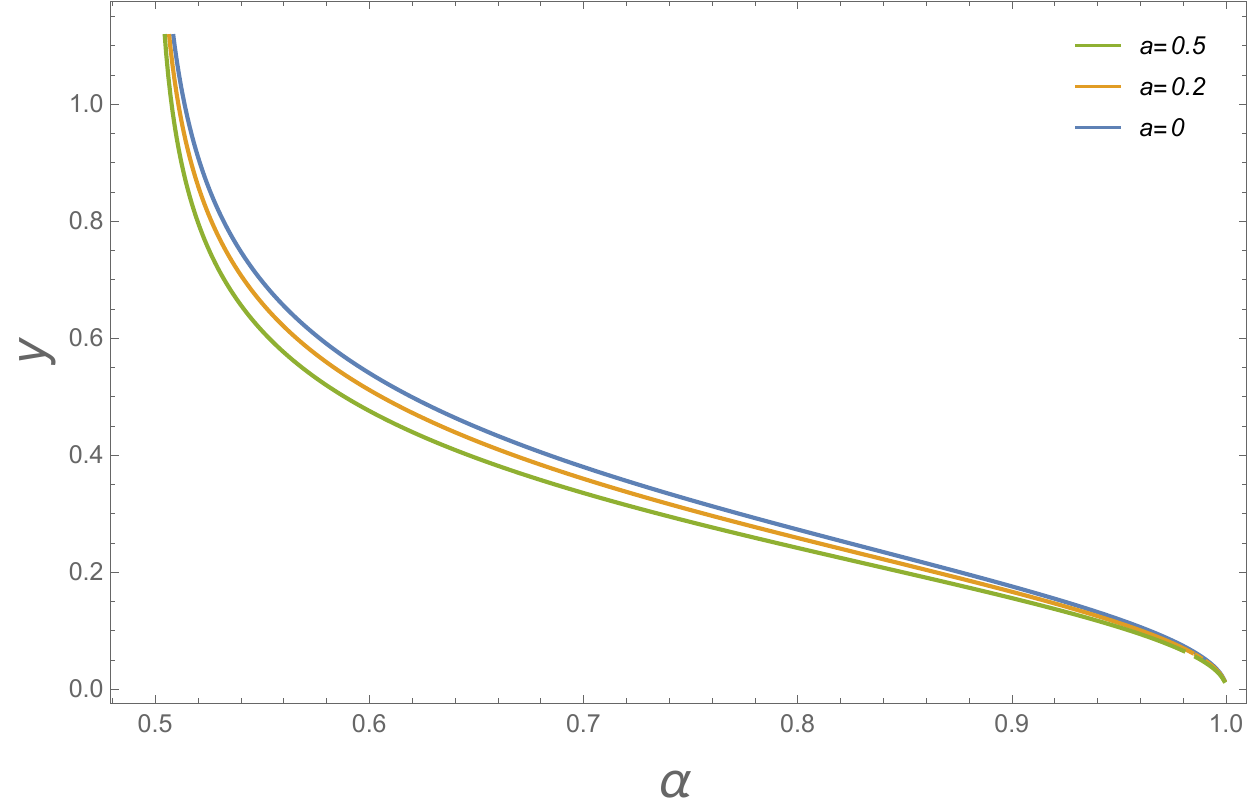}
\par\end{centering}
\caption{\label{fig:2} The separation as a function of $\alpha$. \textbf{Upper:}
the separation $x,z$ as a function of $\alpha$ in the black brane
solution. \textbf{Lower:} the separation $x,y$ as a function of $\alpha$
in the bubble solution.}
\end{figure}

\begin{align}
\frac{du}{dx}= & \sqrt{\mathcal{F}}\sqrt{\frac{\mathcal{B}\mathcal{F}}{\mathcal{B}\left(u_{c}\right)\mathcal{F}\left(u_{c}\right)}\frac{u_{c}^{4}}{u^{4}}-1},\nonumber \\
V_{\mathrm{tot}}^{\parallel}= & \frac{2T_{f}L^{2}}{u_{0}}\alpha\int_{1}^{1/\alpha}dY\frac{Y^{2}\mathcal{B}\sqrt{\mathcal{F}}}{\sqrt{\mathcal{B}\mathcal{F}Y^{4}-\mathcal{B}\left(1\right)\mathcal{F}\left(1\right)}}\nonumber \\
 & -\frac{2T_{f}L^{2}}{u_{0}}\frac{\sqrt{\mathcal{B}\left(Y_{0}\right)\mathcal{F}\left(Y_{0}\right)}}{\alpha}\mathcal{E}_{\parallel}\int_{1}^{1/\alpha}dY\frac{1}{Y^{2}\sqrt{\mathcal{F}}}\frac{1}{\sqrt{\frac{\mathcal{B}\mathcal{F}}{\mathcal{B}\left(1\right)\mathcal{F}\left(1\right)}Y^{4}-1}},\label{eq:25}
\end{align}
and

\begin{align}
\frac{du}{dz}= & \sqrt{\mathcal{F}\mathcal{H}}\sqrt{\frac{\mathcal{B}\mathcal{F}\mathcal{H}}{\mathcal{B}\left(u_{c}\right)\mathcal{F}\left(u_{c}\right)\mathcal{H}\left(u_{c}\right)}\frac{u_{c}^{4}}{u^{4}}-1},\nonumber \\
V_{\mathrm{tot}}^{\perp}= & \frac{2T_{f}L^{2}}{u_{0}}\alpha\int_{1}^{1/\alpha}dY\frac{\mathcal{B}\sqrt{\mathcal{F}\mathcal{H}}Y^{2}}{\sqrt{\mathcal{B}\mathcal{F}\mathcal{H}Y^{4}-\mathcal{F}\left(u_{c}\right)\mathcal{B}\left(u_{c}\right)\mathcal{H}\left(u_{c}\right)}}\nonumber \\
 & -\frac{2T_{f}L^{2}}{u_{0}}\frac{\sqrt{\mathcal{B}\left(Y_{0}\right)\mathcal{F}\left(Y_{0}\right)\mathcal{H}\left(Y_{0}\right)}}{\alpha}\mathcal{E}_{\perp}\int_{1}^{1/\alpha}dY\frac{1}{Y^{2}}\frac{1}{\sqrt{\mathcal{F}\mathcal{H}}}\frac{1}{\sqrt{\frac{\mathcal{B}\mathcal{F}\mathcal{H}}{\mathcal{B}\left(1\right)\mathcal{F}\left(1\right)\mathcal{H}\left(1\right)}Y^{4}-1}},
\end{align}
where we have introduced, 
\begin{figure}[h]
\begin{centering}
\includegraphics[scale=0.33]{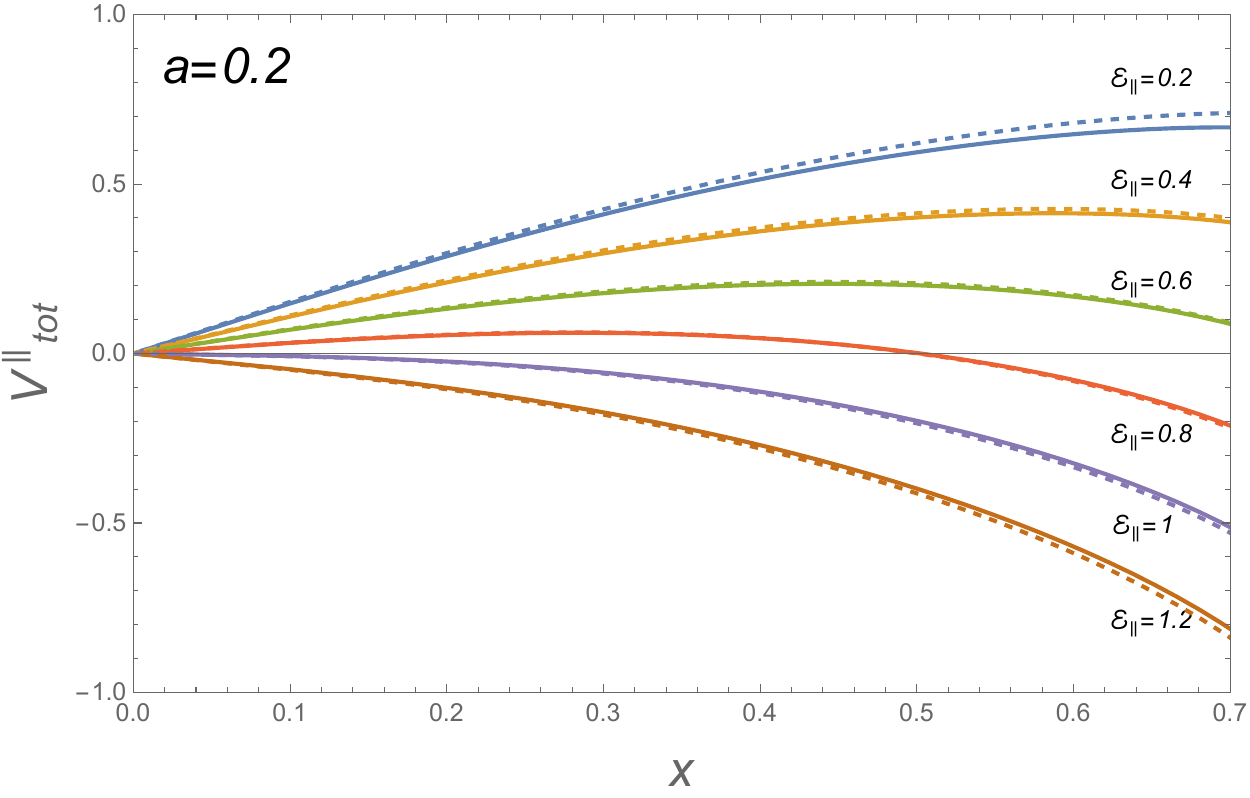}\includegraphics[scale=0.33]{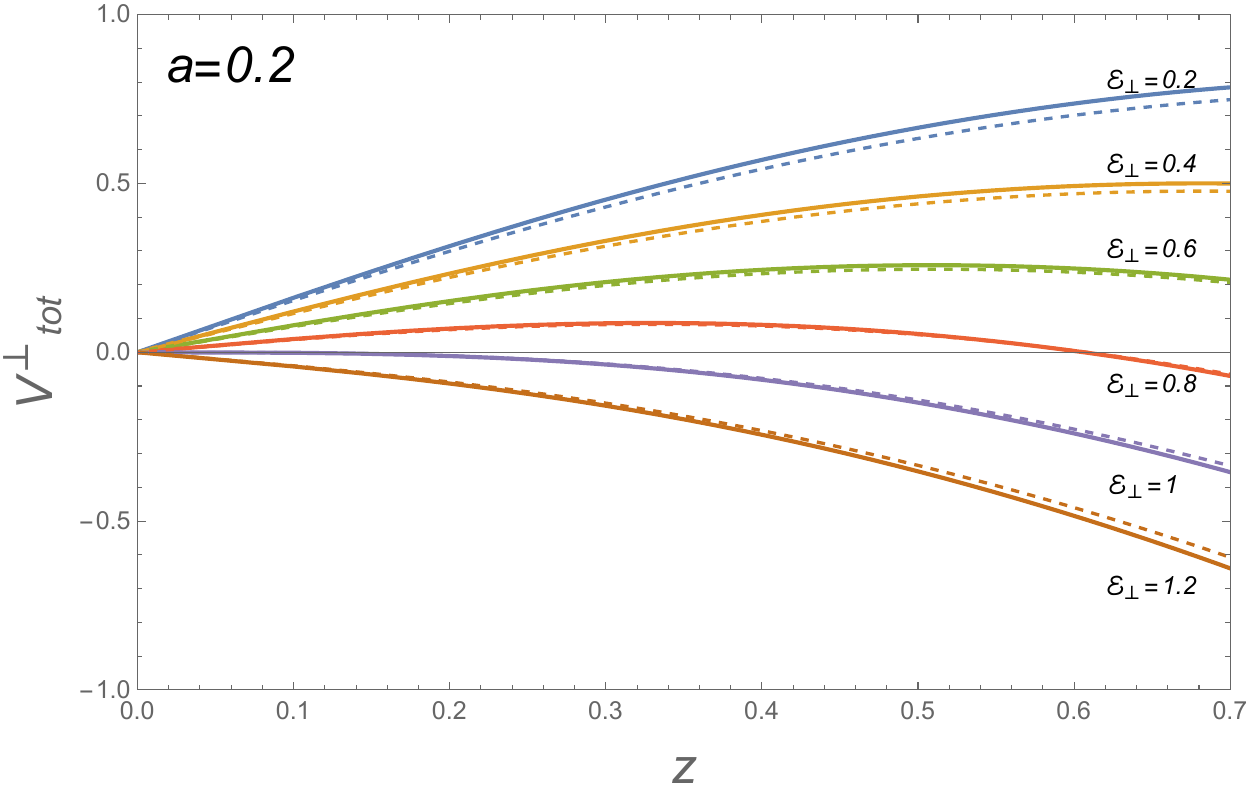}
\par\end{centering}
\begin{centering}
\includegraphics[scale=0.33]{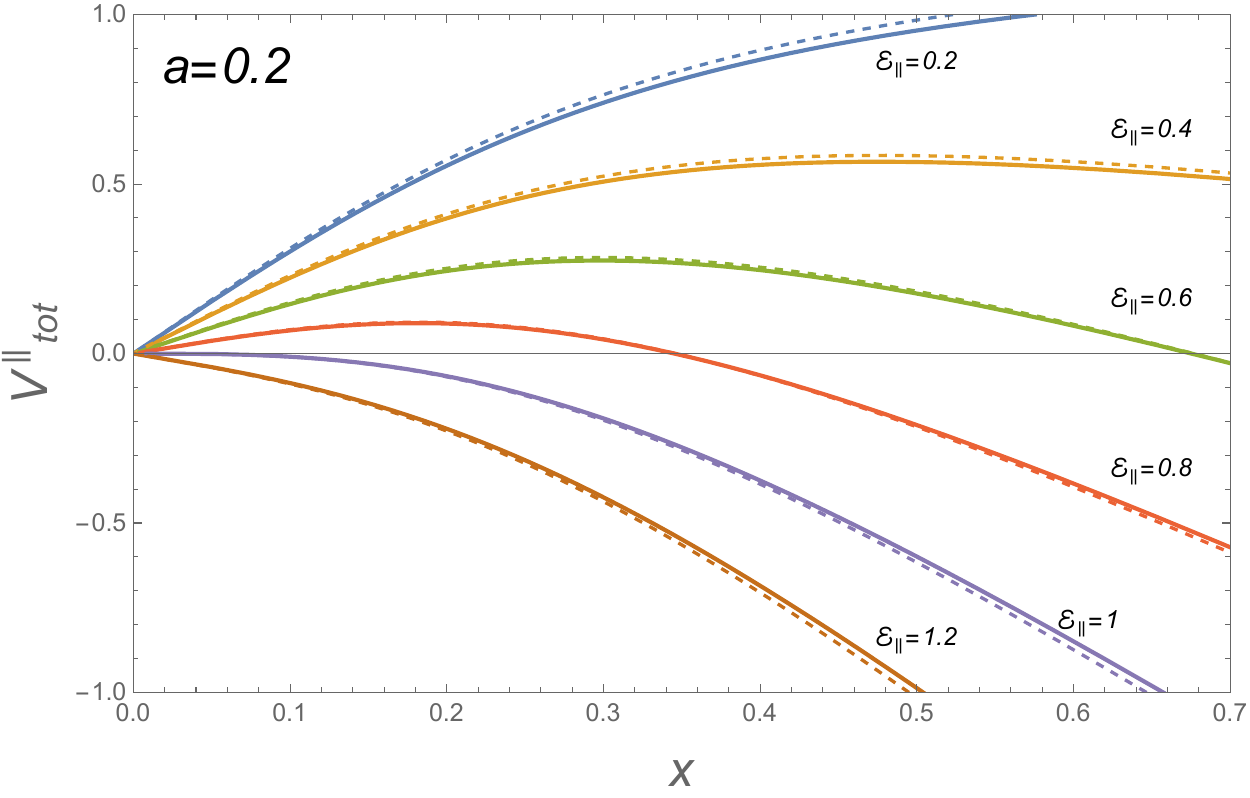}\includegraphics[scale=0.33]{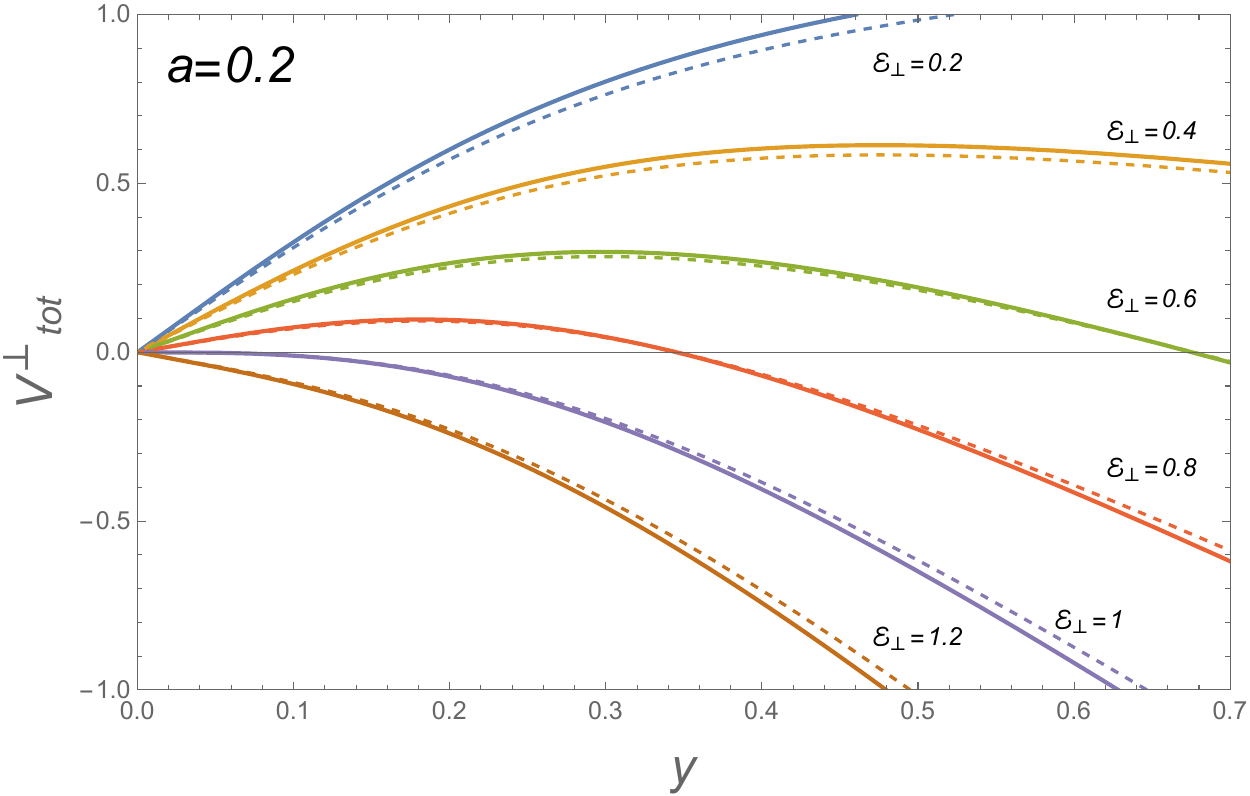}
\par\end{centering}
\caption{\label{fig:3} The holographic potential $V_{\mathrm{tot}}^{\parallel,\perp}$
with parallel ``$\parallel$'' and perpendicular ``$\perp$''
electric field $\mathcal{E}_{\parallel,\perp}$ for $a=0.2$ (solid
lines). \textbf{Upper:} $V_{\mathrm{tot}}^{\parallel,\perp}$ as a
function of $x,z$ in the black brane background. \textbf{Lower: }$V_{\mathrm{tot}}^{\parallel,\perp}$
as a function of $x,y$ in the bubble background. The dashed lines
refer to the isotropic potential ($a=0$) with corresponding electric
field in the black brane and bubble background respectively. The numerical
calculation shows the barrier of the holographic potential $V_{\mathrm{tot}}^{\parallel,\perp}$
is decreased/increased slightly by the anisotropy in parallel and
perpendicular direction, both in the black brane and bubble background.}
\end{figure}

\begin{equation}
Y=\frac{u_{c}}{u},\alpha=\frac{u_{0}}{u_{c}},\beta=\frac{u_{0}}{u_{H,KK}},\mathcal{E}_{\parallel,\perp}=\frac{E_{\parallel,\perp}}{E_{\parallel,\perp}^{c}},T_{f}=\frac{1}{2\pi\alpha^{\prime}}.
\end{equation}
And in the bubble solution, we have,

\begin{align}
\frac{du}{dx}= & \sqrt{\mathcal{F}}\sqrt{\frac{u_{c}^{4}}{u^{4}}-1},\nonumber \\
V_{\mathrm{tot}}^{\parallel}= & \frac{2T_{F}L^{2}}{u_{0}}\int_{1}^{1/\alpha}dY\frac{\alpha}{\sqrt{\mathcal{F}}}\frac{Y^{2}}{\sqrt{Y^{4}-1}}-\frac{2T_{f}L^{2}}{u_{0}}\mathcal{E}_{\parallel}\int_{1}^{1/\alpha}dY\frac{1}{\alpha\sqrt{\mathcal{F}}}\frac{1}{Y^{2}\sqrt{Y^{4}-1}},
\end{align}
and

\begin{align}
\frac{du}{dy}= & \sqrt{\mathcal{F}\mathcal{H}}\sqrt{\frac{\mathcal{H}}{\mathcal{H}\left(u_{c}\right)}\frac{u_{c}^{4}}{u^{4}}-1},\nonumber \\
V_{\mathrm{tot}}^{\perp}= & \frac{2T_{F}L^{2}}{u_{0}}\alpha\int_{1}^{1/\alpha}dY\sqrt{\frac{\mathcal{H}}{\mathcal{F}}}\frac{Y^{2}}{\sqrt{\mathcal{H}Y^{4}-\mathcal{H}\left(1\right)}}\nonumber \\
= & -\frac{2T_{f}L^{2}}{u_{0}}\frac{\sqrt{\mathcal{H}\left(Y_{0}\right)}}{\alpha}\mathcal{E}_{\perp}\int_{1}^{1/\alpha}dY\frac{1}{\sqrt{\mathcal{F}\mathcal{H}}}\frac{1}{Y^{2}\sqrt{Y^{4}\frac{\mathcal{H}}{\mathcal{H}\left(1\right)}-1}}.\label{eq:29}
\end{align}
The behaviors of the separation and total potential presented in (\ref{eq:25})
- (\ref{eq:29}) can be evaluated numerically which are illustrated
in Figure \ref{fig:2}, \ref{fig:3}. The parameters are set as $\frac{T_{f}L^{2}}{u_{0}}=1,\beta=0.5$
in our numerical calculations which refer to the situation of finite
temperature and fixed $M_{KK}$ in the black and bubble background
respectively. As we can see, the numerical calculation shows the separation
is suppressed by the presence of the anisotropy denoted by $a$ and
the potential barrier begins to be vanished for $\mathcal{E}_{\parallel,\perp}\geq1$
with various $a$. Therefore, the critical electric field obtained
from the potential analysis agrees with (\ref{eq:16}) and (\ref{eq:18})
obtained form the D-brane action. Figure \ref{fig:3} also confirms
the behavior of the holographic potential as it is in \cite{key-11,key-12,key-13}
even if the anisotropy is presented, however the influence of the
anisotropy parameter $a$ is too slight to be caught obviously. So,
in order to specify the dependence of the anisotropy in the total
potential, we plot out the behavior of the potential for fixed $\mathcal{E}_{\parallel,\perp}$
by enhancing the affect of the anisotropy with various $a$ in Figures
\ref{fig:4}, \ref{fig:5}, \ref{fig:6}, \ref{fig:7}, which show
that, both in the black brane and bubble background, the barrier of
the potential $V_{\mathrm{tot}}^{\parallel}$ in parallel direction
is suppressed by the presented anisotropy while in the perpendicular
direction, the barrier of potential $V_{\mathrm{tot}}^{\perp}$ is
increased by the anisotropy. And this behavior is in agreement with
the analysis of the holographic quark potential in this system as
in \cite{key-21} since the quark tension is increased/decreased by
the anisotropy in the perpendicular/parallel direction respectively
\footnote{One may additional find the barrier of the potential $V_{\mathrm{tot}}^{\parallel}$
in parallel direction also agrees with the bottom-up approach in \cite{key-21+1}.
So it might be a parallel verification of the holographic Schwinger
effect with anisotropy. }. So this may support the statement of \cite{key-21}, in which the
deconfinement in QCD caused by the extremal anisotropy is predicted
holographically. 
 
\section{Quark pair production}

\begin{figure}
\begin{centering}
\includegraphics[scale=0.3]{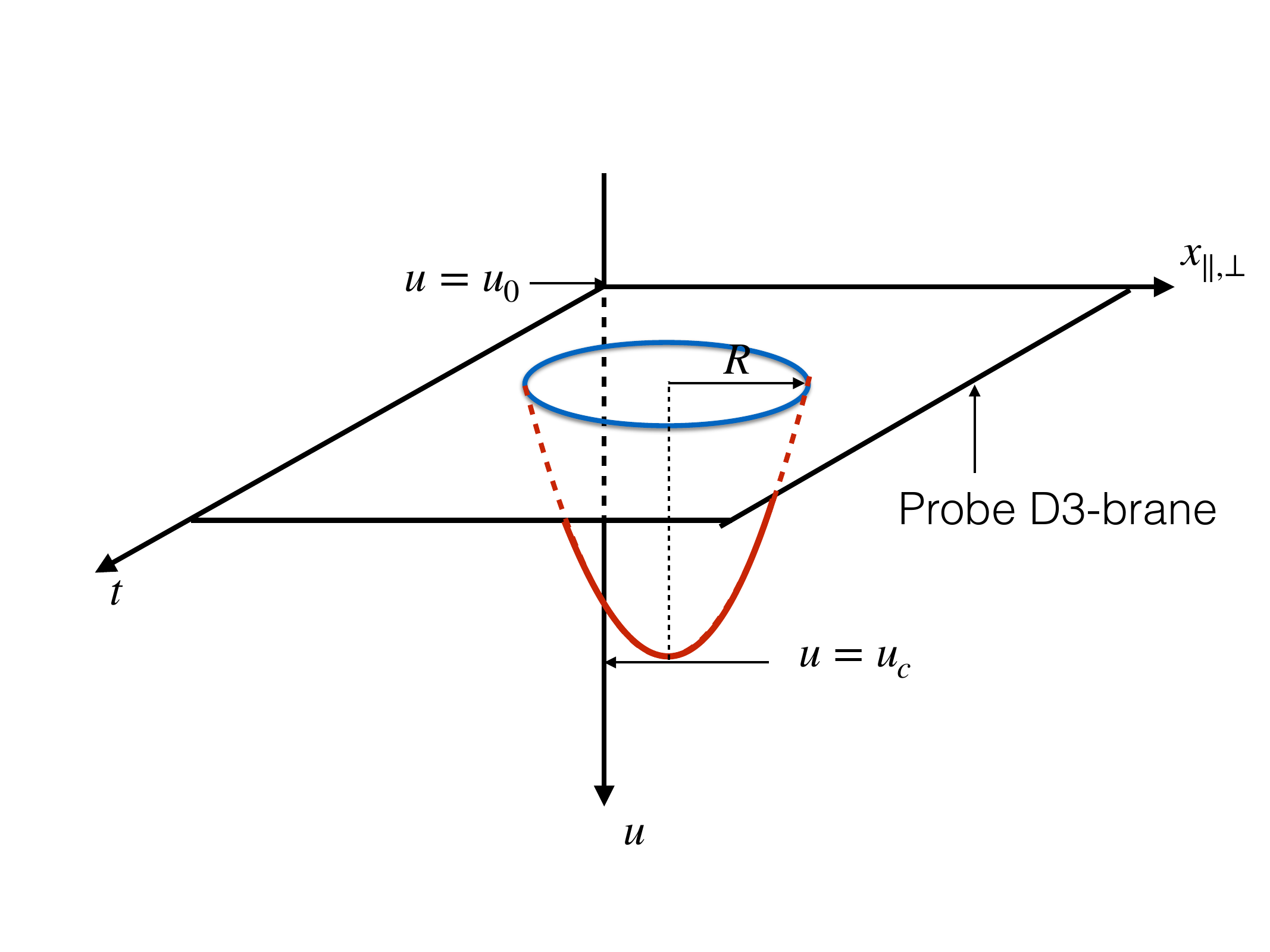}
\par\end{centering}
\caption{\label{fig:8} The circular Wilson loop (blue line) on a probe D3-brane
at $u=u_{0}$ with a fundamental string (red line) stretched in the
bulk.}

\end{figure}

In this section, we will evaluate the pair production rate according
to the AdS/CFT dictionary which is to compute the onshell action of
a fundamental open string in bulk with circular trajectory of each
endpoint in the boundary, as it is illustrated in Figure \ref{fig:8}.
 Since our concern is the radial dependence of string embedding, it
would be convenient to choose the polar coordinates on the probe brane
located at boundary as \cite{key-12,key-16,key-18,key-19}. However
this would be a little tricky due to the anisotropy in the background
geometry. To specify the derivation, let us first take into account
the Wilson loop living in the $\left\{ t,x_{\parallel}\right\} $
plane of the probe D3-brane which is located at $u=u_{0}$. Then impose
the black brane solution with our notation $x_{\parallel}=x$, we
can obtain the induced metric on $\left\{ t,x\right\} $ plane as,

\begin{align}
ds^{2} & =\frac{L^{2}}{u_{0}^{2}}\left[-\mathcal{F}\left(u_{0}\right)\mathcal{B}\left(u_{0}\right)dt^{2}+dx^{2}\right].
\end{align}
Hence the polar coordinates should be introduce as 
\begin{equation}
t=\frac{\bar{t}}{\sqrt{\mathcal{F}\left(u_{0}\right)\mathcal{B}\left(u_{0}\right)}},\bar{t}=\rho\cosh\eta,x=\rho\sinh\eta,\label{eq:31}
\end{equation}
so that

\begin{equation}
ds^{2}=\frac{L^{2}}{u_{0}^{2}}\left(-\rho^{2}d\eta^{2}+d\rho^{2}\right),
\end{equation}
which reduces to a local coordinate transformation. Nonetheless, it
is possible to evaluate the dependence of the anisotropy with this
local coordinate transformation in holography since the asymptotic
behavior of the background geometry near the holographic boundary
is exactly same as the pure AdS spacetime i.e. $\mathcal{F}_{\mathrm{bdry}}=\mathcal{B}_{\mathrm{bdry}}=1,\phi_{\mathrm{bdry}}=0$.
\cite{key-20}. Therefore we can obtain the induced metric on the
string world sheet with the polar coordinates given in (\ref{eq:31})
as,

\begin{equation}
ds^{2}=\frac{L^{2}}{u^{2}}\left[-\rho^{2}d\tau^{2}+\left(1+\frac{u^{\prime2}}{\mathcal{F}}\right)d\rho\right],
\end{equation}
where ``$\prime$'' refers to $du/d\rho$. In this sense the NG
action for such a fundamental string reads,

\begin{align}
S_{NG} & =T_{F}L^{2}\int_{0}^{2\pi}d\eta\int_{0}^{R}d\rho\frac{\rho}{u^{2}}\sqrt{1+\frac{u^{\prime2}}{\mathcal{F}}}\nonumber \\
 & =2\pi T_{F}L^{2}\int_{0}^{x}d\rho\frac{\rho}{u^{2}}\sqrt{1+\frac{u^{\prime2}}{\mathcal{F}}}.\label{eq:34}
\end{align}
Then the equation of motion associated to action is given as,

\begin{equation}
u^{\prime}+\frac{2\rho\mathcal{F}}{u}+\rho u^{\prime\prime}-\frac{\rho u^{\prime2}}{2\mathcal{F}}\frac{d}{du}\mathcal{F}+\frac{u^{\prime3}}{\mathcal{F}}+\frac{2\rho u^{\prime2}}{u}=0,\label{eq:35}
\end{equation}
which is expected to be solved numerically.

For the Wilson loop living in the $\left\{ t,x_{\perp}\right\} $
plane, the NG action with our notation $x_{\perp}=z$ becomes,

\begin{equation}
S_{NG}=2\pi T_{F}L^{2}\int_{0}^{R}d\rho\frac{\mathcal{H}\rho}{u^{2}}\sqrt{1+\frac{u^{\prime2}}{\mathcal{H}\mathcal{F}}},
\end{equation}
where ``$\prime$'' refers to $du/d\rho$ and the local coordinate
transformation is expected to be,

\begin{equation}
t=\bar{t}\sqrt{\frac{\mathcal{H}}{\mathcal{F}\mathcal{B}}},\bar{t}=\rho\cosh\eta,x=\rho\sinh\eta,.
\end{equation}
Then the associated equation of motion is obtained as,

\begin{equation}
\frac{2\rho\mathcal{F}\mathcal{H}}{u}-\rho\mathcal{F}\frac{d}{du}\mathcal{H}+u^{\prime}+\frac{2\rho u^{\prime2}}{u}-\frac{\rho u^{\prime2}}{2\mathcal{F}}\frac{d}{du}\mathcal{F}-\frac{3\rho u^{\prime2}}{2\mathcal{H}}\frac{d}{du}\mathcal{H}+\frac{u^{\prime3}}{\mathcal{F}\mathcal{H}}+\rho u^{\prime\prime}=0.\label{eq:38}
\end{equation}

In the bubble background, let us perform the above discussion again,
then one may find the equation of motion for the fundamental string
is as same as (\ref{eq:35}) and (\ref{eq:38}). Thus what we need
to do next is to solve (\ref{eq:35}) and (\ref{eq:38}) numerically
in order to evaluate the onshell action of the fundamental string.
The boundary condition for (\ref{eq:35}) and (\ref{eq:38}) is illustrated
in Figure \ref{fig:8} which is collected as,

\begin{equation}
u\left(\rho\right)|_{\rho=0}=u_{c},u^{\prime}\left(\rho\right)|_{\rho=0}=0.\label{eq:39}
\end{equation}
Besides, the embedding function must satisfy the constraint as it
is given in \cite{key-15}. Accordingly, in our anisotropic system,
for the Wilson loop living in the $\left\{ t,x_{\parallel}\right\} $
plane, the constraint is, 

\begin{equation}
u^{\prime}\left(\rho\right)|_{\rho=R}=-\sqrt{\mathcal{F}\left(u_{0}\right)\left(\frac{1}{\mathcal{E}_{\parallel}^{2}}-1\right)}.\label{eq:40}
\end{equation}
For the Wilson loop living in the $\left\{ t,x_{\perp}\right\} $
plane, the constraint is, 

\begin{equation}
u^{\prime}|_{\rho=R}=-\sqrt{\mathcal{F}\left(u_{0}\right)\mathcal{H}\left(u_{0}\right)\left(\frac{1}{\mathcal{E}_{\perp}^{2}}-1\right)}.\label{eq:41}
\end{equation}
Altogether, it would be possible to find a numerical solution for
(\ref{eq:35}) (\ref{eq:38}) with the boundary condition (\ref{eq:39})
and the constraints (\ref{eq:40}) (\ref{eq:41}).

\begin{figure}[h]
\begin{centering}
\includegraphics[scale=0.35]{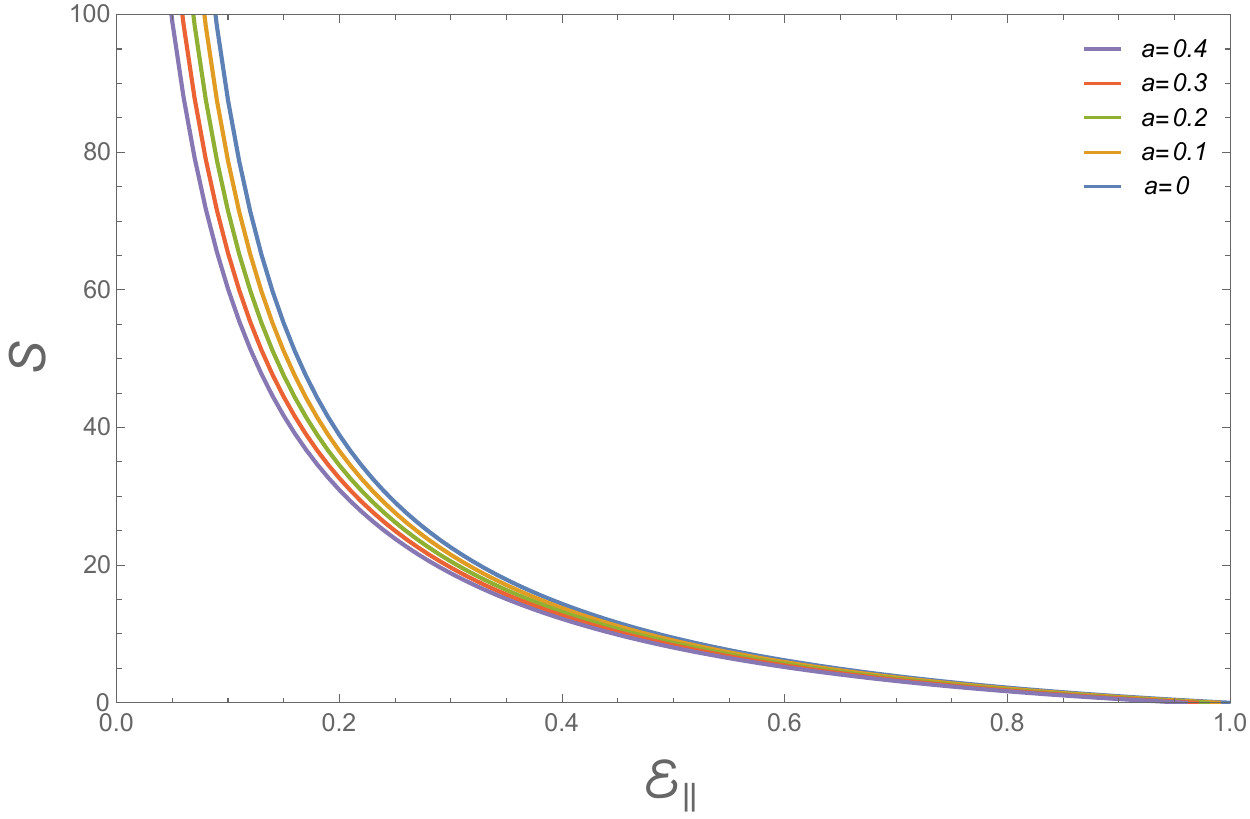}\includegraphics[scale=0.35]{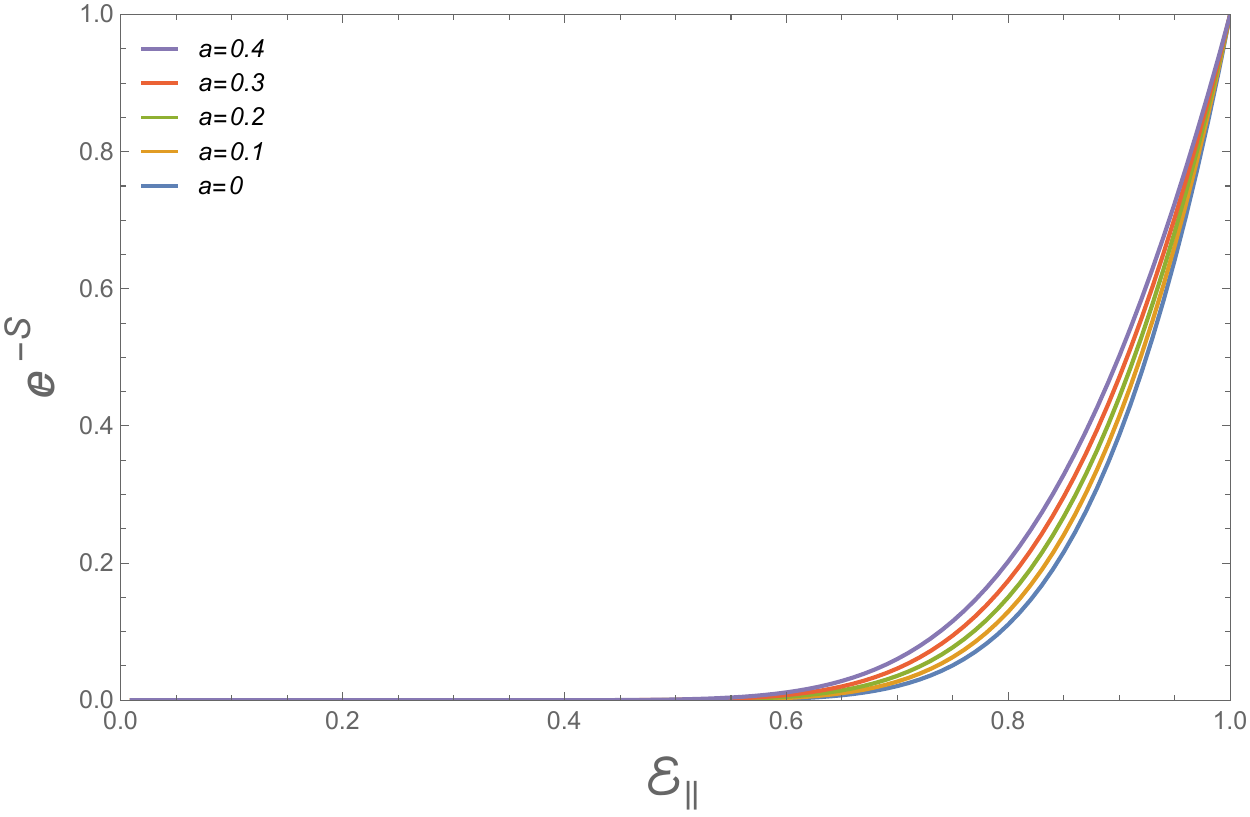}
\par\end{centering}
\begin{centering}
\includegraphics[scale=0.35]{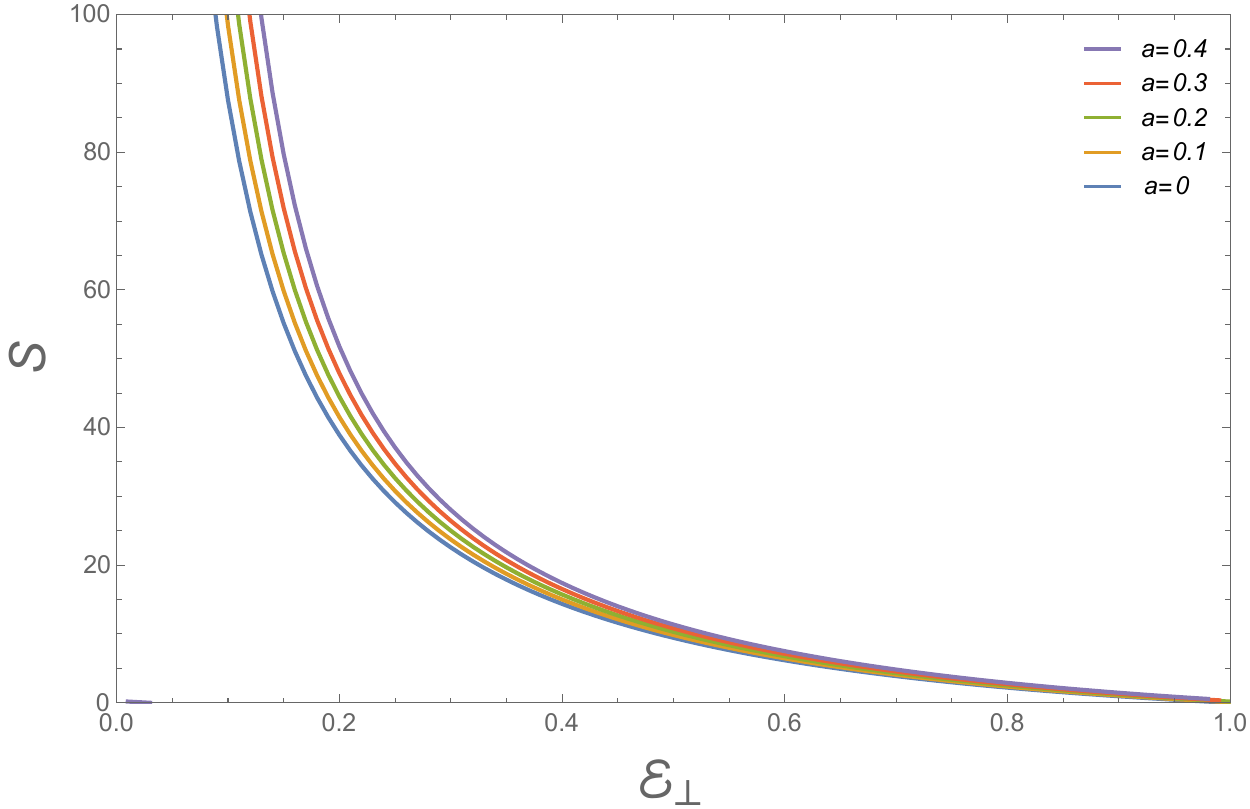}\includegraphics[scale=0.35]{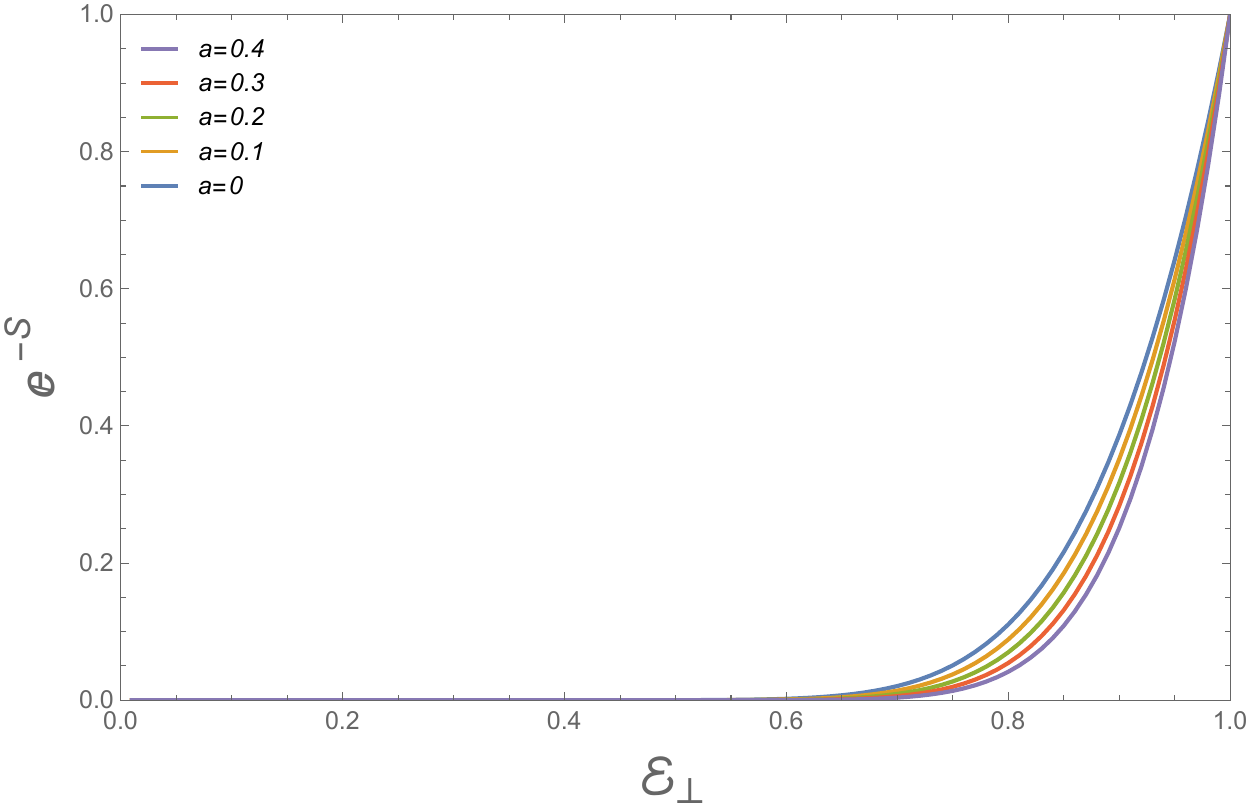}
\par\end{centering}
\caption{\label{fig:9} The production rate as a function of $\mathcal{E}_{\parallel,\perp}$
with the anisotropy denoted by $a$.}
\end{figure}

Afterwards, the pair production rate $\Gamma\sim e^{-S}$ can be evaluated
from the total onshell action $S$ of a fundamental string, which
is given as
\begin{align*}
S= & S_{NG}+S_{B_{2}},\\
S_{B_{2}}= & -2\pi T_{F}B_{01}\int_{0}^{2\pi}d\tau\int_{0}^{R}\rho d\rho=-\pi E_{\parallel,\perp}R^{2}.
\end{align*}
And the behavior of $e^{-S},S$, as a function of $\mathcal{E}_{\parallel,\perp}$,
is numerically computed in Figure \ref{fig:9}. 
 Our numerical calculation displays when the electric field is parallel
(i.e. the case that the Wilson loop living in the $\left\{ t,x_{\parallel}\right\} $
plane), the pair production rate is increased by the anisotropy. However
the pair production rate behaves oppositely in the perpendicular case
(Wilson loop living in the $\left\{ t,x_{\perp}\right\} $ plane).
Interestingly, this conclusion seemingly agrees with the potential
analysis in section 3. Since the potential barrier decreases/increases
in the parallel/perpendicular case by the anisotropy, the production
rate increases/decreases correspondingly in the parallel/perpendicular
case.

\section{Involving the flavor brane}

As most works about D3-D7 approach, it is possible to investigate
the involvement of flavors when a stack of probe flavor D7-branes
is introduced into the bulk. In this section, we will briefly outline
the embedding of the flavor D7-branes, then derive the effective flavored
Lagrangian in dual theory, study the vacuum decay rate for the Schwinger
effect and finally take a look at the V-A curve in this holographic
system.

\subsection{The effective Lagrangian}

To begin with, let us introduce a probe D7-brane to the bulk. While
there are different constructions to embed D7-brane as flavor in the
IIB supergravity background, in this work we would consider the most
widely used configuration of the flavor D7-brane embedding, that is
the flavor D7-brane is parallel to the $N_{c}$ D3-branes. The D-brane
configuration with respect to the black brane and bubble background
is illustrated in Table \ref{tab:3} and \ref{tab:4}. 
\begin{table}
\begin{centering}
\begin{tabular}{|c|c|c|c|c|c|c|c|c|c|c|}
\hline 
Black brane background & 0 & 1 & 2 & 3 & 4 $\left(u\right)$ & 5 & 6 & 7 & 8 & 9\tabularnewline
\hline 
\hline 
$N_{c}$ D3-brane & - & - & - & - &  &  &  &  &  & \tabularnewline
\hline 
$N_{\mathrm{D7}}$ D7-brane & - & - & - &  &  & - & - & - & - & -\tabularnewline
\hline 
$N_{f}$ flavor D7-brane & - & - & - & - & - & - & - & - &  & \tabularnewline
\hline 
\end{tabular}
\par\end{centering}
\caption{\label{tab:3} The D-brane configuration in the black brane background.}

\end{table}
 
\begin{table}
\begin{centering}
\begin{tabular}{|c|c|c|c|c|c|c|c|c|c|c|}
\hline 
Bubble background & 0 & 1 & 2 & 3 $\left(z\right)$ & 4 $\left(u\right)$ & 5 & 6 & 7 & 8 & 9\tabularnewline
\hline 
\hline 
$N_{c}$ D3-brane & - & - & - & - &  &  &  &  &  & \tabularnewline
\hline 
$N_{\mathrm{D7}}$ D7-brane & - & - &  & - &  & - & - & - & - & -\tabularnewline
\hline 
$N_{f}$ flavor D7-brane & - & - & - & - & - & - & - & - &  & \tabularnewline
\hline 
\end{tabular}
\par\end{centering}
\caption{\label{tab:4} The D-brane configuration in the bubble background.}

\end{table}
 According to the gauge-gravity duality, the hypermultiplet is excited
due to the oscillations of an open string connecting the color (D3)
and flavor (D7) branes, thus the massless/massive state corresponds
to the configuration that D7-brane is touched/untouched to the stack
of $N_{c}$ D3-branes, which is the content in the low-energy theory.
Therefore the dual theory will contain the dynamics of flavors when
the action of the flavor brane is taken into account. So the effective
Lagrangian for dual theory is given by the bosonic action for the
flavor D7-brane as\footnote{While there should be a Wess-Zumino term in the D-brane action, it
is vanished according to the supergravity solution in Section 1.},

\begin{equation}
S_{\mathrm{D7}}=-T_{\mathrm{D7}}\int d^{8}xe^{-\phi}\sqrt{-\det\left(g_{ab}+2\pi\alpha^{\prime}F_{ab}\right)},\label{eq:42}
\end{equation}
where $a,b$ runs over the flavor brane, $T_{\mathrm{D7}}$ is the
tension of the D7-brane, $g_{ab}$ refers to the induced metric and
$F_{ab}$ is the gauge field strength on the flavor brane. By picking
up the black brane background (\ref{eq:2}), we can obtain the induced
metric on the flavor D7-brane as,

\begin{equation}
ds_{\mathrm{D7}}^{2}=\frac{L^{2}}{u^{2}}\left(-\mathcal{F}\mathcal{B}dt^{2}+dx^{2}+dy^{2}+\mathcal{H}dz^{2}+\frac{du^{2}}{\mathcal{F}}\right)+L^{2}\mathcal{Z}d\Omega_{3}^{2}.
\end{equation}
When the bubble background (\ref{eq:5}) is imposed, the induced metric
on the flavor D7-brane becomes,

\begin{equation}
ds_{\mathrm{D7}}^{2}=\frac{L^{2}}{u^{2}}\left(-dt^{2}+dx^{2}+\mathcal{H}dy^{2}+\mathcal{F}\mathcal{B}dz^{2}+\frac{du^{2}}{\mathcal{F}}\right)+L^{2}\mathcal{Z}d\Omega_{3}^{2}.
\end{equation}

In addition, we need to turn on the external electric fields to evaluate
electric features in dual theory as \cite{key-23,key-24,key-25,key-26}.
Due to the anisotropy in the background geometry, we set the nonzero
components of the gauge field strength as $F_{01}=E_{\parallel},F_{0u},F_{1u}$
for the case that the electric field is parallel to the $N_{\mathrm{D7}}$
D7-branes in both backgrounds. For the case that the electric field
is perpendicular to the $N_{\mathrm{D7}}$ D7-branes, we turn on the
components of the gauge field strength as $F_{03}=E_{\perp},F_{0u},F_{3u}$
in the black brane background. Notice that in the perpendicular case,
the nonzero components of the gauge field strength should be $F_{02}=E_{\perp},F_{0u},F_{2u}$
in the bubble background since the anisotropy is exhibited by $g_{yy}$
of the metric. Here we use $E_{\parallel,\perp}$ to refer to the
homogeneous external electric field. Keeping these in hand, the effective
action (\ref{eq:42}) could be written as, in the black brane background,

\begin{align}
S_{\mathrm{D7}} & =-T_{\mathrm{D7}}V_{S^{3}}V_{4}L^{8}\int due^{-3\phi/2}\frac{\mathcal{Z}^{3/2}\mathcal{B}^{1/2}}{u^{5}}\sqrt{\xi_{\parallel,\perp}},\label{eq:45}
\end{align}
where,

\begin{align}
\xi_{\parallel} & =1-\left(2\pi\alpha^{\prime}\right)^{2}\frac{u^{4}}{L^{4}}\left[\frac{F_{01}^{2}}{\mathcal{F}\mathcal{B}}+\frac{F_{0u}^{2}}{\mathcal{B}}-\mathcal{F}F_{1u}^{2}\right],\nonumber \\
\xi_{\perp} & =1-\left(2\pi\alpha^{\prime}\right)^{2}\frac{u^{4}}{L^{4}}\left[\frac{e^{\phi}}{\mathcal{F}\mathcal{B}}F_{03}^{2}+\frac{F_{0u}^{2}}{\mathcal{B}}-e^{\phi}\mathcal{F}F_{3u}^{2}\right].\label{eq:46}
\end{align}
And in the bubble background, the effective action (\ref{eq:42})
becomes,

\begin{align}
S_{\mathrm{D7}} & =-T_{\mathrm{D7}}V_{S^{3}}V_{3}\beta_{z}L^{8}\int due^{-3\phi/2}\frac{\mathcal{Z}^{3/2}\mathcal{B}^{1/2}}{u^{5}}\sqrt{\xi_{\parallel,\perp}},\label{eq:47}
\end{align}
where $\beta_{z}$ refers to the size of the compatified direction
$z$ and,

\begin{align}
\xi_{\parallel} & =1-\left(2\pi\alpha^{\prime}\right)^{2}\frac{u^{4}}{L^{4}}\left[F_{01}^{2}+\mathcal{F}\left(F_{0u}^{2}-F_{1u}^{2}\right)\right].\nonumber \\
\xi_{\perp} & =1-\left(2\pi\alpha^{\prime}\right)^{2}\frac{u^{4}}{L^{4}}\left[\mathcal{F}F_{0u}^{2}+e^{\phi}\left(F_{02}^{2}-\mathcal{F}F_{2u}^{2}\right)\right].
\end{align}

\subsection{The electric instability and vacuum decay rate}

\begin{figure}[h]
\begin{centering}
\includegraphics[scale=0.34]{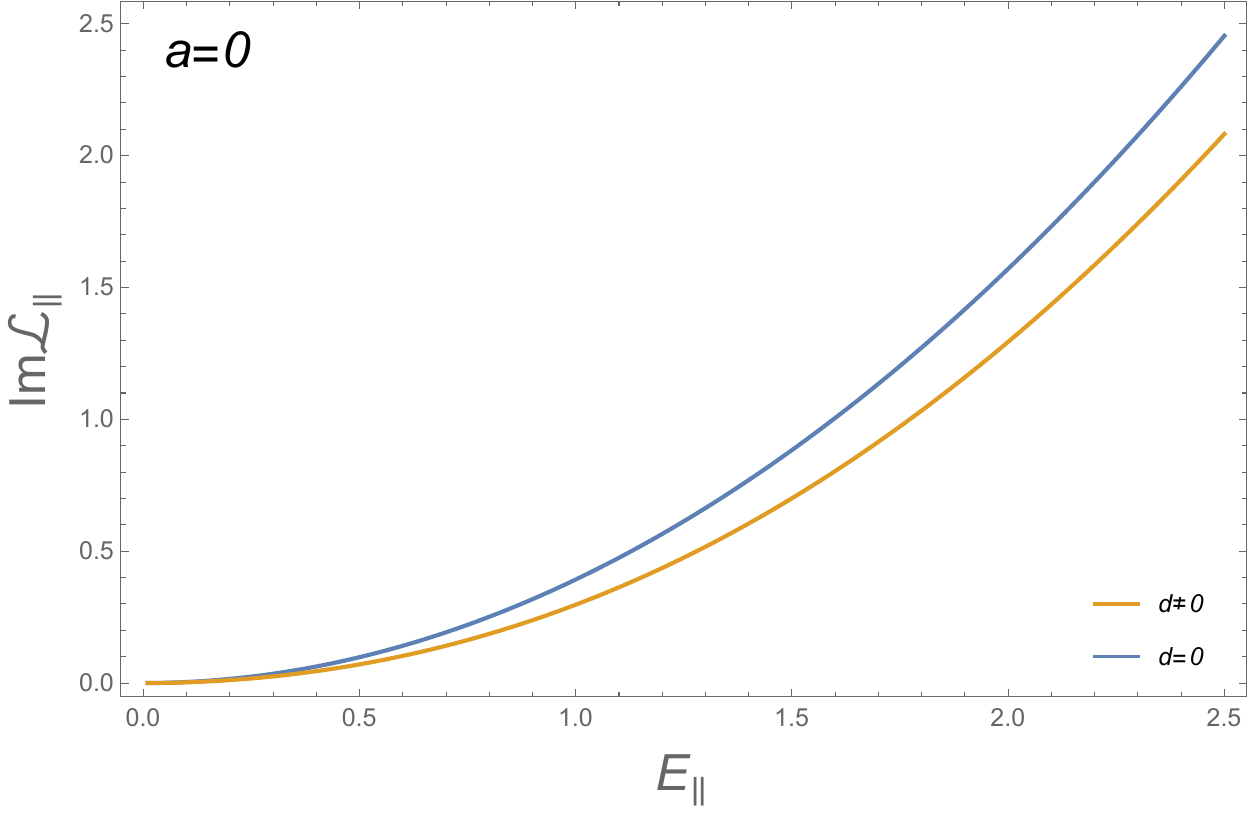}\includegraphics[scale=0.34]{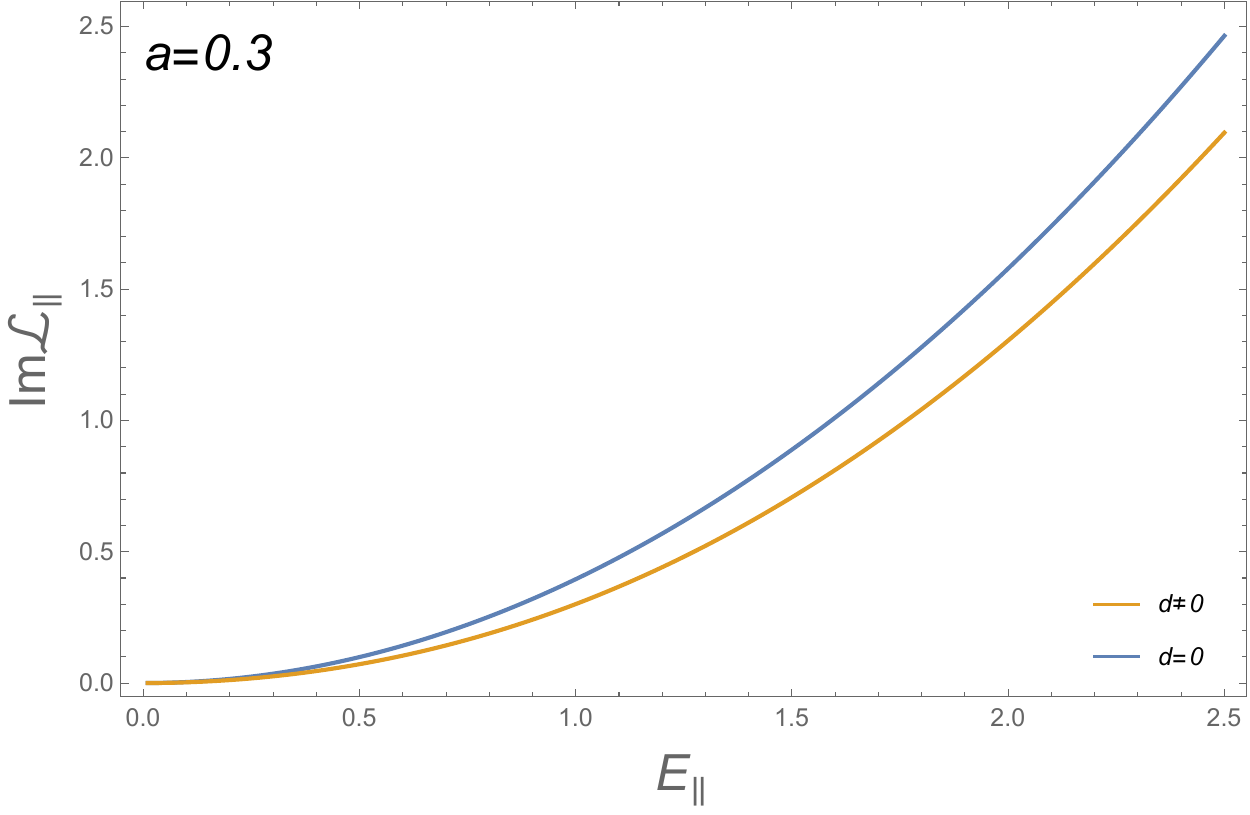}
\par\end{centering}
\begin{centering}
\includegraphics[scale=0.34]{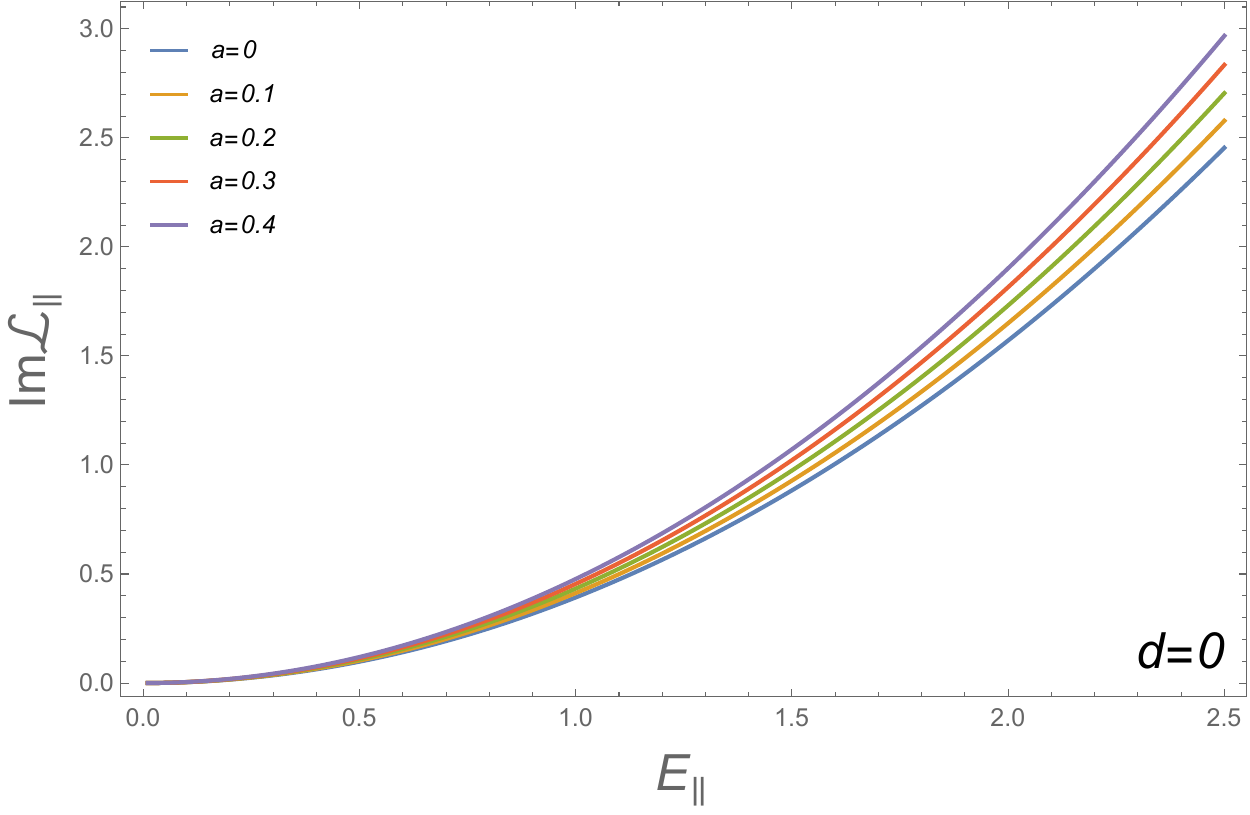}\includegraphics[scale=0.34]{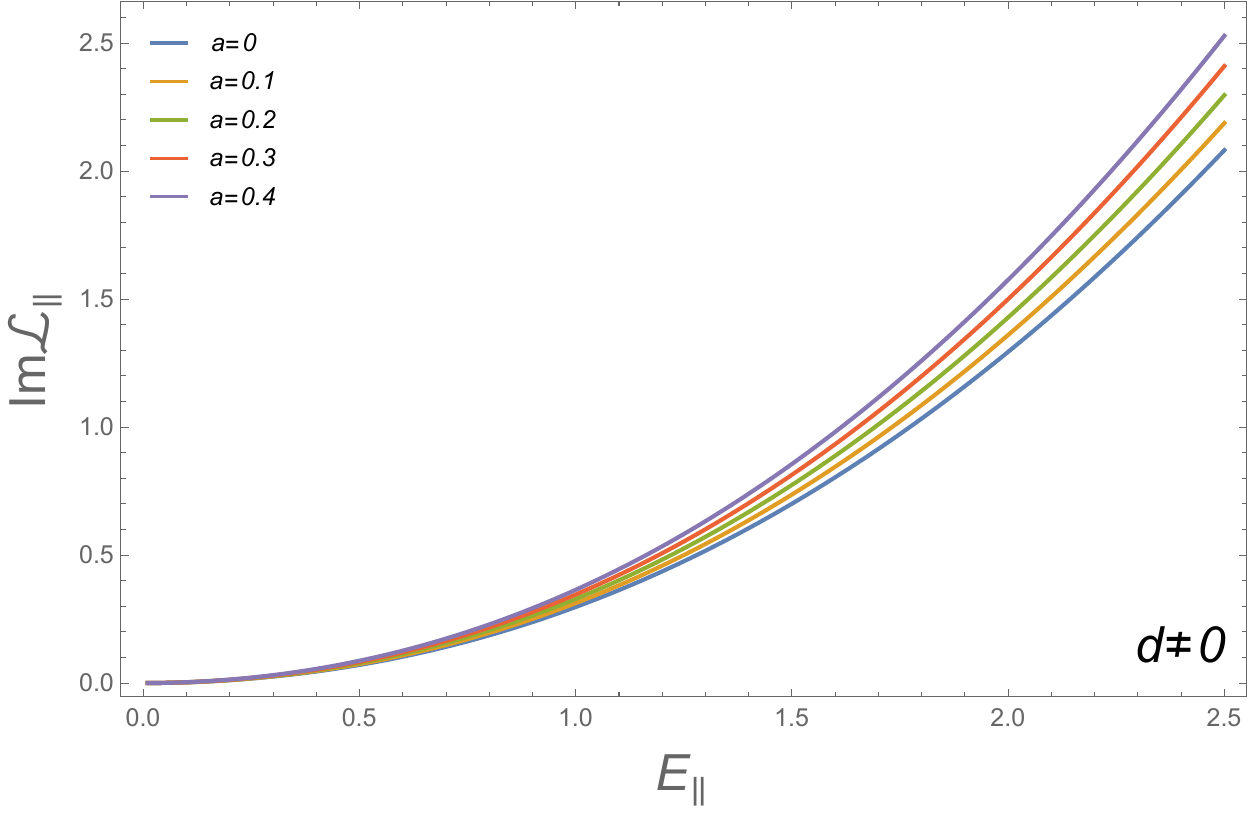}
\par\end{centering}
\caption{\label{fig:10} The imaginary part of the Lagrangian $\mathcal{L}_{\parallel}$
in (\ref{eq:54}) as a function of $E_{\parallel}$ with various $a$
in the black brane background.}
\end{figure}

\begin{figure}[h]
\begin{centering}
\includegraphics[scale=0.34]{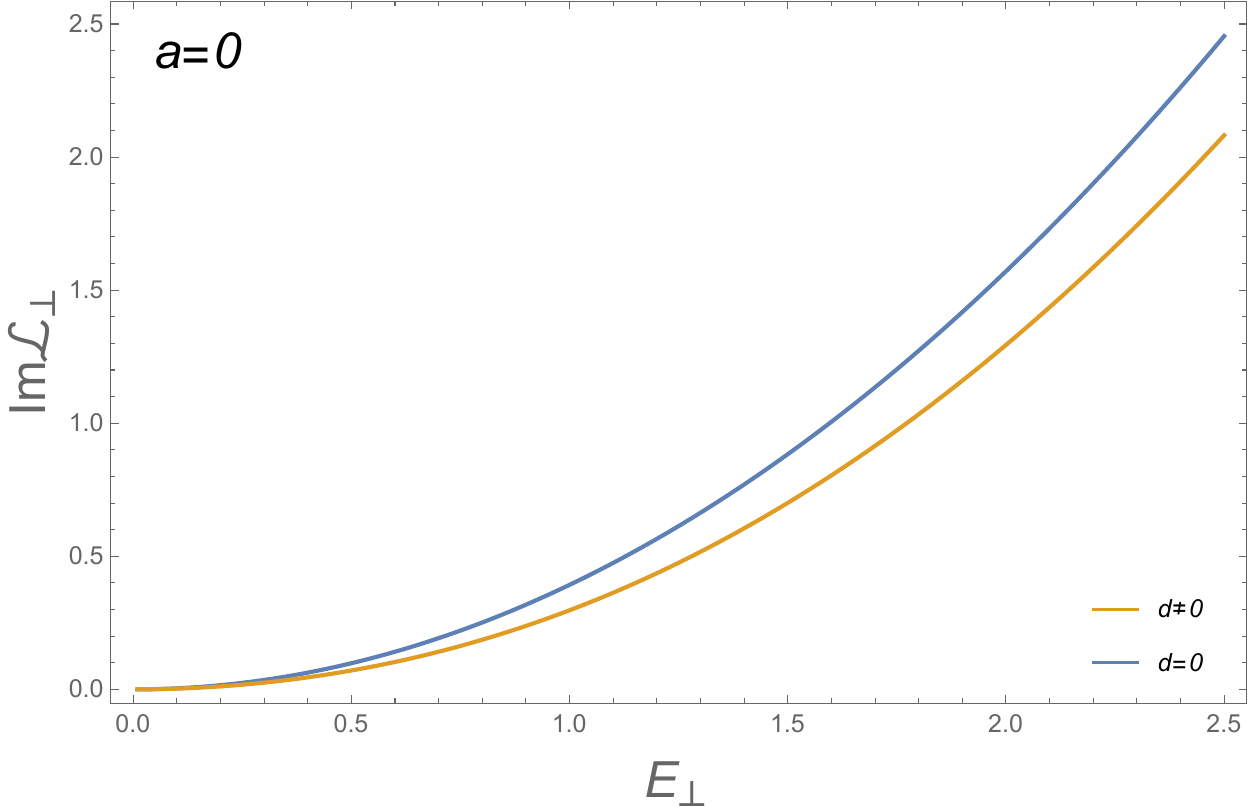}\includegraphics[scale=0.34]{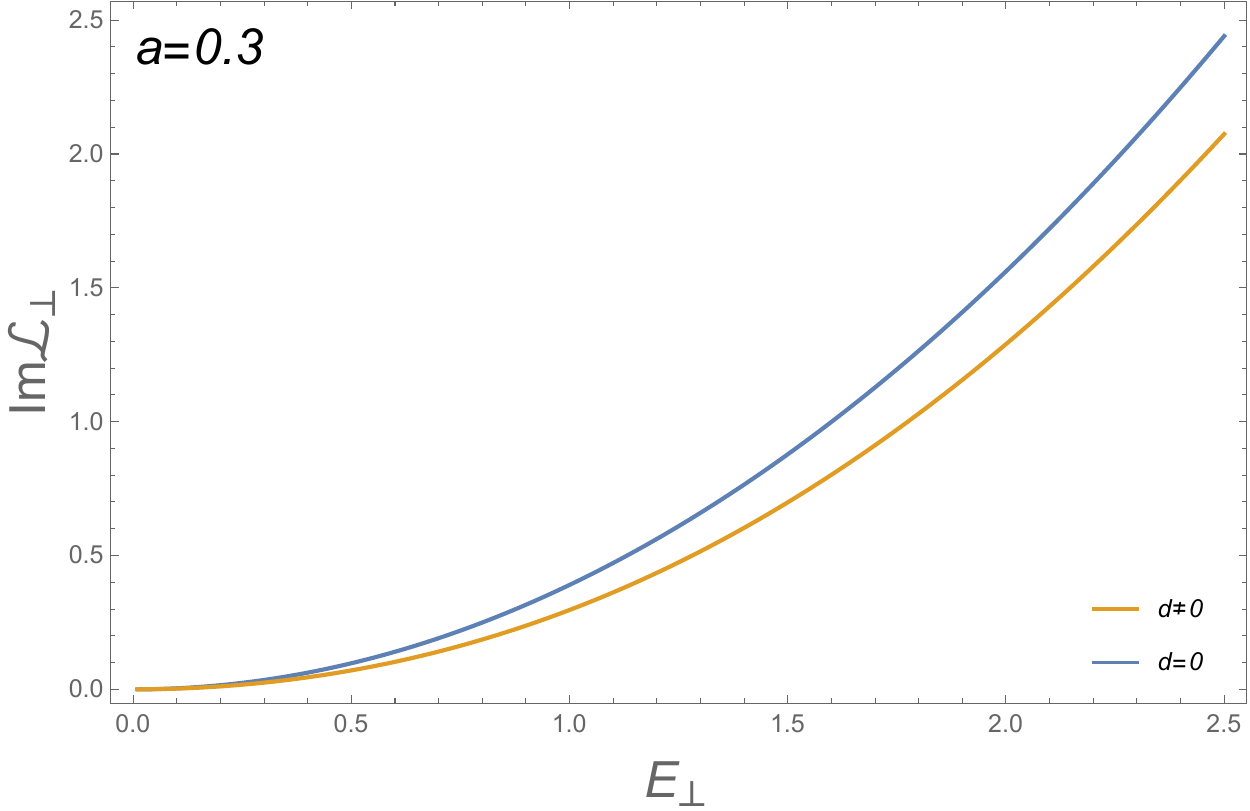}
\par\end{centering}
\begin{centering}
\includegraphics[scale=0.34]{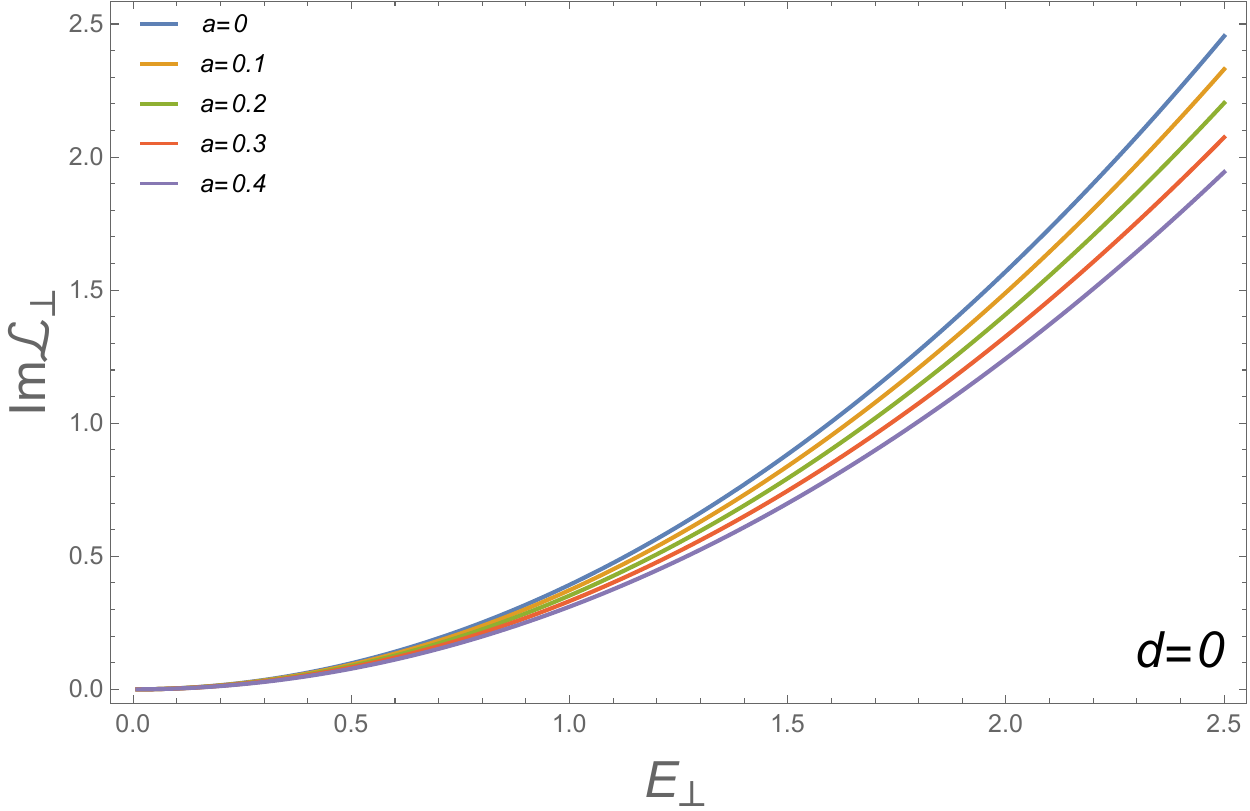}\includegraphics[scale=0.34]{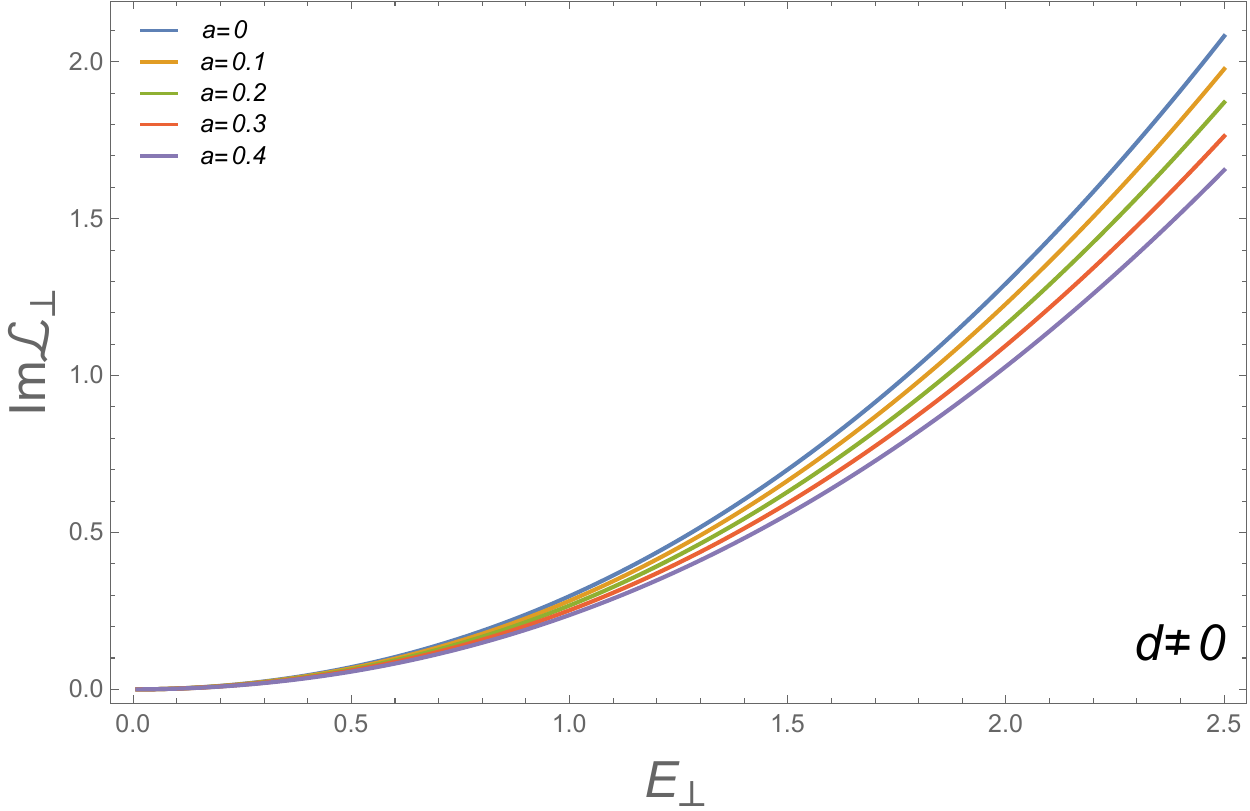}
\par\end{centering}
\caption{\label{fig:11} The imaginary part of the Lagrangian $\mathcal{L}_{\perp}$
in (\ref{eq:54}) as a function of $E_{\perp}$ with various $a$
in the black brane background.}
\end{figure}

The electric instability and vacuum decay rate can be studied by analyzing
the effective Lagrangian in dual theory since it may relate to the
vacuum-to-vacuum amplitude \cite{key-23} as,

\begin{equation}
\left\langle 0\left|U\left(t\right)\right|0\right\rangle =e^{i\mathcal{L}vt},\label{eq:49}
\end{equation}
where $U\left(t\right)$ refers to the time-evolution operator with
external fields, $v$ represents the spatial volume and $\left|0\right\rangle $
refers to the vacuum state without external fields. So if the effective
Lagrangian includes an imaginary part as \footnote{For example, QED has the imaginary part of its Lagrangian up to 1-loop
order as, 
\begin{align*}
\mathrm{Im}\mathcal{L}_{\mathrm{spinor}}^{\mathrm{1-loop}} & =\frac{e^{2}E^{2}}{8\pi^{3}}\sum_{n=1}^{\infty}\frac{1}{n^{2}}\mathrm{exp}\left(-\frac{\pi m^{2}}{eE}n\right),\\
\mathrm{Im}\mathcal{L}_{\mathrm{scalar}}^{\mathrm{1-loop}} & =\frac{e^{2}E^{2}}{16\pi^{3}}\sum_{n=1}^{\infty}\frac{1}{n^{2}}\mathrm{exp}\left(-\frac{\pi m^{2}}{eE}n\right),
\end{align*}
which relates to a single quantum tunneling process in Schwinger effect
where a pair of an electron and a positron is created from the vacuum.},

\begin{equation}
\mathcal{L}=\mathrm{Re}\mathcal{L}+i\frac{\Gamma}{2},\label{eq:50}
\end{equation}
$\Gamma$ could be interpreted as the the vacuum decay rate. Accordingly,
in order to evaluate the vacuum decay rate $\Gamma$ for the Schwinger
effect, we need to derive the imaginary part of the effective Lagrangian
with external electric fields presented in (\ref{eq:45}) (\ref{eq:47})
via holography. To achieve this goal, the equations of motion for
the gauge field on the flavor brane are necessary which are \footnote{The non-vanished components of the gauge field potential can be set
as $A_{0,1}$.}, in the black brane background,

\begin{align}
\partial_{1,3}\left(e^{-3\phi/2}\frac{\mathcal{Z}^{3/2}\mathcal{B}^{1/2}}{u^{5}}\frac{1}{\sqrt{\xi_{\parallel,\perp}}}\frac{\partial\xi_{\parallel,\perp}}{\partial F_{01,03}}\right)+\partial_{u}\left(e^{-3\phi/2}\frac{\mathcal{Z}^{3/2}\mathcal{B}^{1/2}}{u^{5}}\frac{1}{\sqrt{\xi_{\parallel,\perp}}}\frac{\partial\xi_{\parallel,\perp}}{\partial F_{0u}}\right) & =0,\nonumber \\
\partial_{u}\left(e^{-3\phi/2}\frac{\mathcal{Z}^{3/2}\mathcal{B}^{1/2}}{u^{5}}\frac{1}{\sqrt{\xi_{\parallel,\perp}}}\frac{\partial\xi_{\parallel,\perp}}{\partial F_{1u,3u}}\right) & =0.\label{eq:51}
\end{align}
Since nothing depends on $x^{1}$or $x^{3}$ in (\ref{eq:51}), it
leads to $\partial_{1,3}=0$ and two constants defined as,

\begin{align}
\left(2\pi\alpha^{\prime}\right)d & =e^{-3\phi/2}\frac{\mathcal{Z}^{3/2}\mathcal{B}^{1/2}}{2u^{5}}\frac{1}{\sqrt{\xi_{\parallel,\perp}}}\frac{\partial\xi_{\parallel,\perp}}{\partial F_{0u}},\nonumber \\
\left(2\pi\alpha^{\prime}\right)j & =e^{-3\phi/2}\frac{\mathcal{Z}^{3/2}\mathcal{B}^{1/2}}{2u^{5}}\frac{1}{\sqrt{\xi_{\parallel,\perp}}}\frac{\partial\xi_{\parallel,\perp}}{\partial F_{1u,3u}},\label{eq:52}
\end{align}
which is recognized as the electric charge $d$ and current $j$.
Plugging (\ref{eq:52}) back into (\ref{eq:46}) with respect to the
case of ``$\parallel$'' and ``$\perp$'', we can obtain,

\begin{align}
\xi_{\parallel} & =\frac{1-\left(2\pi\alpha^{\prime}E_{\parallel}\right)^{2}\frac{u^{4}}{L^{4}}\frac{1}{\mathcal{F}\mathcal{B}}}{1+\frac{e^{3\phi}u^{6}}{L^{4}\mathcal{Z}^{3}}\left(d^{2}-\frac{1}{\mathcal{F}\mathcal{B}}j^{2}\right)},\nonumber \\
\xi_{\perp} & =\frac{1-\left(2\pi\alpha^{\prime}E_{\perp}\right)^{2}\frac{u^{4}}{L^{4}}\frac{e^{\phi}}{\mathcal{B}\mathcal{F}}}{1+\frac{e^{3\phi}u^{6}}{L^{4}\mathcal{Z}^{3}}\left(d^{2}-\frac{e^{-\phi}}{\mathcal{B}\mathcal{F}}j^{2}\right)},\label{eq:53}
\end{align}
in the black brane background. Remarkably, for any stable configuration,
it demands $\xi_{\parallel,\perp}>0$ since the action (\ref{eq:45})
contains a factor $\sqrt{\xi_{\parallel,\perp}}$. However, general
choice of $E_{\parallel,\perp},d,j$ may lead to negative quantities
for $\xi_{\parallel,\perp}$. Thus it implies there must be a critical
position $u=u_{*},0<u_{*}<u_{H}$ where the denominator and numerator
in $\xi_{\parallel,\perp}$ change their sign together. Then recall
(\ref{eq:49}) (\ref{eq:50}) and suppose the electric field is turned
on at $t=0$ suddenly, the original vacuum without electric current
($j=0$) would become unstable under the external electric field $E_{\parallel,\perp}$.
Therefore the factor $\sqrt{\xi_{\parallel,\perp}}$ could not be
real in the region $u\in\left[0,u_{H}\right]$ which leads to an imaginary
part of the effective Lagrangian as,

\begin{align}
\mathrm{Im}\mathcal{L}_{\parallel} & =\int_{u_{*}}^{u_{H}}due^{-3\phi/2}\frac{\mathcal{Z}^{3/2}\mathcal{B}^{1/2}}{u^{5}}\sqrt{\left[\left(2\pi\alpha^{\prime}E_{\parallel}\right)^{2}\frac{u^{4}}{L^{4}}\frac{1}{\mathcal{F}\mathcal{B}}-1\right]\left(1+\frac{e^{3\phi}u^{6}}{L^{4}\mathcal{Z}^{3}}d^{2}\right)^{-1}},\nonumber \\
\mathrm{Im}\mathcal{L}_{\perp} & =\int_{u_{*}}^{u_{H}}due^{-3\phi/2}\frac{\mathcal{Z}^{3/2}\mathcal{B}^{1/2}}{u^{5}}\sqrt{\left[\left(2\pi\alpha^{\prime}E_{\perp}\right)^{2}\frac{u^{4}}{L^{4}}\frac{e^{\phi}}{\mathcal{B}\mathcal{F}}-1\right]\left(1+\frac{e^{3\phi}u^{6}}{L^{4}\mathcal{Z}^{3}}d^{2}\right)^{-1}},\label{eq:54}
\end{align}
corresponding to the vacuum decay rate $\Gamma$. And the critical
position $u_{*}$ is determined by constraining (\ref{eq:53}), for
parallel case, as

\begin{align}
1-\left(2\pi\alpha^{\prime}E_{\parallel}\right)^{2}\frac{u_{*}^{4}}{L^{4}}\frac{1}{\mathcal{F}\left(u_{*}\right)\mathcal{B}\left(u_{*}\right)} & =0,\nonumber \\
1+\frac{e^{3\phi\left(u_{*}\right)}u_{*}^{6}}{L^{4}\mathcal{Z}\left(u_{*}\right)^{3}}\left[d^{2}-\frac{1}{\mathcal{F}\left(u_{*}\right)\mathcal{B}\left(u_{*}\right)}j^{2}\right] & =0.\label{eq:55}
\end{align}
For perpendicular case, $u_{*}$ is determined by,

\begin{align}
1-\left(2\pi\alpha^{\prime}E_{\perp}\right)^{2}\frac{u_{*}^{4}}{L^{4}}\frac{e^{\phi\left(u_{*}\right)}}{\mathcal{B}\left(u_{*}\right)\mathcal{F}\left(u_{*}\right)} & =0,\nonumber \\
1+\frac{e^{3\phi\left(u_{*}\right)}u_{*}^{6}}{L^{4}\mathcal{Z}\left(u_{*}\right)^{3}}\left[d^{2}-\frac{e^{-\phi\left(u_{*}\right)}}{\mathcal{B}\left(u_{*}\right)\mathcal{F}\left(u_{*}\right)}j^{2}\right] & =0.\label{eq:56}
\end{align}
Using (\ref{eq:54}), the behavior of $\mathrm{Im}\mathcal{L}_{\parallel,\perp}$
as a function of $E_{\parallel,\perp}$ is plotted numerically in
Figure \ref{fig:10} and \ref{fig:11} by solving the constraint of
$E_{\parallel,\perp}$ and $u_{*}$ in (\ref{eq:55}) (\ref{eq:56}).
 
 Our numerical calculation displays the vacuum decay rate is a monotonic
function of $E_{\parallel,\perp}$ which agrees with the intuition
that strong electric field would induce instability. On the other
hand, as the analytical formulas of $\mathcal{F},\mathcal{B},\phi$
describes the situation of high temperature in the dual theory which
is expected to be (like) deconfined quark-gluon plasma, thus there
would not be a definitely critical value of the electric field for
the Schwinger effect in conductive plasma. In addition, this holographic
approach reveals that the vacuum decay rate would be increased/decreased
by the presence of anisotropy with respect to parallel and perpendicular
electric field $E_{\parallel,\perp}$. And this conclusion remains
to agree with the potential analysis in Section 3 and the behaviors
of the pari production rate given in Section 4.

\begin{figure}[h]
\begin{centering}
\includegraphics[scale=0.34]{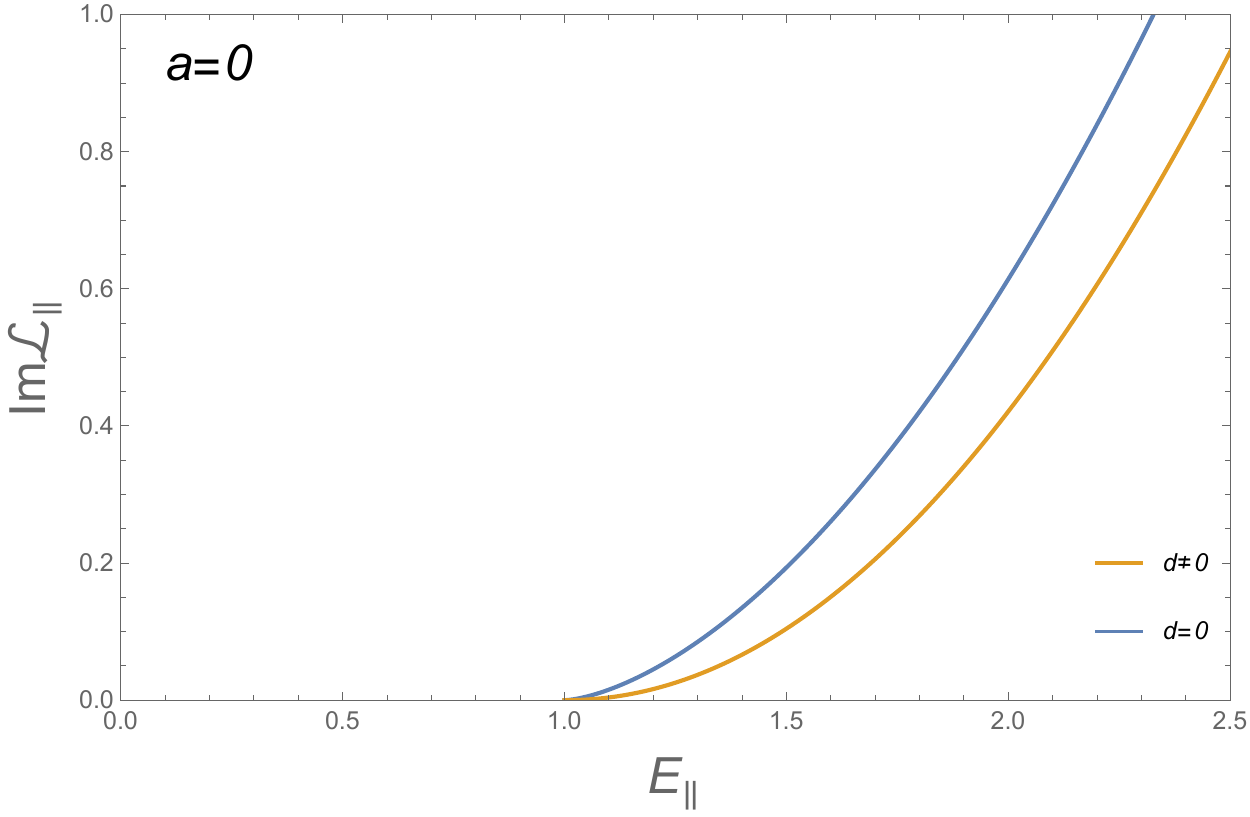}\includegraphics[scale=0.34]{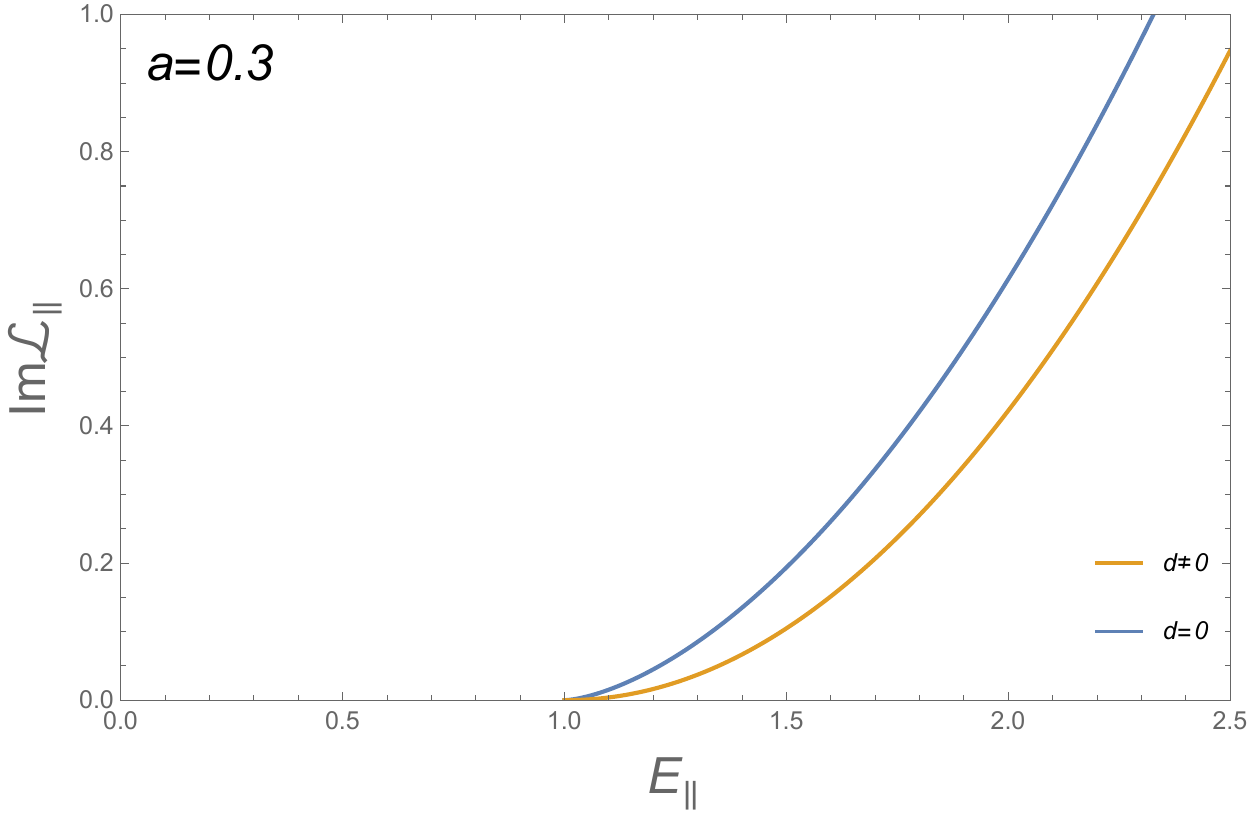}
\par\end{centering}
\begin{centering}
\includegraphics[scale=0.34]{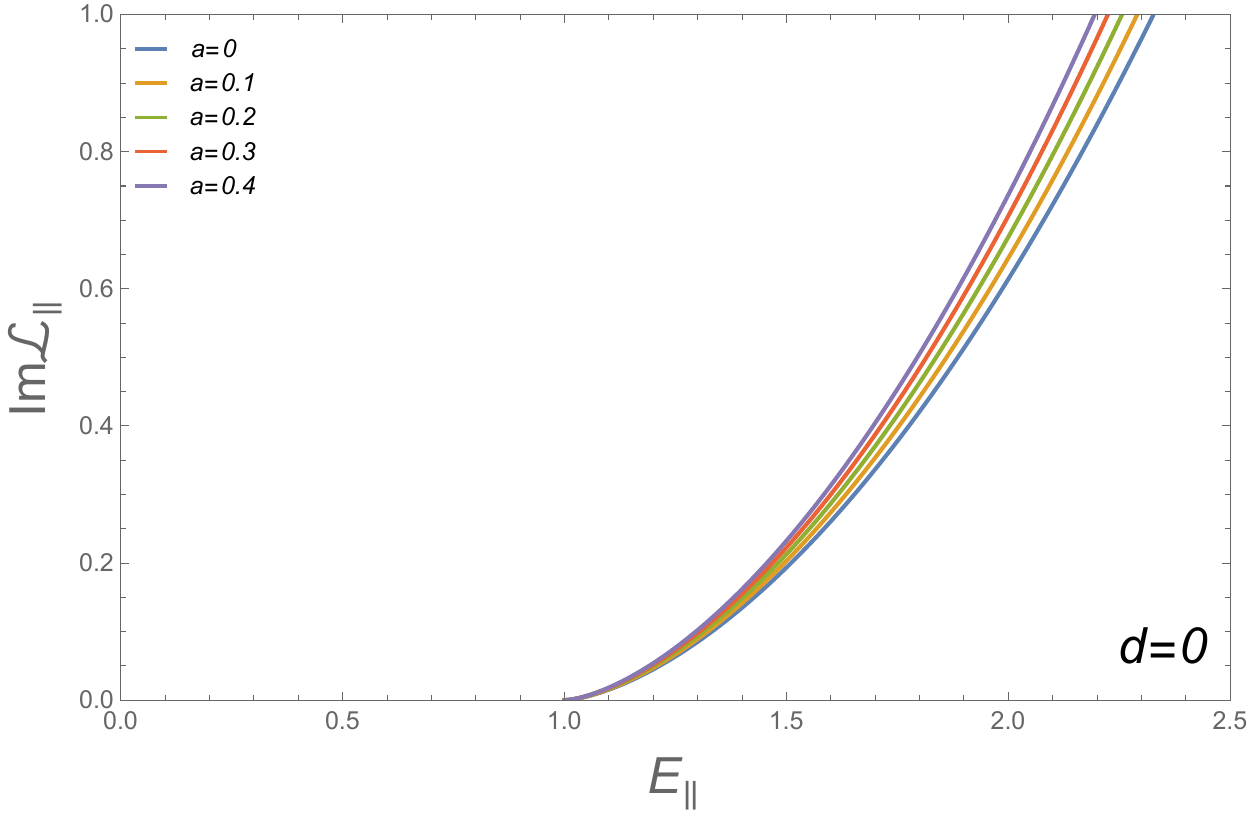}\includegraphics[scale=0.34]{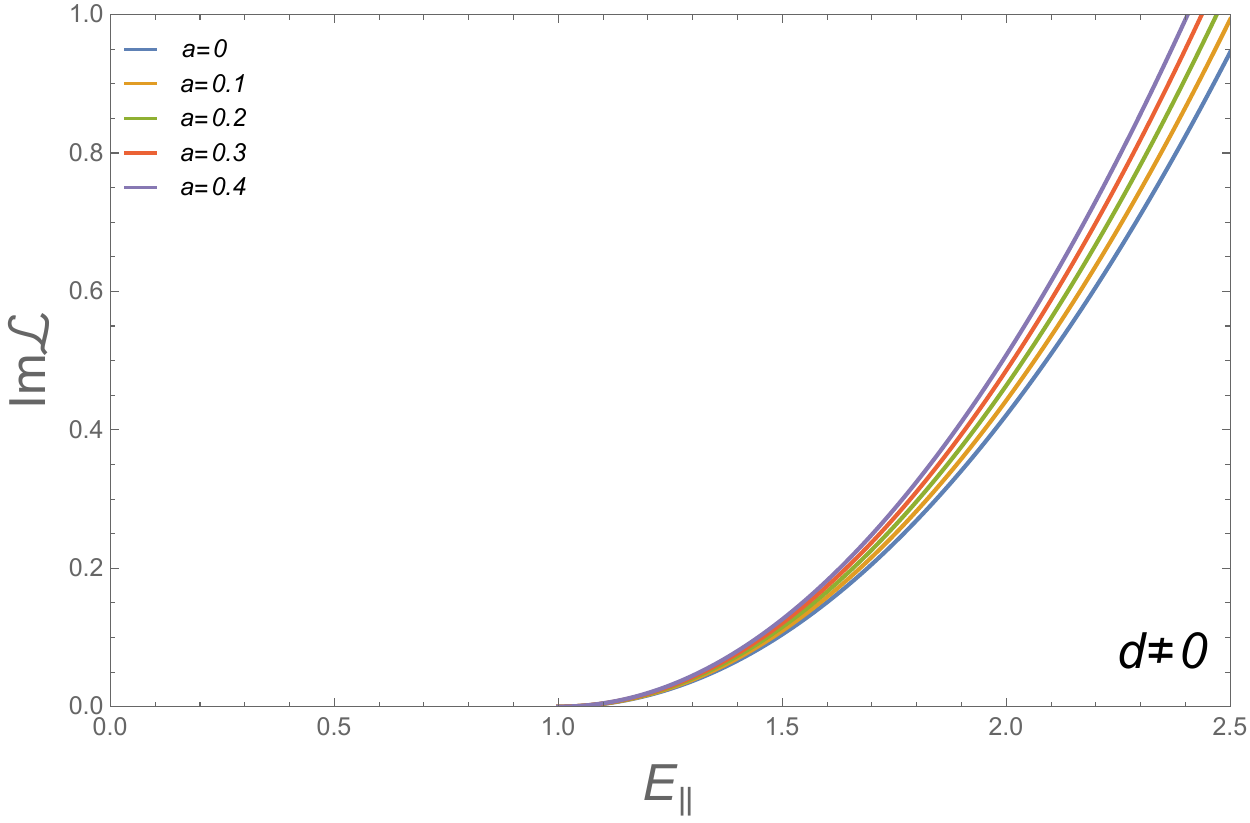}
\par\end{centering}
\caption{\label{fig:12} The imaginary part of the Lagrangian $\mathcal{L}_{\parallel}$
in (\ref{eq:60}) as a function of $E_{\parallel}$ with various $a$
in the bubble brane background.}
\end{figure}

\begin{figure}[h]
\begin{centering}
\includegraphics[scale=0.34]{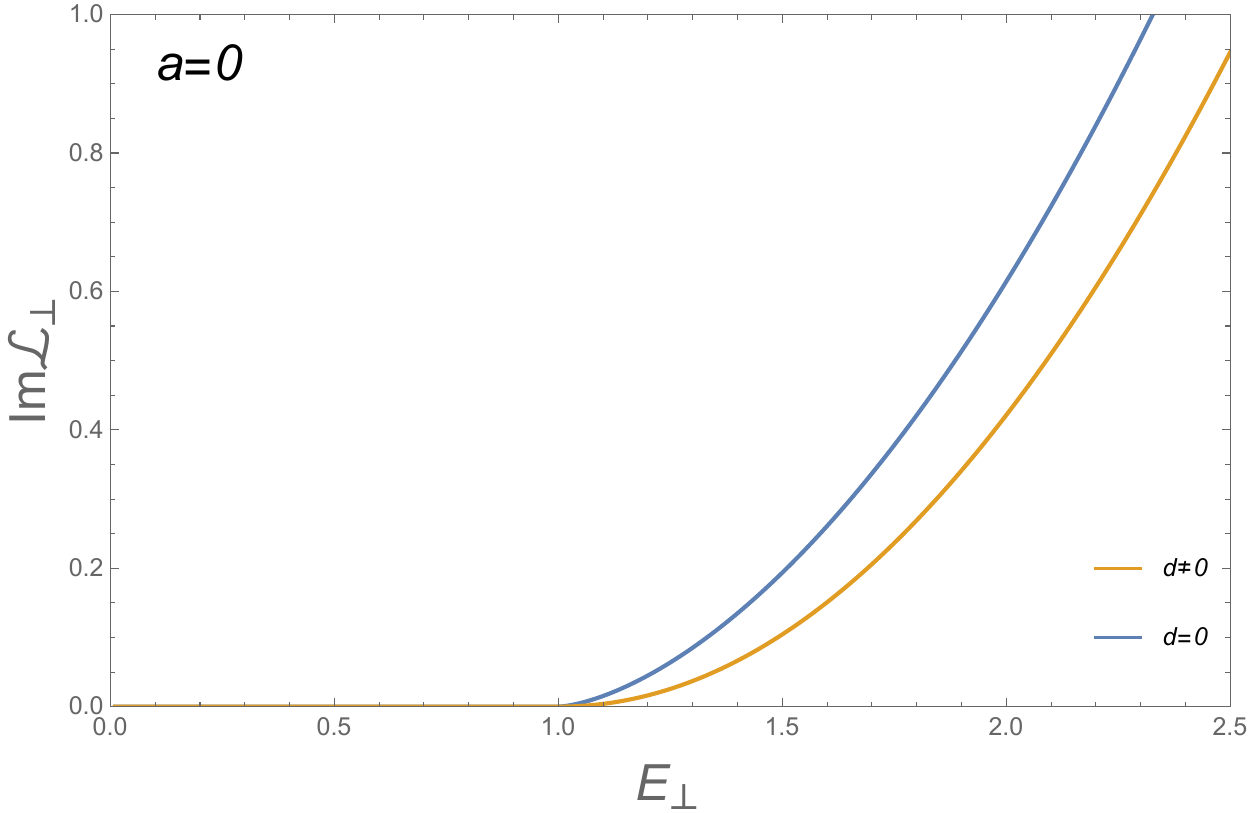}\includegraphics[scale=0.34]{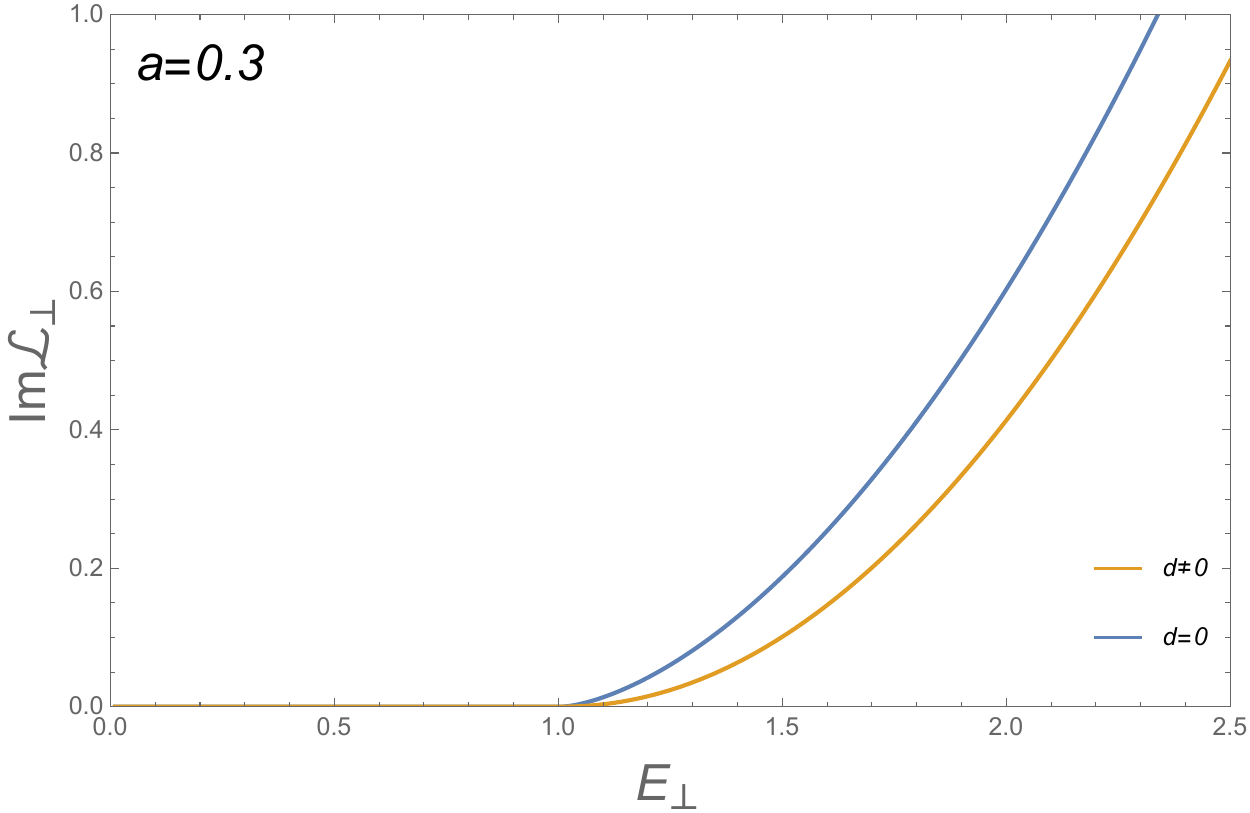}
\par\end{centering}
\begin{centering}
\includegraphics[scale=0.34]{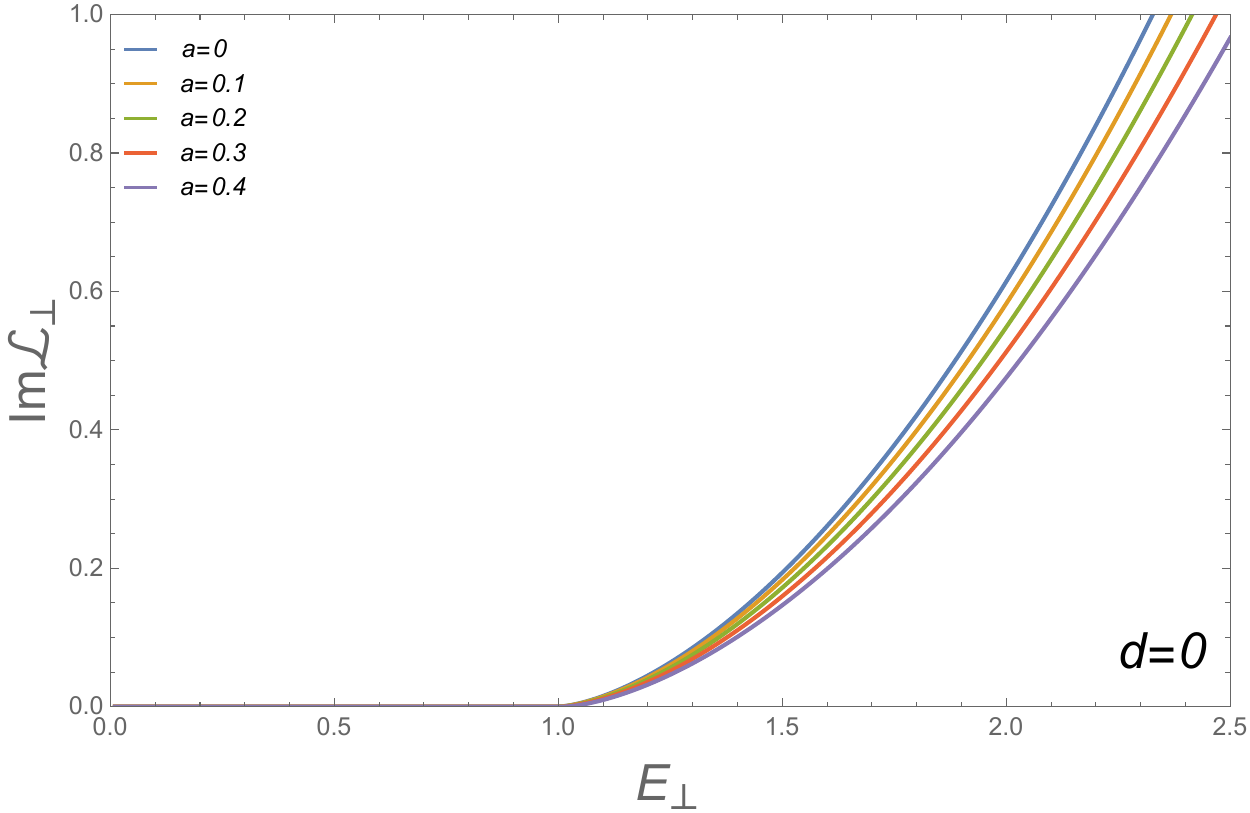}\includegraphics[scale=0.34]{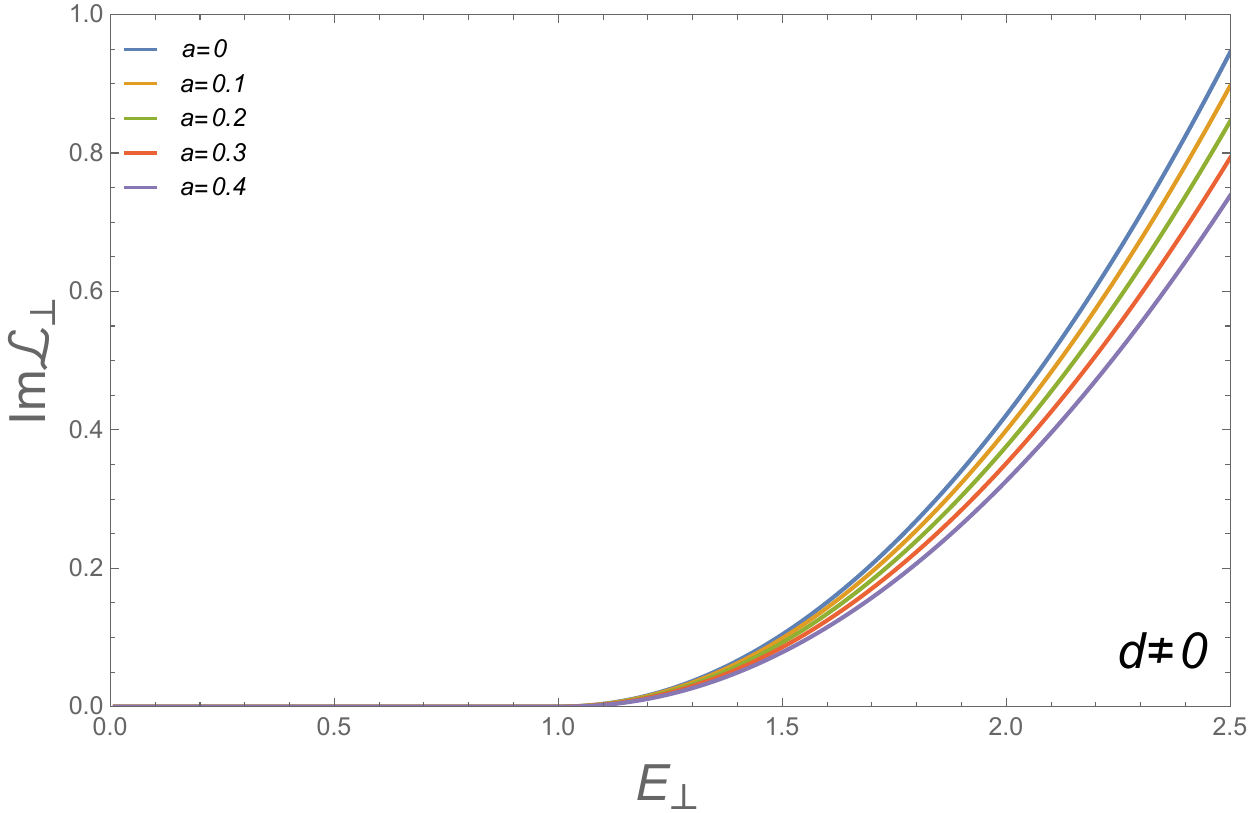}
\par\end{centering}
\caption{\label{fig:13} The imaginary part of the Lagrangian $\mathcal{L}_{\perp}$
in (\ref{eq:60}) as a function of $E_{\perp}$ with various $a$
in the bubble brane background.}
\end{figure}

Similarly, when the bubble background is picked up, the equations
of motion for the gauge field strength can be obtained by varying
action (\ref{eq:47}), then the electric charge $d$ and current $j$
can be achieved by their equations of motions as they are in (\ref{eq:51})
(\ref{eq:52}). As a result, the exact formula for $\xi_{\parallel,\perp}$
in the bubble background is computed as follows,

\begin{align}
\xi_{\parallel} & =\frac{1-\left(2\pi\alpha^{\prime}E_{\parallel}\right)^{2}\frac{u^{4}}{L^{4}}}{1+\frac{e^{3\phi}u^{6}}{\mathcal{F}\mathcal{B}\mathcal{Z}^{3}L^{4}}\left(d^{2}-j^{2}\right)},\nonumber \\
\xi_{\perp} & =\frac{1-\left(2\pi\alpha^{\prime}E_{\perp}\right)^{2}\frac{u^{4}}{L^{4}}e^{\phi}}{1+\frac{e^{3\phi}u^{6}}{\mathcal{F}\mathcal{B}\mathcal{Z}^{3}L^{4}}\left(d^{2}-e^{-\phi}j^{2}\right)}.
\end{align}
Therefore the critical position $u=u_{*}$ must be determined by the
constraints,

\begin{align}
1-\left(2\pi\alpha^{\prime}E_{\parallel}\right)^{2}\frac{u_{*}^{4}}{L^{4}} & =0,\nonumber \\
1+\frac{e^{3\phi\left(u_{*}\right)}u_{*}^{6}}{\mathcal{F}\left(u_{*}\right)\mathcal{B}\left(u_{*}\right)\mathcal{Z}^{3}\left(u_{*}\right)L^{4}}\left(d^{2}-j^{2}\right) & =0,\label{eq:58}
\end{align}
in the parallel case and,

\begin{align}
1-\left(2\pi\alpha^{\prime}E_{\perp}\right)^{2}\frac{u_{*}^{4}}{L^{4}}e^{\phi\left(u_{*}\right)} & =0,\nonumber \\
1+\frac{e^{3\phi\left(u_{*}\right)}u_{*}^{6}}{\mathcal{F}\left(u_{*}\right)\mathcal{B}\left(u_{*}\right)\mathcal{Z}\left(u_{*}\right)^{3}L^{4}}\left[d^{2}-e^{-\phi\left(u_{*}\right)}j^{2}\right] & =0,\label{eq:59}
\end{align}
in the perpendicular case. Afterwards, the vacuum decay rate can be
obtained by evaluating the imaginary part of the effective Lagrangian
in (\ref{eq:47}) which is given as,

\begin{align}
\mathrm{Im}\mathcal{L}_{\parallel} & =\int_{u_{*}}^{u_{KK}}due^{-\frac{3}{2}\phi}\frac{\mathcal{Z}^{3/2}\mathcal{B}^{1/2}}{u^{5}}\sqrt{\left[\left(2\pi\alpha^{\prime}E_{\parallel}\right)^{2}\frac{u^{4}}{L^{4}}-1\right]\left(1+\frac{e^{3\phi}u^{6}}{\mathcal{F}\mathcal{B}\mathcal{Z}^{3}L^{4}}d^{2}\right)^{-1}},\nonumber \\
\mathrm{Im}\mathcal{L}_{\perp} & =\int_{u_{*}}^{u_{KK}}due^{-\frac{3}{2}\phi}\frac{\mathcal{Z}^{3/2}\mathcal{B}^{1/2}}{u^{5}}\sqrt{\left[\left(2\pi\alpha^{\prime}E_{\perp}\right)^{2}\frac{u^{4}}{L^{4}}e^{\phi}-1\right]\left(1+\frac{e^{3\phi}u^{6}}{\mathcal{F}\mathcal{B}\mathcal{Z}^{3}L^{4}}d^{2}\right)^{-1}}.\label{eq:60}
\end{align}
Solving the constraint of $E_{\parallel,\perp}$ and $u_{*}$ in (\ref{eq:58})
(\ref{eq:59}), we plot out the behavior of $\mathrm{Im}\mathcal{L}_{\parallel,\perp}$
in (\ref{eq:60}) as a function of $E_{\parallel,\perp}$ in Figure
\ref{fig:12} and \ref{fig:13}. 
 
 Obviously, the numerical calculation shows us a critical value of
electric field when the bubble background is imposed which is distinct
to the calculations in the black brane background. The reason is that
the bubble background as a soliton solution corresponds to the confinement
in the dual theory \cite{key-29,key-30}. So due to the confinement
below $M_{KK}$, there should not be agreement with the calculations
in the black brane background at small $E_{\parallel,\perp}$. Besides,
Figure \ref{fig:12} and \ref{fig:13} also displays the similar behavior
of the vacuum decay rate depending on the anisotropy as the black
brane case, so it again agrees with our holographic investigation
in Section 3 and 4.

\subsection{The V-A curve}

\begin{figure}[H]
\begin{centering}
\includegraphics[scale=0.34]{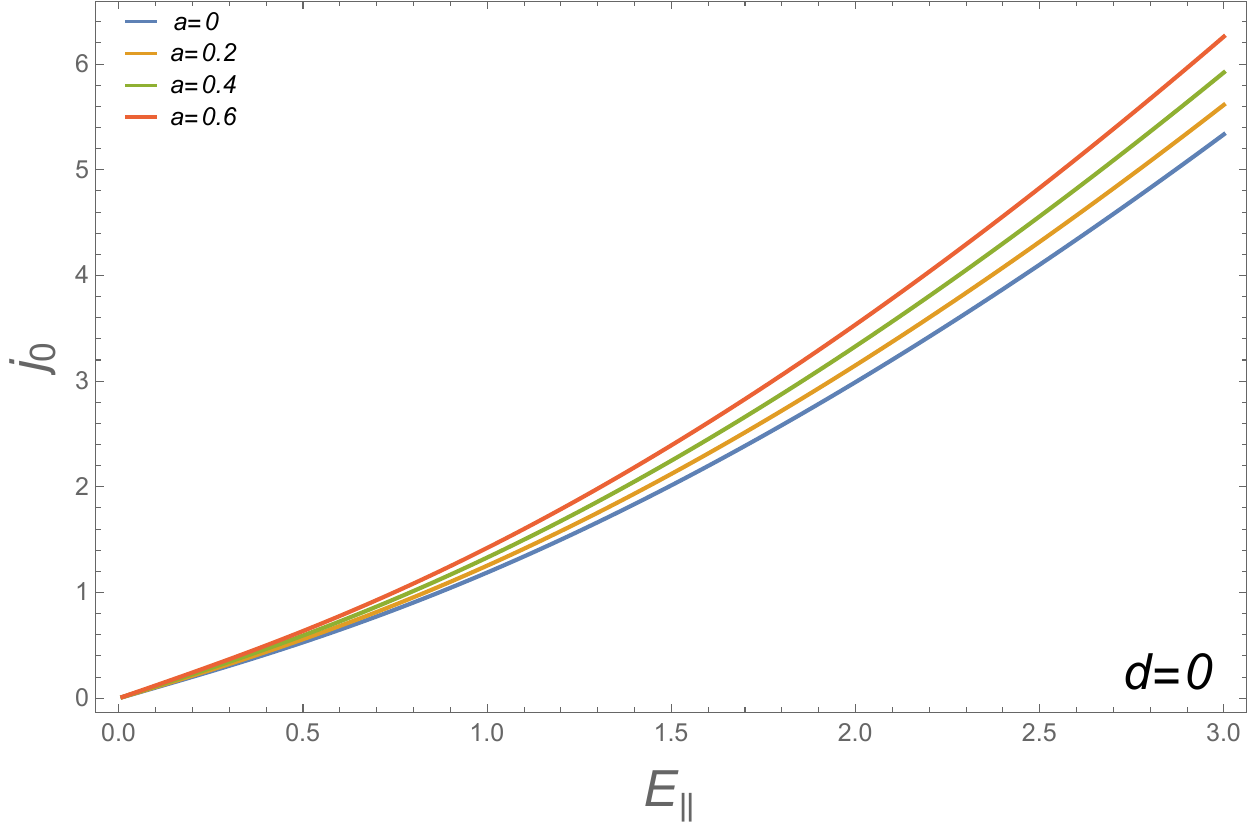}\includegraphics[scale=0.34]{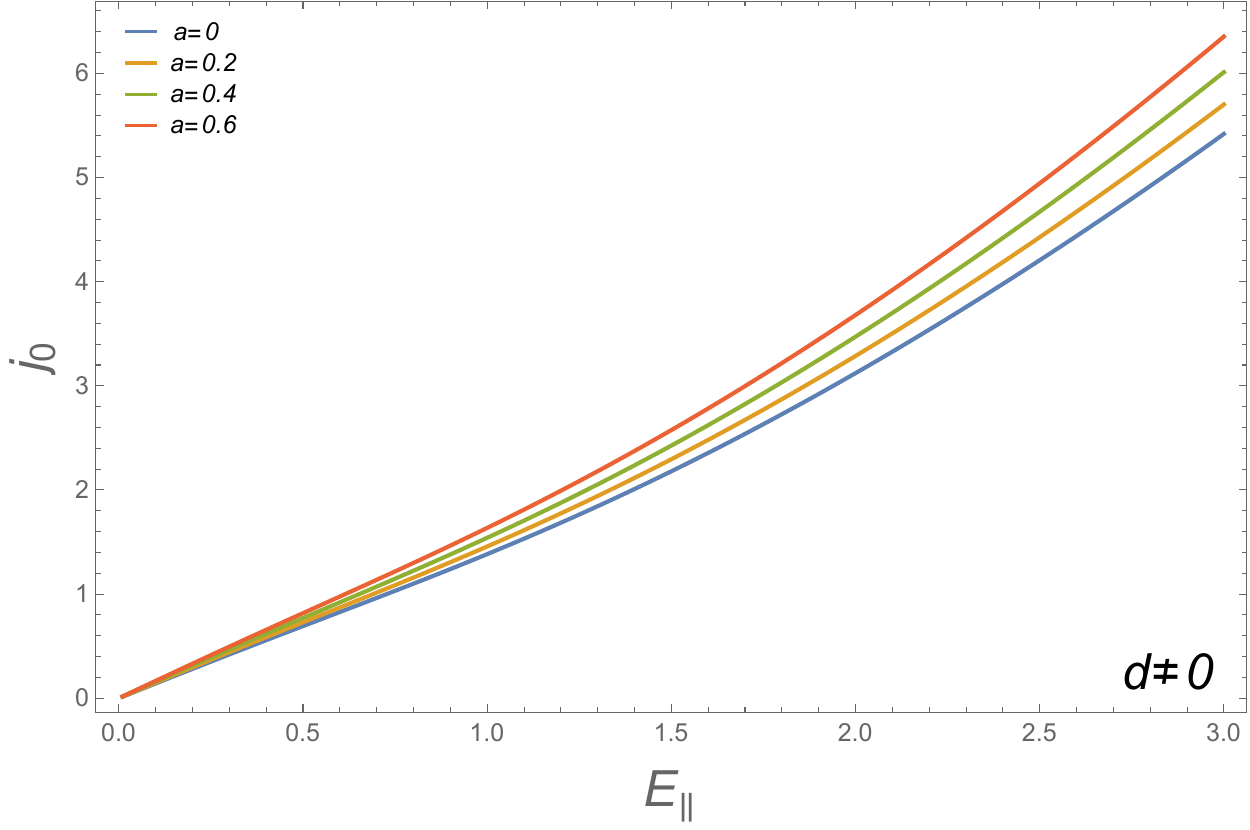}
\par\end{centering}
\begin{centering}
\includegraphics[scale=0.34]{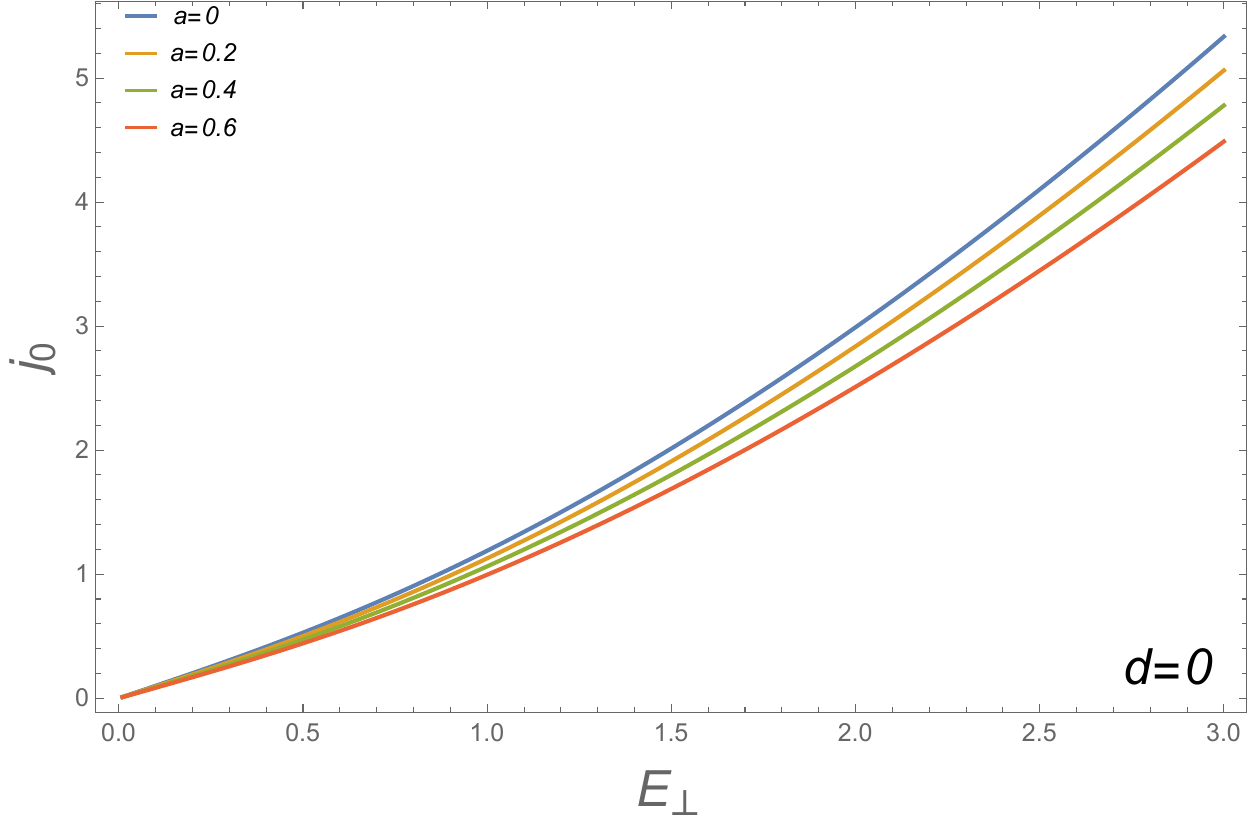}\includegraphics[scale=0.34]{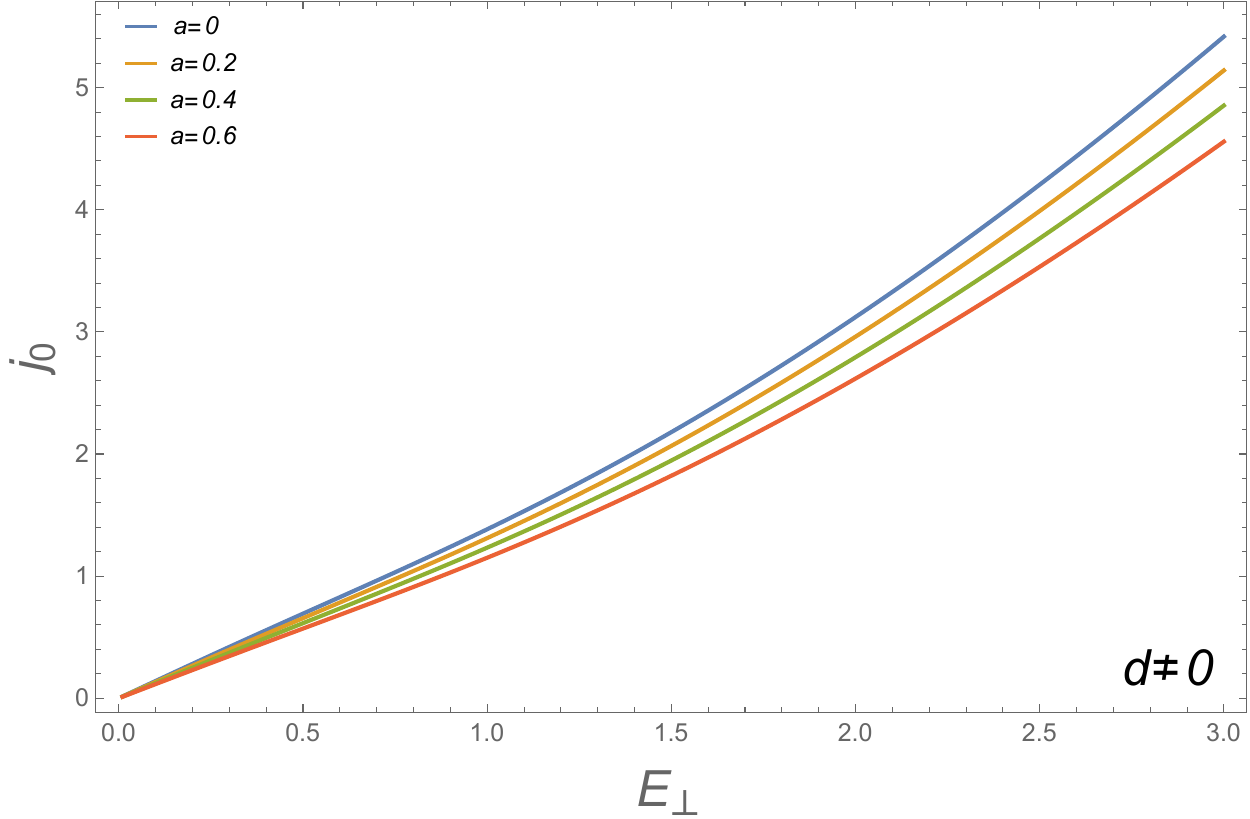}
\par\end{centering}
\caption{\label{fig:14} Relation of $j_{0}$ and electric field $E_{\parallel,\perp}$
with various $a$ in the black brane background.}
\end{figure}

\begin{figure}[H]
\begin{centering}
\includegraphics[scale=0.34]{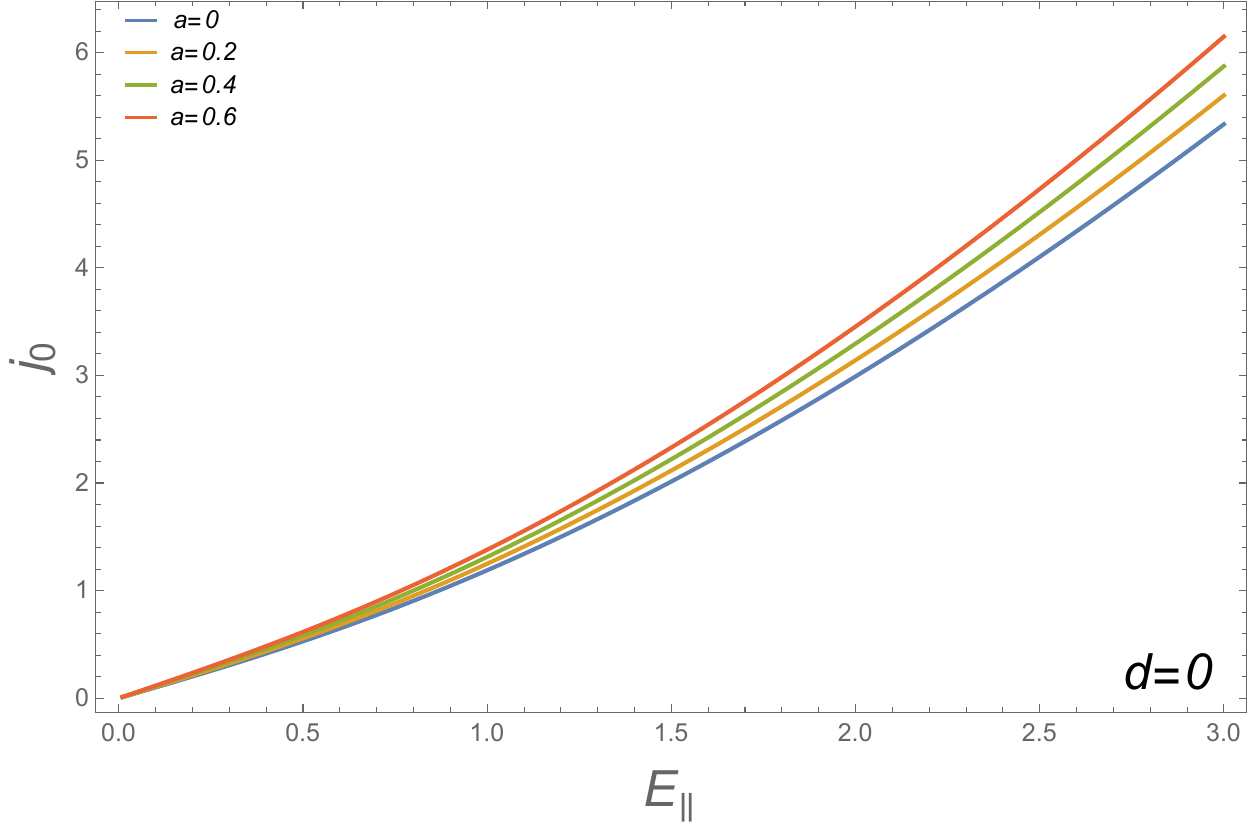}\includegraphics[scale=0.34]{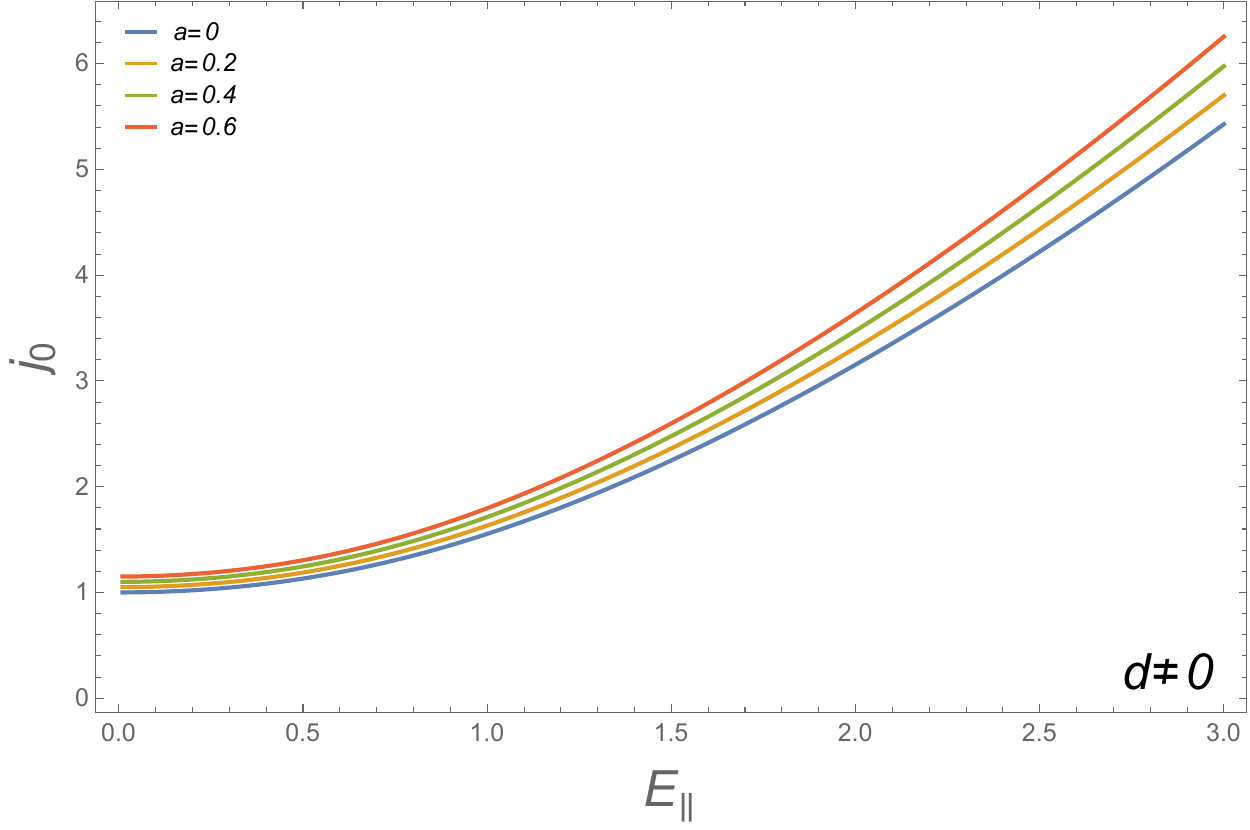}
\par\end{centering}
\begin{centering}
\includegraphics[scale=0.34]{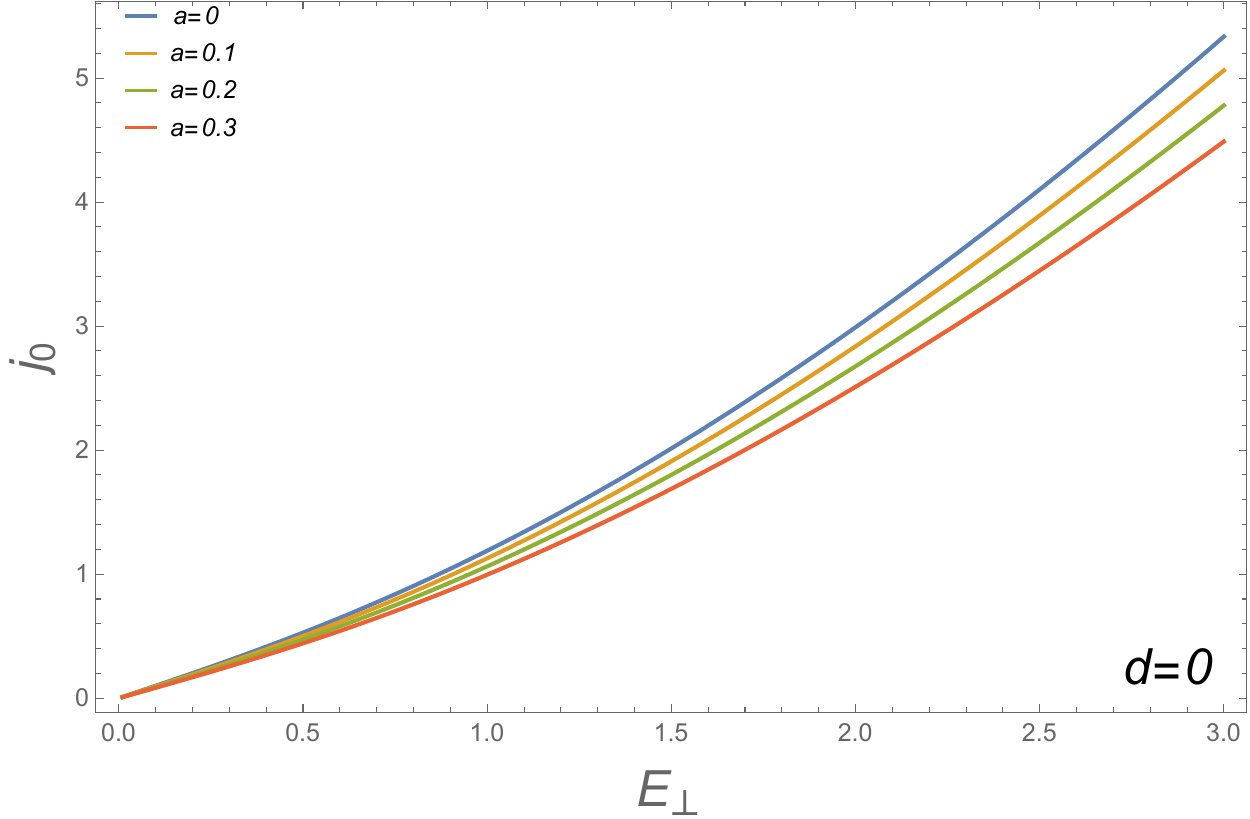}\includegraphics[scale=0.34]{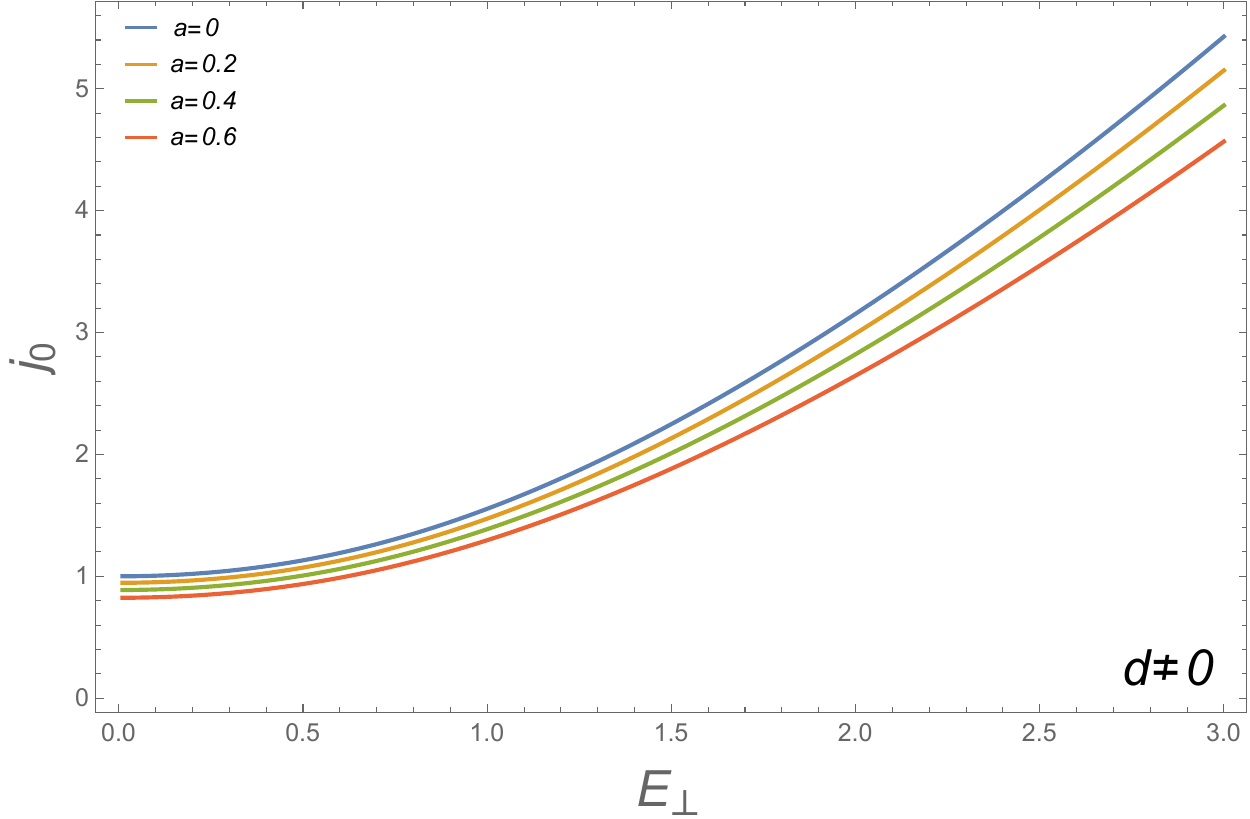}
\par\end{centering}
\caption{\label{fig:15} Relation of $j_{0}$ and electric field $E_{\parallel,\perp}$
with various $a$ in the bubble background.}
\end{figure}

In this section, let us study the V-A curve in this holographic system.
Since the relation of charge $d$, current $j$, critical position
$u_{*}$ and the electric field $E_{\parallel,\perp}$ is constrained
by (\ref{eq:55}) (\ref{eq:56}) (\ref{eq:58}) (\ref{eq:59}), we
solve these equations numerically and respectively by eliminating
$u_{*}$ in order to obtain the relation of current $j$ and electric
field $E_{\parallel,\perp}$. The numerical results are collected
in Figure \ref{fig:14} and \ref{fig:15} in which the stable electric
current $j=j_{0}$ is a function of the electric field $E_{\parallel,\perp}$
with various $a$ and fixed $d,u_{H,KK}$. 
 
 We can see both in the black brane and bubble side, the behavior
of the conductivity $\sigma_{\parallel,\perp}=j_{0}/E_{\parallel,\perp}$
trends to become constant at large electric field which reveals the
feature of DC conductor. And the presence of the anisotropy increases/decreases
$\sigma_{\parallel,\perp}$ respectively which is in qualitative agreement
with the phenomenologically holographic approach in this anisotropic
background \cite{key-27}\footnote{Notice that the notation of ``$\parallel$'' and ``$\perp$''
in our manuscript is opposite to them in \cite{key-27}. In our manuscript,
``$\parallel$'' and ``$\perp$'' refer to that the physical variables
is parallel/perpendicular to the dissolved $N_{\mathrm{D7}}$ D7-branes
in the bulk while they refer to the variables is parallel/perpendicular
to the anisotropic direction in \cite{key-27}.}. However at small $E_{\parallel,\perp}$, our system is nearly insulated
with charge density in bubble side which is consistent with its feature
of confinement.

\section{Summary and discussion}

In this work, we study the Schwinger effect and vacuum electric instability
in the anisotropic IIB supergravity via holography. In gravity side,
two anisotropic gravity solutions are taken into account which are
respectively the black brane solution and the bubble solution (as
the soliton solution in gravity theory). The dual theory in the black
brane background is analyzed at finite temperature while it is expected
to be a three-dimensional confining theory at zero temperature in
the bubble background. Using the AdS/CFT dictionary, we derive the
critical electric field, the separation, the behavior of the holographic
potential and the associated quark pair production both in the the
black brane and the bubble background. However due to the anisotropy
in the background, we also deal with the derivation in the parallel
and perpendicular case respectively. Furthermore, when the D7-brane
as flavor is introduced to the bulk geometry, it is possible to investigate
the vacuum electric instability and the vacuum decay rate for Schwinger
effect by calculating the imaginary part of the effective Lagrangian.
In this sense, we solve the constraints of the charge, current, electric
field then obtain the V-A curve and the conductivity in this holographic
system. Since the analytical formulas of the background geometry is
imposed, our calculation is valid in the limit of high temperature
(black brane case), or in the exactly three-dimensional confining
theory (bubble case).

Overall, our numerical calculation displays consistent results. First,
for fixed electric field, the potential analysis reveals the potential
barrier in parallel case is suppressed by the anisotropy while it
is enhanced in the perpendicular case. And this behavior is in partly
qualitative agreement with the numerical evaluation in the bottom-up
holographic approach of the anisotropic Schwinger effect in \cite{key-21+1}.
Correspondingly, in our current work, the associated quark production
rate increases/decreases in the parallel/perpendicular case, since
potential barrier always hinders the pair production. And this conclusion
agrees with the behavior of the quark potential in our anisotropic
background studied in \cite{key-21}, in which the quark tension increases/decreases
by the presented anisotropy in the perpendicular/parallel case. Second,
when the flavor brane is introduced, we find that while the vacuum
decay rate behaves differently at small electric field, it indicates
similar behaviors of the dependence on the anisotropy as the quark
production rate in the perpendicular/parallel case respectively. So
we believe this is a parallel approach to verify the dependence on
the anisotropy of the production rate via holography. Finally the
V-A curve shows us a straightforward conclusion that is $\sigma_{\perp}/\sigma_{\parallel}$
is always smaller than one with the anisotropy. This conclusion is
in qualitative agreement with the phenomenologically holographic approach
\cite{key-27} in this system due to the interpretation of the transport
property of anisotropic matters. 

Additionally, in this framework, we can obtain the ratio of the perpendicular
and parallel pressure in the dual theory is also smaller than one
i.e. $P_{\perp}/P_{\parallel}<1$ according to (\ref{eq:13}) or \cite{key-20}
at very high temperature or in the associated three-dimensional confining
theory. Comparing $P_{\perp}/P_{\parallel}<1$ with $\sigma_{\perp}/\sigma_{\parallel}<1$
, it may lead to the statement that the pressure in very hot plasma
or dense three-dimensional QCD may be roughly proportionally related
to the conductivity. Afterwards combining the above conclusions together,
we may find a consistent interpretation for the behaviors of the holographic
potential and conductivity: Under the condition $a/T,a/M_{KK}\ll1$
with anisotropy, the parallel direction may be more conductive than
the perpendicular direction due to $\sigma_{\perp}/\sigma_{\parallel}<1$
(i.e. the conductivity in the perpendicular/parallel direction is
decreased/increased by the anisotropy respectively). Accordingly,
it implies the virtual pairs of charged particles would be easier
to be pulled apart to become real charged particles in the parallel
direction than in the perpendicular direction under an external electric
field. Therefore, the Schwinger effect with parallel electric field
would be more likely to appear than with perpendicular electric field
intuitively due to the presence of the anisotropy, which is embodied
consistently as that the barrier of the holographic potential is decreased/increased
with  parallel/perpendicular electric field by the anisotropy because
the potential barrier always hinders the Schwinger effect. In this
sense, all the behaviors of the holographic potential in the presence
of the anisotropy are basically due to the anisotropic bulk metric
in our holographic setup. Thus we believe this holographic approach
could provide a very powerful way to explore the non-perturbative
Schwinger effect in the presence of anisotropy.

\section*{Acknowledgements}

We would like to thank Xun Chen and Jing Zhou for helpful discussion.
This work is supported by the National Natural Science Foundation
of China (NSFC) under Grant No. 12005033, the research startup foundation
of Dalian Maritime University in 2019 under Grant No. 02502608 and
the Fundamental Research Funds for the Central Universities under
Grant No. 3132022198.

\section*{Appendix: The analytical formulas for the anisotropic background}

In general, the functions $\mathcal{F},\mathcal{B},\phi$ presented
in (\ref{eq:2}) and (\ref{eq:5}) are non-analytical functions which
must be determined by the equations of motion associated to the supergravity
action (\ref{eq:1}). The details to solve $\mathcal{F},\mathcal{B},\phi$
numerically can be reviewed in \cite{key-20}. Here we only collect
the relevant formulas used in this work. We focus on the analytical
formulas of $\mathcal{F},\mathcal{B},\phi$ which can be obtained
in the high temperature limit $T\rightarrow\infty$, or $\delta t_{E}\sim T^{-1}\rightarrow0$
in the black brane solution (\ref{eq:2}) as (up to the leading order
terms of $a$),

\begin{align}
\mathcal{F}\left(u\right) & =1-\frac{u^{4}}{u_{H}^{4}}+a^{2}\hat{\mathcal{F}}_{2}\left(u\right)+\mathcal{O}\left(a^{4}\right),\nonumber \\
\mathcal{B}\left(u\right) & =1+a^{2}\hat{\mathcal{B}}_{2}\left(u\right)+\mathcal{O}\left(a^{4}\right),\nonumber \\
\phi\left(u\right) & =a^{2}\hat{\phi}_{2}\left(u\right)+\mathcal{O}\left(a^{4}\right),\tag{A-1}\label{eq:61}
\end{align}
where

\begin{align}
\hat{\mathcal{F}}_{2}\left(u\right) & =\frac{1}{24u_{H}^{2}}\left[8u^{2}\left(u_{H}^{2}-u^{2}\right)-10u^{4}\log2+\left(3u_{H}^{4}+7u^{4}\right)\log\left(1+\frac{u^{2}}{u_{H}^{2}}\right)\right],\nonumber \\
\hat{\mathcal{B}}_{2}\left(u\right) & =-\frac{u_{H}^{2}}{24}\left[\frac{10u^{2}}{u_{H}^{2}+u^{2}}+\log\left(1+\frac{u^{2}}{u_{H}^{2}}\right)\right],\nonumber \\
\hat{\phi}_{2}\left(u\right) & =-\frac{u_{H}^{2}}{4}\log\left(1+\frac{u^{2}}{u_{H}^{2}}\right).\tag{A-2}\label{eq:62}
\end{align}
Here $u_{H}$ refers to the event horizon in the black brane solution.
We note that the functions in (\ref{eq:61}) ( \ref{eq:62}) are in
fact a series of $a/T$ once the relation of $u_{H}$ and $T$ is
imposed. So the high temperature analysis of $\mathcal{F},\mathcal{B},\phi$
refers to the condition $T\gg a$ in the original solution exactly. 

Since the bubble background (\ref{eq:5}) is obtained under the double
Wick rotation, it would reduce to the replacement $T\rightarrow M_{KK}/\left(2\pi\right),\delta t_{E}\rightarrow\delta z$
where $\delta z$ refers to the period of $z$. Note that in the bubble
background we would replace $u_{H}$ by $u_{KK}$ because it now refers
to the bottom of the Cigar-like bulk instead of a horizon. In this
sense, the high temperature limit $T\rightarrow\infty$, or $\delta t_{E}\sim T^{-1}\rightarrow0$
in the original black brane solution corresponds to the limit of dimension
reduction $M_{KK}\rightarrow\infty$ or $\delta z\rightarrow0$ in
the bubble solution. Namely, in this limit, the size of the compactified
direction $z$ trends to be zero in (\ref{eq:5}). In a word, by replacing
$u_{H}$ by $u_{KK}$, if we remain to employ analytical formulas
of $\mathcal{F},\mathcal{B},\phi$ given in (\ref{eq:61}) (\ref{eq:62})
in the bubble background, the dual theory is effectively three-dimensional
below $M_{KK}$ with $M_{KK}\rightarrow\infty$ which means the dual
theory would become exactly a three-dimensional theory. And in this
work, we use the above formulas for functions $\mathcal{F},\mathcal{B},\phi$,
thus it means all our calculations are valid with the anisotropy of
$\mathcal{O}\left(a^{2}\right)$ under the condition $T,M_{KK}\gg a$.

\end{document}